\documentclass{aa}
\usepackage{graphicx}
\usepackage{natbib}
\usepackage{aalongtable}
\usepackage{txfonts}

\def\NrPulsarsSourceList{191 }
\def\NrPulsars{187 } 
\def\NrNewDrifters{42}
\def\NrConvinedPulsars{30 }
\def\NrPulsarsModulation{170}
\def\NrDrifters{68 }
\def\NrCandidates{6 }
\def\NrPulsarsSNR{106 }
\def\NrDriftersSNR{57 }
\def\NrCandidatesSNR{5 }
\def\DriftPercentageSNR{54}
\def\CandidatePercentageSNR{5}

\def\AgeDriftNonDriftperc{0.03\% }
\def\AgeCohNondriftperc{0.4\% }
\def\AgeCohNoCohperc{8\% }
\def\BNonDriftperc{50\% }
\def\BNonCohpers{20\% }

\def\MMinDrNonDr{50\% }
\def\MMinCohNonDr{6\% }
\def\MMinCohNonCoh{5\% }

\def\AgeDriftNonDriftpercALL{0.03\% }
\def\AgeCohNondriftpercALL{0.1\% }

\newcommand{\be}{\begin{equation}}
\newcommand{\ee}{\end{equation}}

\newcommand{\degrees}[1]{\ensuremath{#1^\circ}}

\def\floatsep{30pt}            % V-space between floats
\def\textfloatsep{30pt}        % V-space between floats and text [tb]
\def\intextsep{30pt}           % V-space between floats and text [h]

\begin{document}

\title{The subpulse modulation properties of pulsars at 21 cm}
\titlerunning{The subpulse modulation properties of pulsars at 21 cm}
\author{P. Weltevrede\inst{1} \and R.~T. Edwards\inst{1,3} \and B.~W. Stappers\inst{2,1}}

\date{}
\institute{Astronomical Institute ``Anton Pannekoek'', 
        University of Amsterdam,
        Kruislaan 403, 1098 SJ Amsterdam, The Netherlands 
  \and
   Stichting ASTRON, Postbus 2, 7990 AA Dwingeloo, The Netherlands
  \and 
   CSIRO Australia Telescope National Facility, PO Box 76, Epping NSW 1710,  Australia    }
\offprints{\\P. Weltevrede, \email{wltvrede@science.uva.nl}}

\abstract{ We present the results of a systematic, unbiased search for
subpulse modulation of \NrPulsars pulsars performed with the
Westerbork Synthesis Radio Telescope (WSRT) in the Netherlands at an
observing wavelength of 21 cm. Using new observations and archival
WSRT data we have expanded the list of pulsars which show the drifting
subpulse phenomenon by \NrNewDrifters, indicating that at least one in
three pulsars exhibits this phenomenon. The real fraction of pulsars
which show the drifting phenomenon is likely to be larger than some
55\%. The majority of the analysed pulsars show subpulse modulation
(\NrPulsarsModulation), of which the majority were not previously
known to show subpulse modulation and \NrConvinedPulsars show clear
systematic drifting.  The large number of new drifters we have found
allows us, for the first time, to do meaningful statistics on the
drifting phenomenon. We find that the drifting phenomenon is
correlated with the pulsar age such that drifting is more likely to
occur in older pulsars. Pulsars which drift more coherently seem to be
older and have a lower modulation index. There is no significant
correlation found between $P_3$ and other pulsar parameters (such as
the pulsar age), as has been reported in the past. There is no
significant preference of drift direction and the drift direction is
not found to be correlated with pulsar parameters.
None of the four complexity parameters predicted by different emission
models (\citealt{jg03}) are shown to be inconsistent with the set of
modulation indices of our sample of pulsars. Therefore none of the
models can be ruled out based on our observations. We also present
results on some interesting new individual sources like a pulsar which
shows similar subpulse modulation in both the main- and interpulse and
six pulsars with opposite drift senses in different components.

 }

\maketitle

\section{Introduction}

Despite the fact that explaining the emission mechanism of radio
pulsars has proved very difficult, this field has the advantage that
we have very detailed knowledge about the emission mechanism from
observations. We know from the very high observed brightness
temperatures that the radio emission must be coherent, we know what
kind of magnetic field strengths are involved and even the orientation
of the magnetic axis, rotation axis and the line of sight can be
derived from observations. Furthermore if one can detect single pulses
one can see that the pulses of some pulsars consist of subpulses and
for some pulsars these subpulses drift in successive pulses in an
organized fashion through the pulse window (\citealt{dc68,sspw70}). If
one plots a so-called ``pulse-stack'', a plot in which
successive pulses are displayed on top of one another, the
drifting phenomenon causes the subpulses to form ``drift bands'' (an
example is shown in the left panel of Fig. \ref{Classes_fig}). This
complex, but highly regular intensity modulation in time is known in
great detail for only a small number of well studied pulsars.

Because the properties of the subpulses are most likely determined by
the emission mechanism, we learn about the physics of the emission
mechanism by studying them. That drifting is linked with the emission
mechanism is suggested by the fact that drifting is affected by
``nulls'' (e.g. \citealt{th71,vkr+02,jv04}), where nulling is the
phenomenon whereby the emission mechanism is switched off for a number
of successive pulses. Another complex phenomenon is drift mode changes
where the drift rate switches between a number of discrete values. For
some pulsars there are observationally determined rules describing
which drift mode changes are allowed from which drift mode
(e.g. \citealt{wf81,rwr05}). It has been found that the nulls of PSR
B2303+30 are confined to a particular drift mode (\citealt{rwr05}),
which further strengthens the link between drifting and the emission
mechanism.

Another characteristic feature of the emission mechanism is that when
one averages the individual pulses, the resulting pulse profile is
remarkably stable over time (\citealt{hmt75}). Explaining the various
shapes of the pulse profiles of different pulsars and their
dependence on observing frequency has proven to be very complicated,
so not surprisingly an explanation that is fully
consistent with the overwhelmingly detailed complex behavior of
individual (sub)pulses, the nulling phenomenon and the
polarization of individual pulses (e.g. \citealt{edw04}) seems to be
far away.  In this paper we describe trends of the subpulse modulation
we find for a large sample of pulsars. By doing this we determine
observationally what the important physical parameters are for
subpulse modulation, which could help formulating an emission model
which is fully consistent with the observations.

There are a few types of models that attempt to explain the drifting
phenomenon. The most well known model is the sparking gap model
(\citealt{rs75}), which has been extended by many authors
(e.g. \citealt{cr80,fr82,gs00,gmg03,qlz+04}) making it the most
developed model for explaining the drifting phenomenon.  These models
explain the drifting phenomenon by the generation of the radio
emission via a rotating ``carousel'' of discharges which circulate
around the magnetic axis due to an $\mathbf{E}\times\mathbf{B}$
drift. In the carousel model it is expected that all pulsars should
have some sort of circulation time. For PSR B0943+10
(\citealt{dr99,dr01,ad01}) and possibly PSR B0834+06 (\citealt{ad05})
a tertiary subpulse modulation feature has been detected from the
fluctuation properties and viewing geometry. This periodicity has been
interpreted as related to the carousel modulation period (i.e. the
circulation time $\hat{P}_3$), supporting the interpretation of the
drifting subpulses being caused by a rotating carousel of
sub-beams. The circulation times of these pulsars, as well as the more
indirectly derived circulation times of PSR B0809+74
(\citealt{vsr+03}) and PSR B0826$-$34 (\citealt{ggk+04}) are
consistent with the sparking gap model (\citealt{gmg03}). A different
geometry of the polar cap of PSR B0826$-$34 is proposed by
\citealt{elg+05}. In their interpretation the carousel changes drift
direction, something what would be inconsistent with the sparking gap
model.

These models still have problems, like explaining the subpulse phase
steps which are observed for some pulsars. Two clear examples of
pulsars that show subpulse phase steps are PSR B0320+39 and PSR
B0809+74 as found by \cite{esv03} and \cite{es03c}. We find that the
new drifter PSR B2255+58 also shows a phase step.

Non-radial pulsations of neutron stars were originally proposed as the
origin of the radio pulses of pulsars (\citealt{rud68}) and later as a
possible origin of the drifting subpulses (\citealt{dc68}). Recently
this idea was revised by \cite{cr04}.  This model gives a natural
explanation for observed subpulse phase steps, nulls and mode
changes. This model can be tested, although there are many
complications, by exploring average beam geometries. Although this
model can explain phase steps, it cannot explain the curvature of the
drift bands of many pulsars (see Sect. \ref{SubpulsePhase} for
details). In this model it is also difficult to explain pulsars with
opposite drift senses in different components, because drifting is a
simply a beat between the pulse period and the pulsation
time. Bi-drifting is recently observed for PSR J0815+09
(\citealt{mlc+04}). In this paper we show a number of other pulsars
with opposite drift senses in different components\footnote{PSRs
B0450+55, B1540$-$06, B0525+21, B1839$-$04, B2020+28, the outer
components of B0329+54 and possibly PSR B0052+51. Also PSR B1237+25 is
a known example.}. For PSR B1839$-$04 we observe that the two
components have mirrored drift bands (i.e. the components drift in
phase) like PSR J0815+09, something we do not know for the other
pulsars.  In the sparking gap model bi-drifting can be explained if
these pulsars have both an inner annular gap and an inner core gap
(\citealt{qlz+04}).

A feedback model is proposed by \cite{wri03} as a natural mechanism
for both the sometimes regular and sometimes chaotic appearance of
subpulse patterns. In this model the outer magnetosphere interacts
with the polar cap and the observed dependency of conal type on pulse
period (\citealt{ran93}) and angle between the rotation and magnetic
axis (\citealt{ran90}) follows naturally.

Up to now most observational literature on the drifting phenomenon has
been focused on describing individual very interesting drifting
subpulse pulsars. The focus of this paper will not only be the
individual systems, but also the properties common to the pulsars that
show drifting, an approach started by \cite{bac81}, \cite{ash82} and
\cite{ran86}. In the work of \cite{bac81} 20 pulsars were studied for
their subpulse behavior at 430 MHz and 9 were observed to be
drifting. In the work of \cite{ash82} the single pulse properties of
nine new drifters are described and the properties common to 28
drifters in a sample of 52 pulsars are analysed. This sample consists
of both their own results and a few previously published results. Most
observations were obtained at or near 400 MHz, but some at higher
frequencies.

In the work of \cite{ran86} all the, then published, single pulse
properties are combined and described in the light of her empirical
theory. Because understanding the drifting phenomenon is considered
important for unraveling the mysteries of the emission mechanism
of radio pulsars, we decided that it was time to start this more
general and extensive observational program on the drifting
phenomenon.

The main goals of this unbiased search for pulsar subpulse modulation
is to determine what percentage of the pulsars show the drifting
phenomenon and to find out if these drifters share some physical
properties. As a bonus of this observational program new individually
interesting drifting subpulse systems are found. In this paper we
focus on the 21 cm observations and in a subsequent paper we will
focus on lower frequency observations and the frequency dependence of
the subpulse modulation properties of radio pulsars.

The list of pulsars which show the drifting phenomenon is slowly
expanding in time as more sufficiently bright pulsars are found by
surveys (e.g. \citealt{lwf+04}), but we have successfully chosen a
different approach to expand this list much more rapidly. The reason
that we have found so many new drifting subpulse systems is twofold:
we have analyzed a large sample of pulsars of which many were not
known to show this phenomenon, and we used a sensitive detection
method. Previous studies of drifting subpulses often used tracking of
individual subpulses through time, an analysis method that requires a
high signal-to-noise (S/N) ratio because it requires the detection of
single pulses. This automatically implies that this kind of analysis
can only be carried out on a limited number of pulsars. Analyzing the
integrated Two-Dimensional Fluctuation Spectrum (2DFS; \citealt{es02})
and the Longitude-Resolved Fluctuation Spectrum (LRFS;
\citealt{bac70b}) allows us to detect drifting subpulses even when the
S/N is too low to see single pulses. This method was already
successfully used with archival Westerbork Synthesis Radio Telescope
(WSRT) data by \cite{es03b} to find drifting subpulses in millisecond
pulsars.

By using the technique described above combined with the high
sensitivity of the WSRT we have analyzed a large sample of \NrPulsars
pulsars. An important aspect when calculating the statistics of
drifting is that one has to be as unbiased as possible, so we have
selected our sample of pulsars based only on the predicted S/N in a
reasonable observing time. While this sample is obviously still
luminosity biased, it is not biased towards well-studied pulsars,
pulse profile morphology or any particular pulsar characteristics as
were previous studies (e.g.  \citealt{ash82}, \citealt{bac81} and
\citealt{ran86} and references therein).  Moreover, all the
conclusions in this paper are based on observations at a single
frequency.

The paper is organized such that we start by explaining the technical
details of the observations and data analysis. After that the details
of the individual detections are described and in table
\ref{Table_section} all the details of our measurements can be
found. After the individual detections the statistics of the drifting
phenomenon are discussed followed by the summary and conclusions. In
appendix \ref{Figures_ref} are the plots for all the pulsars in our
source list. They can also be found in appendix \ref{Figures_ref2},
but there they are ordered by appearance in the text. Note that the
astro-ph version is missing the appendices due to file size
restrictions. Please download appendices from {\tt
http://www.science.uva.nl/$\sim$wltvrede/21cm.pdf}.

\section{Observations and data analysis}

\subsection{Source list}

All the analyzed observations were collected with the WSRT in the
Netherlands. The telescope is located at a latitude of \degrees{52}.9
in the north, meaning that not all pulsars are visible for the WSRT.
Only catalogued\footnote{\tt
http://www.atnf.csiro.au/research/pulsar/psrcat/ \rm} pulsars with a
declination (J2000) above -\degrees{30} were included in our source
list.

This list of pulsars that are visible to the WSRT was sorted on the
observation duration required to achieve a signal-to-noise (S/N) ratio
of 130. Of this list we selected the first \NrPulsarsSourceList
pulsars, which required observations less then half an hour in
duration. The S/N ratio of a pulsar observation can be predicted with
the following equation (\citealt{dtws85})
\begin{equation}
\label{SNR_formula}
\mathrm{S/N}=\frac{\eta_Q S
  G}{T_\mathrm{sys}+T_\mathrm{sky}}\sqrt{\frac{\Delta\nu\,
    t_\mathrm{obs}n_p(P_0-w)}{w}}
\end{equation}
where $\eta_Q$ is the digitization efficiency factor, $S$ the mean
flux density of the pulsar, $G$ the gain of the telescope,
$T_\mathrm{sys}$ the system temperature, $T_\mathrm{sky}$ the sky
temperature, $\Delta\nu$ the bandwidth of the pulsar backend,
$t_\mathrm{obs}$ the observation duration, $n_p$ the number of
polarizations that are recorded, $P_0$ the barycentric pulse period of
the pulsar and $w$ the width of the pulse profile.

All observations were conducted with the 21 cm backend at WSRT, which
has the following receiver system parameters: $\eta_Q=1$, $G=1.2$
K/Jy, $T_\mathrm{sys}=27$ K, $T_\mathrm{sky}=6$ K (which is the
average of the entire sky), $\Delta\nu=80$ MHz and $n_p=2$.  It is
required that the pulsars have an integrated pulse profile with a
predicted S/N ratio of 130, so the required observation duration in seconds is
\begin{equation}
t_\mathrm{obs} \geq \left(\frac{283\;\mathrm{mJy}}{S_{1400}}\right)^2\frac{w}{P_0-w}
\end{equation}
where $S_{1400}$ is the flux of the pulsar at our observation
frequency of 1400 MHz and $w$ the FWHM of the pulse profile. Those
pulsars lacking the necessary parameters ($S_{1400}$ and $w$) in the
catalog were excluded from the sample, because in such cases it
was not possible to evaluate $t_\mathrm{obs}$.

The sensitivity to detect drifting subpulses does not only depend on
the S/N ratio of the observation, but also on obtaining a large
number of pulses. This is because the observation should contain
enough drift bands to be able to identify the drifting phenomenon. Our
second requirement on the minimum observation length was therefore
that the observations should contain at least one thousand pulses, so
some pulsars had to be observed for longer than was required to get a
S/N ratio of 130. To make sure that the statistics on the
drifting phenomenon is not biased on pulse period, it is important to
include these long period pulsars in the source list.

\begin{figure*}[htb]
%%\resizebox{0.99\hsize}{!}{\includegraphics[angle=0]{stack4.ps}}
%%%\resizebox{0.49\hsize}{!}{\includegraphics[angle=0]{stack4.eps}}
\rotatebox{270}{\resizebox{0.5195\hsize}{!}{\includegraphics[angle=0]{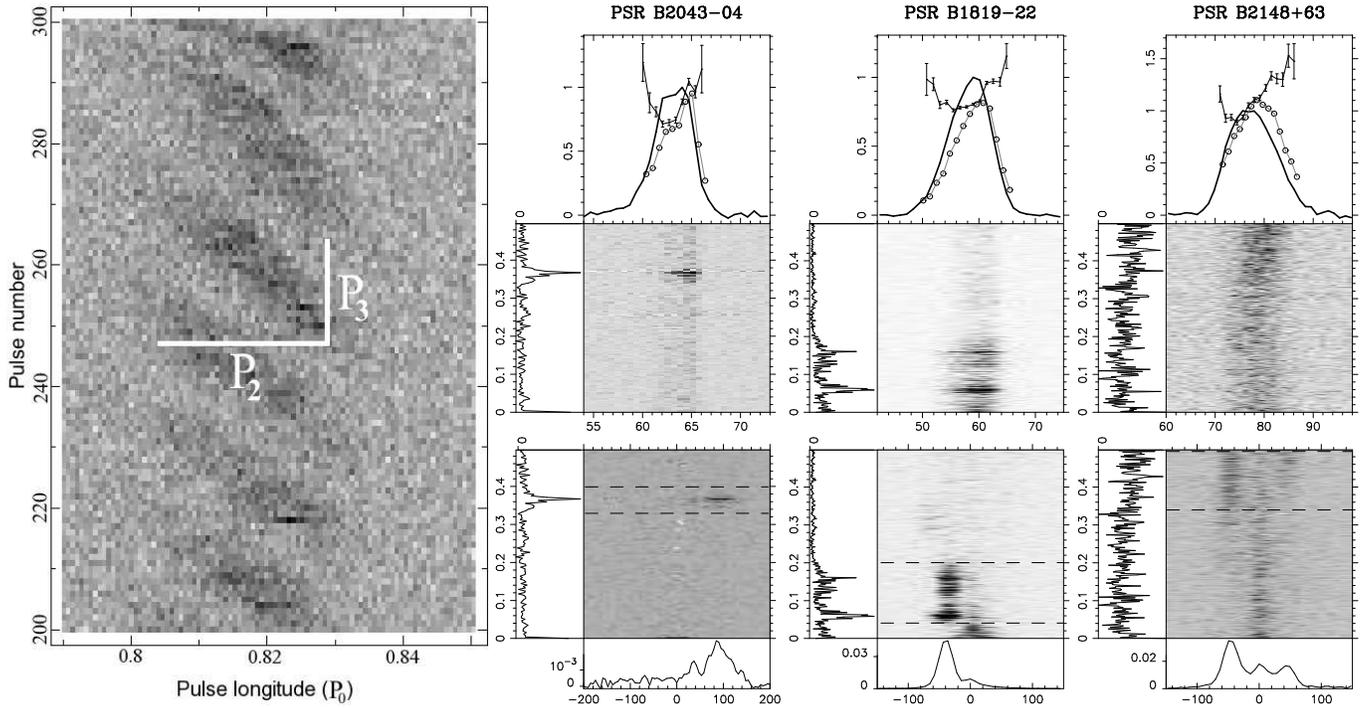}}}
\rotatebox{270}{\resizebox{0.5195\hsize}{!}{\includegraphics[angle=0]{200309086_nozap.ps}}}
\rotatebox{270}{\resizebox{0.5195\hsize}{!}{\includegraphics[angle=0]{200404322.ps}}}
\rotatebox{270}{\resizebox{0.5195\hsize}{!}{\includegraphics[angle=0]{200309093.ps}}}
%\rotatebox{270}{\resizebox{0.51\hsize}{!}{\includegraphics[angle=0]{200309086_nozap.ps}}}
%\rotatebox{270}{\resizebox{0.51\hsize}{!}{\includegraphics[angle=0]{/scratch2/wltvrede/observations/results/21cm_paper_new/200404322.ps}}}
%\rotatebox{270}{\resizebox{0.51\hsize}{!}{\includegraphics[angle=0]{/scratch2/wltvrede/observations/results/21cm_paper_new/200309093.ps}}}
\caption{\label{Classes_fig}The left panel shows a pulse-stack of one
hundred successive pulses of PSR B1819$-$22.  Two successive drift
bands are vertically separated by $P_3$ and horizontally by $P_2$.
The products of our analysis are shown for three pulsars.  The top
panel shows the integrated pulse profile (solid line), the
longitude-resolved modulation index (solid line with error bars) and
the longitude-resolved standard deviation (open circles). Below this
panel the LRFS is shown with on its horizontal axis the pulse
longitude in degrees, which is also the scale for the abscissa of the
plot above.  Below the LRFS the 2DFS is plotted and the power in the
2DFS is vertically integrated between the dashed lines, producing the
bottom plots. Both the LRFS and 2DFS are horizontally integrated,
producing the side-panels of the spectra. See the main text for
further details about the plots.}
\end{figure*}

Archival data was used if available and the sample of pulsars was
completed with new observations. The best WSRT data available was
chosen, so for a number of pulsars the data greatly exceed the minimum
S/N and the number of pulses requirement. This does not bias our
sample of observations toward well-studied pulsars, because all the
observations are long enough to provide a good chance to detect the
drifting phenomenon. We have observations of all the sources except
the millisecond pulsar B1821$-$24, because of the high time resolution
required and the associated data storage problems.  The observations
of PSR B1823$-$13, B1834$-$06 and J1835$-$1020 failed, and therefore
are not included in this paper.

\subsection{Calculation of the pulse-stacks}

All the observations presented in this paper were made at an
observation wavelength of 21 cm spread out over the last five
years. The signals of all fourteen 25-meter dishes of the WSRT were
added together by taking into account the relative time delays between
them and processed by the PuMa pulsar backend (\citealt{vkv02}). In
order to reduce the effects of interference, badly affected frequency
channels were excluded. The frequency channels were then added
together in an offline procedure after dedispersing them by using
previously published dispersion measures.

To study the single pulse behavior of pulsars one usually
converts the one-dimensional de-dispersed time series into a
two-dimensional pulse longitude versus pulse number array
(pulse-stack). An example is shown in the left panel of Fig.
\ref{Classes_fig}, where one hundred successive pulses are plotted on
top of one other. The pulse number is plotted vertically and the time
within the pulses (i.e. the pulse longitude) horizontally. The
off-pulse region is used to remove the baseline from the pulsar
signal, making the average noise level zero.

To correct for the pulse longitude shift of successive pulses
the TEMPO software package\footnote{\tt
http://pulsar.princeton.edu/tempo/ \rm} was used.
Because the pulse period ($P_0$) of the pulsar is not exactly equal to
an integer number of time sample intervals, each pulse (as it appears
in the binned sequence) is effectively shifted by a constant amount
modulo one bin.  This induces, as noted by \citealt{vam98}, a periodic
longitude shift of successive pulses.  Following \cite{es03b}, we have
compensated for this longitude shift of each pulse, and thereby
avoiding artificial features appearing in the spectra that are
derived from the pulse-stacks. All pulse longitudes in this paper
have an arbitrary offset because absolute alignment was not necessary
for our analysis.

In the left panel of Fig. \ref{Classes_fig} one can see a sequence of
100 pulses of one of the new drifters we have found which clearly
shows the drifting phenomenon. Drifting means that the subpulses drift
in longitude from pulse to pulse and thereby the pulsar emission
shows diagonal intensity bands in the pulse-stack (drift bands). The
drift bands are characterized by two numbers: the horizontal
separation between them in pulse longitude ($P_2$) and the vertical
separation in pulse periods ($P_3$). The drift bands of this
pulsar are clearly seen by eye in the pulse-stack and the values $P_2$
and $P_3$ could in principle be measured directly, but in many cases
of the newly discovered drifters the drift bands are not visible to
the eye. To be able to detect the drifting phenomenon in as many
pulsars as possible, all the pulse-stacks were analyzed in a
systematic way as described in the next two subsections.

\subsection{Processing of the pulse-stacks}
\label{Processing_section}

In Fig. \ref{Classes_fig} the products of our method of analysis are
shown for three pulsars and in this section it is explained how
these plots are generated from the pulse-stack and how one can
interpret them.

The first thing that is produced from the pulse-stack is the
integrated pulse profile. This is simply done by vertically
integrating the pulse-stack, i.e. adding the bins with the same pulse
longitude in the successive pulses:
\begin{equation}
\mu_i=\frac{1}{N}\sum_{j=0}^{N-1}S_{ij}
\end{equation}
Here $\mu_i$ is the average intensity at longitude bin $i$, and
$S_{ij}$ is the signal at pulse longitude bin $i$ and pulse number $j$
in the pulse-stack and $N$ is the number of pulses. 
In Fig. \ref{Classes_fig} the solid line in the top panels corresponds
to the integrated pulse profile $\mu_i$ (which is normalized to
the peak intensity). On the horizontal axis is the pulse longitude in
degrees and the value can be read from the horizontal axis of the
panel below which is aligned with the top panel.

The first basic method to find out if there is some kind of subpulse
modulation is to calculate the longitude-resolved variance $\sigma^2_i$
\begin{equation}
\label{sigma_equation}
\sigma^2_i=\frac{1}{N}\sum_{j=0}^{N-1}(S_{ij}-\mu_{i})^2
\end{equation}
and the longitude-resolved modulation index  $m_i$
\begin{equation}
m_i=\frac{\sigma_i}{\mu_i}
\label{modulation_equation}
\end{equation}
The modulation index is a measure of the factor by which the intensity
varies from pulse to pulse and could therefore be an indication
for the presence of subpulses. In Fig. \ref{Classes_fig} the open
circles in the top panel is the longitude-resolved standard deviation
$\sigma_i$ and the solid line with error bars corresponds to the
longitude-resolved modulation index $m_i$.

The detection of a modulation index does not give information about
whether the subpulse modulation occurs in a systematic or a disordered
fashion. The first step in detecting a regular intensity variation is
to calculate the Longitude Resolved Fluctuation Spectrum (LRFS;
\citealt{bac70b}).  The pulse-stack is divided into blocks of 512
successive pulses\footnote{\label{ShorterTransforms} For a few
observations with a low number of pulses shorter transforms were
used.} and the Discrete Fourier Transform (DFT) was performed on these
blocks to calculate the LRFS (for details of the analysis we refer to
\citealt{es02,es03b}). The fluctuation power spectra of the different
blocks were then averaged to obtain the final spectrum.

In Fig. \ref{Classes_fig} the LRFS of the three pulsars are shown
below the pulse profile plots. 
The units of the vertical axis are in cycles per period (cpp), which
corresponds to $P_0/P_3$ in the case of drifting (where $P_3$ is the
vertical drift band separation). The horizontal axis is the pulse
longitude in degrees, which is aligned with the plot above. The power
in the LRFS is horizontally integrated, producing the side panel.
If the emission of the pulsar is modulated with a period $P_3$, then a
distinct region of the LRFS will show an excess of power (i.e. a
feature) in the corresponding pulse longitude range. The LRFS can be
used to see at which pulse longitudes the pulsar shows subpulse
modulation and with which periodicities. The grayscale in the
LRFS corresponds to the power spectral density. Under Parseval's
theorem, the summed LRFS is identical to Eq. 4 (\citealt{es03b}), so
integrating the LRFS vertically gives the longitude resolved variance
(the open dots in the plot above the LRFS).

The detection of a modulation index suggests that there is subpulse
modulation and by analyzing the LRFS it can be determined if this
modulation is disordered or (quasi-)periodic. However from the
LRFS one cannot determine if the subpulses are drifting over a certain
longitude range, because to calculate the LRFS only DFTs along
vertical lines in the pulse-stack are performed. To determine if the
subpulses are drifting, the Two-Dimensional Fluctuation Spectrum
(2DFS; \citealt{es02}) is calculated. The procedure is similar to
calculating the LRFS, but now we select one or more pulse longitude
ranges between which the DFT is not only calculated along vertical
lines, but along lines with various slopes. The effect is that the
pulse longitude information that we had in the LRFS is lost, but we
gain the sensitivity to detect periodic subpulse modulation in the
horizontal direction (i.e. if there also exists a preferred $P_2$
value). Following the same procedure used while calculating the LRFS,
the pulse-stack is divided in blocks of 512 successive
pulses\footnotemark[4] and the spectra of the
different blocks were then averaged to obtain the final spectra.

In Fig. \ref{Classes_fig} the 2DFS is plotted below the LRFS. The
vertical axis has the same units as the LRFS, but now the units of the
horizontal axis are also cycles per period, which corresponds to
$P_0/P_2$ in the case of drifting (where $P_2$ is the horizontal drift
band separation in time units). The power in the 2DFS is
horizontally and (between the dashed lines) vertically integrated,
producing the side and bottom panels in Fig. \ref{Classes_fig}. These
panels are only produced to make it easier to see by eye what the
structure of the feature is.

From the pulse-stack in Fig. \ref{Classes_fig} one can see that two
successive drift bands of PSR B1819$-$22 are vertically separated by
$P_3\approx 18.0 P_0$ and horizontally by $P_2\approx0.025 P_0$.
Instead of measuring drifting directly from the pulse-stack, we use
the 2DFS. From both the 2DFS and LRFS of this pulsar we see that there
are multiple drift features. This is because PSR B1819$-$22 is a drift
mode changer (i.e. the drift bands have different slopes in different
parts of the observation). We note that only one drift mode is seen in
the short stretch of pulses shown in the pulse-stack in
Fig. \ref{Classes_fig}. For PSR B1819$-$22 one can see the main
feature in the LRFS around 0.056 cpp, which corresponds to the $P_3$
value we see in the plotted pulse-stack. In the 2DFS of this pulsar we
see the main feature at the same vertical position as in the LRFS
(corresponding to the same $P_3$ value) and because the feature is
offset from the vertical axis we know that the subpulses drift. From
the horizontal position of the feature in the 2DFS we see that
$P_2\approx -P_0/40=-0.025P_0$, which corresponds well with the $P_2$
measured directly from the pulse-stack shown.

In this paper we use the convention that $P_3$ is always a positive
number and $P_2$ can be either positive or negative. A negative value
of $P_2$ means that the subpulses appear earlier in successive pulses,
which is called negative drifting in the literature. The tabulated
signs of $P_2$ in this paper therefore correspond to the drift
direction, such that a positive sign corresponds to positive
drifting. To comply with this convention, all the plotted 2DFSs in
this paper are in fact flipped about the vertical axis compared with
the definition of the 2DFS in \cite{es02}.

\subsubsection{Interference}

To reduce the effect of interference on the LRFS and 2DFS the spectrum
of the off-pulse noise was subtracted from the LRFS and 2DFS if a
large enough off-pulse longitude interval was available. Interference
will in general not be perfectly removed by this procedure, however
any artificial features produced by interference can easily be
identified because it will not be confined to a specific pulse
longitude range. In Fig. \ref{Classes_fig} the spectra of PSR
B2043$-$04 shows interference with a $P_0/P_3\simeq0.372$. In the
spectra as shown in appendix \ref{Figures_ref} and \ref{Figures_ref2},
the channels containing interference are set to zero, thereby
improving the visual contrast of the plots.

In this paper the modulation index is not directly derived from the
pulse-stack (Eqs. \ref{sigma_equation} and \ref{modulation_equation}),
but from the LRFS. This is done by vertically integrating the LRFS,
which gives the longitude resolved variance
(Eq. \ref{sigma_equation}). The advantage of this method is that by
excluding the lowest frequency bin the effect of interstellar
scintillation (which at this observing frequency has typical low
frequencies) can be removed from the modulation index (for details of
the analysis we refer to \citealt{es02,es03b}). After exclusion of the
lowest frequency bins the variance is overestimated by
$m^2_\mathrm{scint}$, where $m^2_\mathrm{scint}$ is the modulation
index induced by the scintillation (see Eqs. 20-22 of
\citealt{es04}). The longitude resolved modulation index and standard
deviation are corrected accordingly.

\subsection{Analysis of the drift features}
\label{Analysis_section}

If a feature is seen in the 2DFS and we make sure it is not associated
with interference, $P_2$ and $P_3$ can be measured and its
significance determined. The drift feature will always be smeared out
over a region in the 2DFS. This could be because there is not one
fixed value of the drift rate throughout the observation due to
random slope variations of the drift bands, drift mode changes or
nulling. But the feature is also broadened if the drifting is not
linear (i.e. subpulse phase steps or swings) and because of subpulse
amplitude windowing (\citealt{es02}). 

Because of all these effects it is impossible to fit one
specific mathematical function to all the detected features, so it is
more practical to calculate the centroid of a rectangular region in
the 2DFS containing the feature. The advantage of this procedure is
that no particular shape of the feature has to be assumed. The
centroids are defined as:
\begin{eqnarray}
\nonumber P_0/P_3 &=& \frac{1}{F}\sum_{k,l} y_l F_{kl}\\
P_0/P_2 &=& \frac{1}{F}\sum_{k,l} x_k F_{kl}
\end{eqnarray}
Here $k$ and $l$ are indices for the horizontal and vertical position
within the region in the 2DFS containing the feature, $x_k$ and $y_l$
are the corresponding axis values of bin $(k,l)$, $F_{kl}$ is the
power in that bin and $F=\sum_{k,l} F_{kl}$ is the total power in the
selected region of the 2DFS. An uncertainty on the position of the
centroid (hence on $P_2$ and $P_3$) can be estimated by using the
power in a region in the 2DFS that only contains noise. However we
have found that in most cases the uncertainty is dominated by the more
or less arbitrary choice of what exact region in the 2DFS is selected
around the feature. To estimate this uncertainty we have therefore
calculated the centroid for slightly different regions containing the
feature. 

Because the analyzed pulse sequences are sometimes relatively
short, there is the possibility that the occasional occurrence of
strong (sub)pulses are dominating the spectra and therefore lead to
misleading conclusions. 
To estimate what kind of ``random'' fluctuation features one can
expect from a given pulse sequence, we have randomized the order of
the pulses and then passed this new pulse-stack through our software
to determine the magnitude of any features which could be attributed
to strong (sub)pulses. Any actual drifting will lose coherence in this
process and thus we can use the randomized results when estimating the
significance of drift features when there is no well defined $P_3$.  

For most pulsars there is power along the vertical axis in the
2DFS. Therefore the centroid of a larger region around a drift feature
usually results in a centroid located closer to the vertical axis,
hence in a larger absolute value of $P_2$. In most cases this causes
the uncertainty on $P_2$ to be asymmetric around its most likely value
and therefore it is useful to tabulate the uncertainty in both signs.

If the centroid of the feature in the 2DFS is significantly offset
from the vertical axis, it means that $P_0/P_2\neq0$ and that drifting
can be associated with the feature (i.e. there exists a preferred
drift direction of the subpulses from pulse to pulse). Drifters are
defined in this paper as pulsars which have at least one feature in
its 2DFS which show a significant finite $P_2$. It must be noted that
drift bands are often very non-linear and therefore the magnitude of
$P_2$ is probably of little meaning. Also we want to emphasize
that not only an offset from the vertical axis indicates drifting, but
any asymmetry about the vertical axis indicates drifting-like
behavior. Any frequency dependence along the vertical axis indicates
structured subpulse modulation, either quasi-periodic or as a low
frequency excess. Pulsars with a low frequency excess show an excess
of power in their spectra toward long frequencies (i.e. the spectra
are ``red'').  

The drifters are classified into three classes and an example of each
class is shown in Fig. \ref{Classes_fig}. One can see that PSR
B2043$-$04 has a vertically narrow drift feature in its spectra,
meaning that $P_3$ has a stable and fixed value throughout the
observation. We will call these pulsars the \em coherent drifters \rm
class (class Coh in table \ref{Table_section}). The criterion used for
this class is that the drift feature has a vertical extension smaller
than 0.05 cpp.  The pulsars that show a vertical, broad, {\em diffuse}
drift feature are divided into two classes, depending on whether the
feature is clearly separated from the alias borders ($P_0/P_3=0$
and $P_0/P_3=0.5$). In table \ref{Table_section} the pulsars in
class Dif are clearly separated from the alias border and the pulsars
in class Dif$^\ast$ not.
In Fig. \ref{Classes_fig} PSR B1819$-$22 is a diffuse (Dif) drifter
and PSR B2148+63 is a Dif$^\ast$ drifter. 

Besides the drifters there is also a class of pulsars which show
longitude stationary subpulse modulation (class Lon in table
\ref{Table_section}). These pulsars show subpulse modulation with a
$P_3$ value, but without a finite $P_2$ value. Because it is not clear
if these pulsars should be counted as drifters or as non-drifters,
they are excluded from the statistics. 

\begin{figure*}[htb]
\begin{center}
\rotatebox{270}{\resizebox{0.42\vsize}{!}{\includegraphics[angle=0]{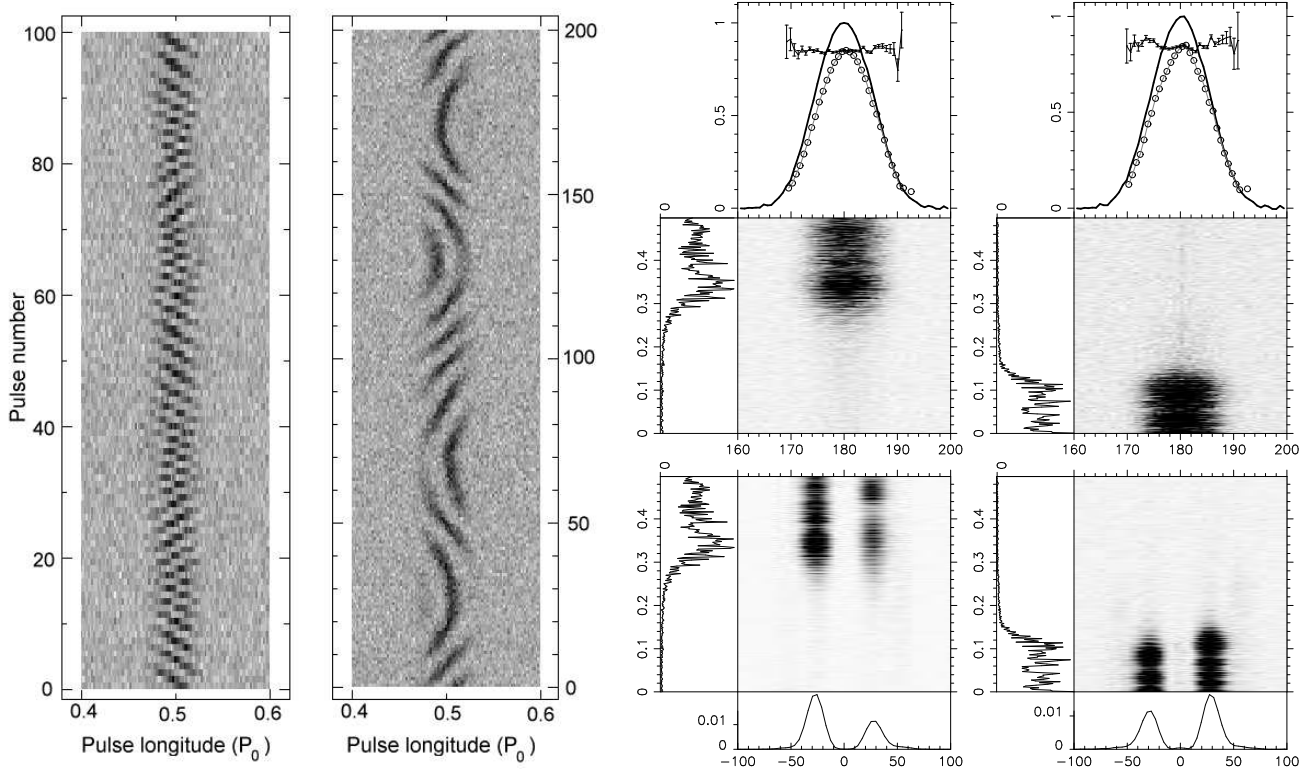}}}\hspace{3mm}
\rotatebox{270}{\resizebox{0.42\vsize}{!}{\includegraphics[angle=0]{fakefast.ps}}}
\rotatebox{270}{\resizebox{0.42\vsize}{!}{\includegraphics[angle=0]{fakeslow.ps}}}
\end{center}
\caption{\label{FakeFig}A section of the pulse-stacks and the derived
spectra of the two artificially generated pulse sequences to
illustrate possible effects of pulsars that constantly change
their alias order. The drifting in the left pulse-stack constantly
crosses the $P_3=2P_0$ alias border and the right pulse-stack
constantly changes its apparent drift direction via longitude
stationary subpulse modulation. For the explanation of the plots we
refer to Fig. \ref{Classes_fig} and the main text.}
\end{figure*}

For many pulsars we find that the magnitude of $P_2$ exceeds the
pulse width. This means in the case of regular drifting that in a
single pulse only one subpulse is visible and that the drifting will
manifest itself more as an amplitude modulation rather than as a phase
modulation. An illustrative example of a regular drifter with a large
$P_2$ is PSR B0834+06 (\citealt{ad05}). Whether the drifting manifests
itself more as an amplitude modulation or as a phase modulation will
largely depend on the viewing geometry. Also the presence of pulse
sequences without organized drifting, longitude stationary subpulse
modulation or drift reversals will result in a large $P_2$ value.

It must be noted that the calculation of the 2DFS is an averaging
process. This is what makes it a powerful tool to detect drifting
subulses in low S/N observations, but at the same time this implies
that different pulse-stacks can produce similar 2DFS. For instance a
feature that is split and shows a horizontal separation can be caused
by drift reversals, but also by subpulse phase steps or swings (see
\citealt{esv03} for a pulse sequence of PSR B0320+39 and the resulting
2DFS). Note also that with only the LRFS it is impossible to identify
complex drift behavior like drift reversals or subpulse phase steps.

The details of each observation can be found in table
\ref{Table_section}, including the classification we made, the
measured $P_2$ and $P_3$ values, the modulation index and the
detection threshold for the modulation index. It must be emphasized
that the $P_2$ and $P_3$ values are average values during the
observation. Especially when the pulsar is a drift mode changer a
different observation may lead to different values for $P_2$ and
$P_3$. The plots of all the pulsars can be found in appendix
\ref{Figures_ref}. For some of the pulsars the 2DFS for two different
pulse longitude ranges are shown if useful. The plots of these pulsars
come after the plots of the pulsars with only one 2DFS plot. The same
plots can also be found in appendix \ref{Figures_ref2}, but there they
are ordered by appearance in the text instead of ordered by name.

\subsubsection{Drift reversals}

As can be seen from Fig. \ref{Classes_fig}, the subpulses of PSR
B1819$-$22 appear to drift toward the leading part of the pulse
profile. In the sparking gap model (e.g. \citealt{rs75}), every
subpulse is associated with one emission entity close to the surface
of the star. These entities (the sparks) move around the magnetic
axis, causing the subpulses to drift through the pulse window. Because
the emission entities are only sampled once per rotation period of the
star, it is very difficult to determine if the subpulses in one drift
band correspond to the same emission entity for successive pulses. For
instance for PSR B1819$-$22 we do not know if the emission entities
drift slowly toward the leading part of the pulse profile (not
aliased) or faster toward the trailing part of the pulse profile
(aliased).

The physical conditions of the pulsar probably determines what the
physical drift rate of the emission entities are, rather than the
observed (possibly aliased) drift rate. This could already be a
serious problem if one wants to correlate physical parameters of
pulsars to the observed drift rate of coherent drifters, for which we
at least know they stay in the same alias mode. If a feature in the
2DFS is not clearly separated from the alias borders, the power in
that feature could consist of drift bands in different alias modes
(i.e. the apparent drift direction could be changing constantly during
the observation). In that case the measured value of $P_3$ using the
centroid of the feature is related to the true drift rate of the
emission entities in a complicated way, depending on what fraction of
the time the pulsar spends in which alias mode. Therefore it is
expected that it will be very hard to find a correlation between
physical pulsar parameters and the $P_3$ values of the pulsars in the
Dif$^\ast$ class, so it will be useful to classify the pulsars
depending on whether the features in the 2DFS are clearly separated
from the alias borders. Inspection of the pulse-stacks with strong
enough single pulses reveals that some of the pulsars in the
Dif$^\ast$ class change their drift direction during the
observation. This is further evidence to indicate the value of
considering the Dif and Dif$^\ast$ classes separately.

To illustrate this we have artificially generated two pulse sequences
of a pulsar that has a variable rotation period of the emission
entities in two different scenarios (see Fig. \ref{FakeFig}). In both
scenarios the emission entities are simulated to drift from the
trailing to the leading edge of the pulse profile with a variable
drift rate. In the left sequence the vertical drift band separation
$P_3$ is close to $2P_0$ and in the right sequence the $P_3$ period is
much larger. In the left sequence the subpulses around the first pulse
appear to drift toward the leading edge of the pulse profile. As time
increases, we speed up the rotation of the emission entities, which
causes $P_3$ to become smaller. Around pulse number 15 the Nyquist
border $P_3=2P_0$ is reached and the drift pattern becomes a
check-board like pattern. As time further increases the emission
entities are still speeding up, causing clear drift bands to reappear
with an opposite apparent drift direction (around pulse 25). After
this the rotation of the emission entities is gradually slowed down to
the initial value, causing the drift bands to change apparent drift
direction again around pulse 50. The same cycle is repeated for the
next pulses. In this simulation the carousel rotation period is set to
vary with about 40\% around its mean value. 

The resulting spectra of this pulse sequence are also shown in
Fig. \ref{FakeFig}. The LRFS shows that the subpulse modulation is
extended toward the $P_3=2P_0$ alias border and the 2DFS shows a
feature that is split by the vertical axis, because there are two
apparent drift directions in the pulse sequence. As can be seen in the
bottom panel, there is more power in the left peak. This corresponds
to more power being associated with negative drifting (drifting toward
the leading edge of the pulse profile). This is also directly visible
in the pulse-stack. A good example of a known pulsar that shows this
kind of drift behavior is PSR B2303+30 (e.g. \citealt{rwr05}) and its
2DFS (see Fig. \ref{B2303+30}) indeed shows a very clear double peaked
feature.

\begin{figure}[t]
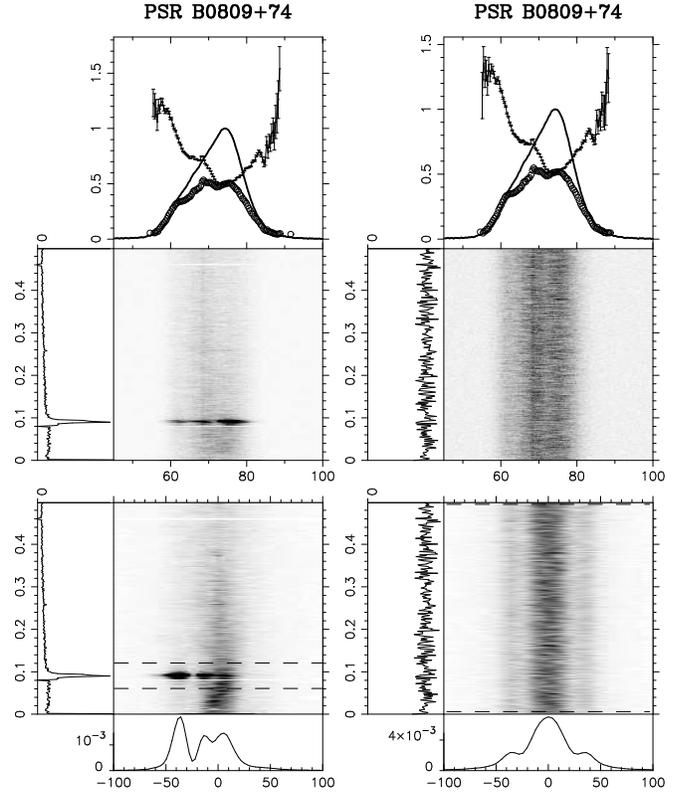

\begin{center}
\rotatebox{270}{\resizebox{!}{0.49\hsize}{\includegraphics[angle=0]{200301433.ps}}}
\rotatebox{270}{\resizebox{!}{0.49\hsize}{\includegraphics[angle=0]{200301433_scrambled.ps}}}
\end{center}
\caption{\label{ScrambleFig}This figure shows the spectra of the
well known drifter PSR B0809+74 (left) as well as the results from the
same data after putting the pulses in a random order (right).}
\end{figure}

In the second scenario of Fig. \ref{FakeFig} the pulse-stack shows
drift bands with a much larger $P_3$. Because of possible aliasing
this does not directly imply that the emission entities are rotating
slower. In fact, we have chosen the entities to rotate faster than in
the first scenario. This causes the driftbands to be aliased and the
drift bands appear to drift in the opposite direction to the emission
entities (which again are simulated to drift toward the leading edge
of the pulse profile). As time increases the rotation of the emission
entities is sped up. Because of aliasing the drift rate appears to
decrease until around pulse 25 the drifting has become longitude
stationary ($P_3=\infty$). The emission entities are now rotating so
fast that in one pulse period time they exactly reappear at the pulse
longitude of another drift band. When the rotation period of the
emission entities is speeded up even further, the drift bands are
changing their apparent drift direction again as can be seen around
pulse 50. Now the rotation period of the emission entities is slowed
down again until around pulse 100 the initial conditions are reached
again. After this the same cycle is repeated. A clear example of a
known pulsar that shows this kind of drift reversals is PSR B0826$-$34
(\citealt{ggk+04,elg+05}).

The LRFS of this sequence shows that the subpulse modulation is
extended toward the horizontal axis and the 2DFS shows again a feature
that is split by the vertical axis. As can be seen in the bottom
panel, there is more power in the right peak, which corresponds to more
power associated with positive drifting. In the pulse-stack there are
indeed more drift bands that drift toward the trailing edge of the
pulse profile than in the opposite direction.

\subsubsection{Non linear drift bands}

In the left panel of Fig. \ref{ScrambleFig} the spectra of the
well known regular drifter PSR B0809+74 are shown. The drift feature
in the 2DFS shows a clear horizontal structure caused by the non
linear drift bands of PSR B0809+74 at this observing frequency
(\citealt{es03c,pw86,wbs81}). If drift bands are not straight, there
is no one unique value of $P_2$ that is associated with the drifting
and therefore the drift feature in the 2DFS will be more complex than
just one peak. This makes $P_2$ an ill-defined parameter. However this
is no shortcoming of the 2DFS, it is a shortcoming of the whole
concept of $P_2$ under curved driftbands.  In this paper $P_2$ is just
a rough measure of the presence of drift, its direction and the
magnitude of the slope in a overall mean sense only.

If the drift bands are non linear, the subpulses will have a pulse
longitude dependent spacing. This pulse longitude dependent spacing is
described by the so-called ``modulation phase profile'' or ``phase
envelope'' (\citealt{es02}).

\subsubsection{Quasiperiodic subpulses}

For most drifters the vertical drift band separation $P_3$ is a much
better defined parameter than the horizontal separation $P_2$,
e.g. the drift feature in the 2DFS of PSR B0809+74 (left panel of
Fig. \ref{ScrambleFig}) is much sharper in the vertical direction than
in the horizontal direction. There is however the possibility that
single pulses show regular spaced subpulses, while there is no memory
for where the subpulses appear in successive pulses. In that case
there is a $P_2$ value, but $P_3$ is undefined. One can simulate such
a scenario by putting the pulses of a regular drifter in a random
order. This is done for PSR B0809+74 in the right panel of
Fig. \ref{ScrambleFig}.

One can see, first of all, that the longitude-resolved standard
deviation and modulation index are independent on the ordering of the
pulses, as expected from Eqs. \ref{sigma_equation} and
\ref{modulation_equation}. Secondly, the spectra do not show any
vertical structure anymore. This indicates that, as expected, $P_3$
has become undefined. The 2DFS is symmetric about the vertical axis, so
there is no preferred drift direction anymore. The horizontal subpulse
separation is however still visible in the 2DFS as two vertical bands
at $\pm37$ cpp, the same horizontal position as the largest peak in
the 2DFS in the left panel of Fig. \ref{ScrambleFig}.

We have found two pulsars that shows these kind of features: PSRs
B2217+47 and B0144+59. In the carousel model this phenomenon could be
explained by a highly variable circulation time that causes the alias
order to change constantly. However there is no evidence that this
phenomenon is related to the same origin as the drifting subpulses, as
it only shows that the subpulses appear quasiperiodic.

\section{Individual detections}

\subsection{Coherent drifters (class ``Coh'')}
The coherent drifters are the pulsars which show a drift feature in
their 2DFS over a small $P_3$ range (smaller than 0.05 cpp). First the
pulsars that were already known to have drifting subpulses are
described followed by the pulsars that were not known to show this
phenomenon.

\subsubsection{Known drifters}
{\bf B0148$-$06}: Both components of the pulse profile of this
pulsar\footnote{\label{TooLongObs}This pulsar is not in our source
list, because the required observation length is too long. Therefore
this pulsar is not included in the statistics.}  have the same drift
sign and $P_3$ varies slightly during the observation (see
Fig. \ref{B0148-06}). The drift bands are clearly visible in the
pulse-stack and the drifting is most prominent in the leading
component. The feature in the leading component seems to consist
of different drift modes. This is all consistent with results
reported by \cite{bhmm85},
who discovered the drifting subpulses at 645 MHz.\\
{\bf B0320+39}: This pulsar is known to show very regular drifting
subpulses (\citealt{iks82}), which is confirmed by the very narrow
drift feature in our observation (Fig. \ref{B0320+39}). \cite{ikl+93}
have shown that drifting at both 102 and 406 MHz occurs in two
distinct pulse longitude intervals and that the energy contribution in
the drifting subpulses is less at higher frequencies, especially in
the leading component. We see that this trend continues at higher
frequencies, because the feature is most prominent in the trailing
component at 1380 MHz.  The drift feature in the 2DFS shows a clear
horizontal structure caused by a subpulse phase step in the drift
bands (as is also seen for instance for PSR B0809+74 at this
frequency). At both 328 MHz and 1380 MHz the drift bands of PSR
B0320+39 are known to show a phase step (\citealt{esv03} and
\citealt{es03c} respectively). This is the same observation as used by
\cite{es03c}.\\
{\bf B0809+74}: The drift feature in the 2DFS (Fig. \ref{B0809+74})
shows clearly horizontal structure, which is caused by a phase step of
the drifting subpulses. This phase step is only present at high
frequencies and is consistent with \cite{wbs81} at 1.7 Ghz,
\cite{pw86} at 1420 MHz and \cite{es03c} at 1380 MHz (this is the same
data as used by \citealt{es03c}). The drift rate is affected by nulls
(\citealt{th71}) and detailed analysis of this phenomenon allowed
\cite{vsr+03} to conclude that the drift is not aliased for this
pulsar.\\
{\bf B0818$-$13}: This pulsar has a clear drift feature
(Fig. \ref{B0818-13}) that contains almost all power in the 2DFS. The
drift feature has a horizontal structure like observed for PSR
B0809+74 and PSR B0320+39. From the modulation phase profile it
follows that the drift bands make a smooth swing of about
\degrees{200} in subpulse phase in the middle of the pulse profile.
A decrease of the drift rate in the middle of the pulse profile has
been reported at 645 MHz by \cite{bmhm87} and this effect appears to
be more pronounced in our observation at 1380 MHz. The longitude
resolved modulation index shows a minimum at the position of the
subpulse phase swing (consistent with \citealt{bmhm87}), something
that is also observed for the phase steps of PSR B0320+39 and
B0809+74. The subpulse phase swing is clearly visible in the
pulse-stack of this pulsar. This phase swing seems to be related
to the complex polarization behavior of the single pulses as observed
by \cite{edw04}. For this pulsar nulling was shown to interact with
the drift rate by \cite{la83} and from detailed analysis of this
phenomenon \cite{jv04} concluded that the drift of this pulsar is
aliased.\\
{\bf B1540$-$06}: The 2DFS of the two halves of the pulse profile
show opposite drift directions (Fig. \ref{B1540-06}). The drift sense
of the leading part of the pulse profile is consistent with the
positive drifting of this pulsar as has been reported by \cite{ash82}
at 400 MHz.\\
{\bf B2045$-$16}: Only the 2DFS of the trailing component is plotted
in Fig. \ref{B2045-16}, because we do not detect any features in the
leading component. However, drifting has been observed previously in
both components (e.g. \citealt{os77b}) and is observed to be broad
with $P_3$ values between 2 and 3$P_0$ (\citealt{os77b} at 1720 MHz,
\citealt{nuwk82} at at 1.4 and 2.7 GHz and \citealt{th71} at low
frequencies). The second component in our observation has a clear
narrow drift feature. This could indicate that this pulsar is a drift
mode changer and that our observation was too short\footnote{Our
observation was too short to contain the required minimum of one
thousand pulses. Because we have detected drifting, we did not
reobserve this pulsar.} to detect the drift rate variations.\\
{\bf B2303+30}: This pulsar is known to drift close to the alias
border (e.g. \citealt{so75} at 430 MHz). This pulsar has a clear
double-peaked feature in its 2DFS exactly at the alias border
(Fig. \ref{B2303+30}), which suggest that the apparent drift
sense changes during the observation because the alias changes during
the observation. The change of drift sense can be seen by eye in the
pulse-stack and also in the pulse-stacks shown in
\cite{rwr05}. These authors show that, besides this
$P_3\!\!\approx\!\!2P_0$ `$B$' drift mode, there is occasionally also a
$P_3\!\!\approx\!\!3P_0$ `$Q$' drift mode if the S/N conditions are
good. There is no evidence for this weaker `$Q$'-mode in our
observation.\\
{\bf B2310+42}: The two components of this pulsar are clearly drifting
at the alias border (Fig. \ref{B2310+42}). The drift feature in
the trailing component is clearly double peaked, so the alias mode is
probably changing during the observation. The leading component shows
the same feature at the alias border, although much weaker. The
dominant drift sense is consistent with the positive drifting found by
\cite{ash82} at 400 MHz. We also find that the low frequency excess of
the leading component is clearly drifting with two signs as well ($P_2
= \degrees{65}\pm35$, $P_3 = 21\pm3P_0$ and $P_2 =
-\degrees{65}\pm15$, $P_3 = 70\pm10P_0$), but no significant drift is
detected in the low frequency excess of the trailing component.\\
{\bf B2319+60}: This pulsar shows a clear drift component in the 2DFS
of the trailing component of the pulse profile and a less clear drift
component in the leading component (Fig. \ref{B2319+60}). No
significant drifting is detected in the 2DFS of the center part of the
pulse profile (which is not plotted). It has been found that this
pulsar is a drift mode changer and that the allowed drift mode
transitions follow certain rules (\citealt{wf81} at 1415 MHz). In the
2DFS there is only evidence for one stable drift mode, but drifting is
detected over the whole $P_3$ range.  Drift mode changes can been seen
in the pulse-stack.  The drift mode we see in the 2DFS is the `$A$'
drift mode, which was found by \cite{wf81} to be the most common drift
mode. The many nulls in the pulse-stack probably smears out the
feature over a large $P_3$ range. There is a sharp $P_3 \simeq
130P_0$ feature in the second component, which could be related to
nulling or mode changes.\\

\subsubsection{New drifters (class ``Coh'')}
{\bf B0149$-$16}: The 2DFS of this pulsar (Fig. \ref{B0149-16})
shows a weak drift feature.\\
{\bf B0609+37}: Almost all power in the 2DFS of this pulsar is in a
 well confined drifting feature (Fig. \ref{B0609+37}), meaning that the
subpulse drifting is very organized and stable.\\
{\bf B0621$-$04}: A strong and very coherent $P_3\simeq2P_0$
feature is seen in the LRFS and 2DFS of this pulsar
(Fig. \ref{B0621-04}), which is also seen in other observations.
The feature is significantly offset from the vertical axis, so this
pulsar shows very stable drifting subpulses. Only the 2DFS of the
trailing peak is plotted.\\
{\bf J1650$-$1654}: This pulsar shows a very clear drift feature in
its 2DFS (Fig. \ref{J1650-1654}). The feature seems to show some
vertical structure, which could be because of drift rate
variations. The feature also seems to show some horizontal structure
like PSR B0818$-$13, which could indicate that the drift bands are
curved or show a phase step. This would be consistent with the minimum
in the modulation index in the middle of the pulse profile. However
this observation is too short to state if this effect is due to a
systematic drift rate change across the pulse profile or due to random
variations.\\
{\bf B1702$-$19}: The pulse profile of this pulsar shows an interpulse
 (\citealt{blh+88}). The main pulse shows a drift feature and the 2DFS
 of the interpulse shows a feature with the same $P_3$ value
 (Fig. \ref{B1702-19}), but no significant offset from the vertical
 axis could be detected for this feature. This is not the only pulsar
 to show correlations in emission properties between the main- and
 interpulse. PSR B1822$-$09 exhibits an anti-correlation between the
 intensity in the main- and the interpulse (e.g. \citealt{fwm81,fw82}) and
 also a correlation in the subpulse modulation
 (e.g. \citealt{gjk+94}). That pulsar also shows drifting in the
 main pulse in the `$B$'-mode and longitude stationary subpulse
 modulation in both the main- and interpulse in the `$Q$'-mode with the
 same $P_3$.  Also PSR B1055$-$52 is known to show a main
 pulse-interpulse correlation. For that pulsar a correlation between
 the intensity of the main and interpulse has been found by
 \cite{big90a}. However PSR B1702$-$19 is the first pulsar that shows
 a correlation between the drifting subpulses in the main pulse and
 the subpulse modulation in the interpulse. \\
{\bf B1717$-$29}: A very narrow drift feature is seen in both the
LRFS and the 2DFS of this pulsar (Fig. \ref{B1717-29}). Because the
very low S/N of this observation it was impossible to measure a
significant modulation index using the whole $P_3$ range of the
LRFS. By only using the frequencies in the LRFS that contains the
drift feature it was possible to significantly determine the
modulation index corresponding to the drift feature. This drifting is
confirmed in another observation.\\
{\bf B1839$-$04}: Both components of the pulse profile of this pulsar
are drifting (Fig. \ref{B1839-04}). The drift bands are clearly
visible in the pulse-stack to the eye and both components have an
opposite drift sense.  The slope of the drift bands in the two
components are mirrored and the drift bands of the two components are
also roughly in phase. So when a drift band is visible in one
component, it is also visible (although mirrored) in the other
component. This ``bi-drifting'' subpulse behavior is also observed for
PSR J0815+09, which also has opposite drift senses in different
components (\citealt{mlc+04}). This ``bi-drifting'' could be a sign
that these pulsars have both an inner annular gap and an inner core
gap (\citealt{qlz+04}), but also ``double imaging'' could be
responsible (\citealt{esv03}). Note also that the second harmonic of
the drift feature is visible in the 2DFS of especially the trailing component.\\
{\bf B1841$-$04}: This pulsar has a weak, definite drift feature
in its 2DFS (Fig. \ref{B1841-04}),  which is also visible in the LRFS.\\
{\bf B1844$-$04}: There is a weak detection of a narrow drift feature
in the 2DFS of this pulsar, which is also visible in the LRFS
(Fig. \ref{B1844-04}).\\
{\bf J1901$-$0906}: The trailing component of this pulsar shows a
clear and narrow drift feature in its 2DFS, which is not detected in
the leading component (Fig. \ref{J1901-0906}). The 2DFS of the leading
component has a drift feature with a different $P_3$ which is also
present in the right component ($P_2 = -\degrees{8.2}\pm9.2$, $P_3 =
6.8\pm0.5 P_0$). The drifting can be seen by eye in the
pulse-stack. The different measured $P_3$ values in the two
components could indicate that this pulsar is a drift mode changer.\\
{\bf B2000+40}: Although this observation is contaminated by
interference, clear drifting is detected in the leading component. The
rest of the pulse profile (mostly in the trailing component) is also
drifting (Fig. \ref{B2000+40}). The feature in the leading component
shows horizontal structure which could be caused by drift reversals or
more likely by a subpulse phase jump or swing.\\
{\bf B2043$-$04}: This pulsar has a very clear and narrow drifting
component in its 2DFS (Fig. \ref{B2043-04}). The feature is 
perhaps extended toward the alias border, but this
is not significant. Almost all power in the 2DFS is in the drift
feature.\\

\subsection{Diffuse drifters (classes ``Dif'' and ``Dif$^\ast$'')}

These pulsars show a drift feature over an extended $P_3$ range
(larger than 0.05 cpp). If the drift feature is clearly separated from
both alias borders ($P_0/P_3 = 0$ and $P_0/P_3=0.5$), the pulsar is
classified as Dif. However if it is not the pulsar is classified as
Dif$^\ast$ in table \ref{Table_section}. In this section the pulsars
in the latter class are indicated with an asterisk next to their
name. Note that not all drift features in the spectra have peaks which
are offset from the vertical axis, but they must be asymmetric about
the vertical axis. \\

\subsubsection{Known drifters}

{\bf B0031$-$07}: This pulsar shows a broad drifting feature in its
2DFS (Fig. \ref{B0031-07}). Three drift modes have been found for this
pulsar by \cite{htt70} at 145 and 400 MHz. In our observation most
power in the 2DFS is due to the `$A$'-mode drift ($P_3=12P_0$). The
slope of the drift bands change from band to band
(e.g. \citealt{vj97}), causing the feature to extend vertically in the
2DFS. The `$B$'-mode drift ($P_3=6P_0$) is also visible in our
observation, but there is no feature corresponding to `$C$'-mode drift
($P_3=4P_0$). This is consistent with the multifrequency study of
\cite{smk05}.\\
{\bf B0301+19$^\ast$}: The trailing component shows a broad drift feature in
its 2DFS (Fig. \ref{B0301+19}), but no drifting is detected in the
leading component. This pulsar is observed to have straight drift
bands in both components of the pulse-stack (\citealt{ss73} at 430
MHz).  The feature in the trailing component is reported to be broader
than in the leading component (\citealt{brc75}, also at 430 MHz),
probably because drifting subpulses appear more erratic in the
trailing component. The feature we see is also broad and may even
be extended to the alias border.\\
{\bf B0329+54$^\ast$}: The power in the LRFS of this pulsar peaks toward
$1/P_3=0$, as reported by \cite{th71} for low frequencies
(Fig. \ref{B0329+54}). Drifting is detected in four of the five
components. The third component (the right part of the central peak)
has a broad drift feature and the subpulses have a positive drift
sense, something that is also reported by \cite{tmh75} at 400 MHz.
Besides these known features we find that the first component (left
peak) and the fourth component (the bump between the central and
trailing peak) are also drifting with a positive drift senses:
$P_2=\degrees{750}\pm900$ and $\degrees{147}\pm110$ and $P_3=
2.9\pm1.7$ and $4.2\pm1.0$ respectively. The second component (the
left part of the central peak) has an opposite drift sense:
$P_2=-\degrees{175}\pm150$ and $P_3=3.0\pm1.7$. The last component
shows no significant drifting. The 2DFS of the second and third
component are shown in Fig. \ref{B0329+54}.  The difference between
the values of $P_3$ in the different components seems not to be 
significant.\\
{\bf B0628$-$28$^\ast$}: Sporadic drifting with a positive drift sense has
been reported for this pulsar by \cite{ash82} at 400 MHz, but the
$P_2$ and $P_3$ values could not be measured. The positive drift sense
is confirmed in our observation as a clear excess of power in the right
half of the 2DFS (Fig. \ref{B0628-28}). The feature in the 2DFS is not
separated from either alias borders. Most power in both the LRFS and
2DFS is in the lower half.\\
{\bf B0751+32$^\ast$}: The 2DFS of the leading component of the pulse
profile of this pulsar\footnotemark[5] shows drifting
over the whole $P_3$ range with a negative $P_2$ value
(Fig. \ref{B0751+32}). This can clearly be seen in the bottom plot,
which shows an excess of power in the left half. This confirms the
drifting as reported by \cite{bac81} at 430 MHz. Both components also
show a strong $P_3=70\pm10P_0$ feature. This feature shows negative
drifting in the leading component, but no significant drifting in the
trailing component.\\
{\bf B0823+26$^\ast$}: Only the pulse longitude range of the main pulse is
shown in Fig. \ref{B0823+26} and the 2DFS of the main pulse shows a
clear broad drift feature. \cite{bac73} found that at 606 MHz this
pulsar shows drifting in bursts, but the drift direction is different
for different bursts. In our observation there seems to exist a clear
preferred subpulse drift direction, so this pulsar is classified
as a drifter.\\
{\bf B0834+06$^\ast$}: The 2DFS of both components have a weak drift feature
at the alias border (Fig. \ref{B0834+06}). This confirms the drifting
detected by \cite{sspw70} at low frequencies. The circulation time of
this pulsar ($\hat P_3$) has been measured by \cite{ad05}.\\
{\bf B1133+16$^\ast$}: The 2DFS of both components of this pulsar
(Fig. \ref{B1133+16}) show a very broad drifting feature with the
same drift sign consistent with other data we have analyzed. The
trailing component shows also a long period drift feature ($P_2 =
\degrees{160}\pm100$ and $P_3 = 33\pm3P_0$). This drifting is
consistent with the drifting found by \cite{now96} at 430 and 1418 MHz
and by \cite{bac73,tmh75} at low frequencies.\\
{\bf B1237+25$^\ast$}: The 2DFS of the outer components of the pulse profile
are clearly drifting with opposite drift sign (they are plotted in
Fig. \ref{B1237+25}). The 2DFS of the three inner components (which
are not plotted) all show drifting with a positive drift sense (except
the middle one which does not show significant drifting). The values
are $P_2 = \degrees{26}\pm18$, $P_3 = 2.7\pm1 P_0$ and $P_2 =
\degrees{17}\pm9$, $P_3 = 2.7\pm1 P_0$ respectively. This drifting is
consistent with \cite{pw86} at 408 and 1420 MHz.\\
{\bf J1518+4904}: This millisecond pulsar\footnote{This pulsar is not
in our source list, because the flux is not in the catalog. Therefore
this pulsar is not included in the statistics.} has a clear broad
drift feature (Fig. \ref{J1518+4904}). This pulsar was already known
to be a drifter (\citealt{es03b}  at 1390 MHz), showing that
drifting is not an phenomenon exclusive to slow pulsars.\\
{\bf B1642$-$03$^\ast$}: Drifting is observed to occur in bursts in this
pulsar with both drift senses (\citealt{th71} at 400 MHz) and
also \citealt{tmh75} report that there is no preferred drift sense at
400 MHz. The 2DFS of our observation (Fig. \ref{B1642-03}) reveals a
broad drift feature with a preferred drift sense, so this pulsar
is classified as a drifter. The alias border seems to be crossed on
both sides, because the feature is extended over the whole $P_3$ range
and seems double peaked. \\
{\bf B1822$-$09$^\ast$}: For this pulsar a correlation in the subpulse
modulation between the main- and interpulse has been found (see the
text of the coherent drifter PSR B1702$-$19 for details). There are no
features in the spectra of our observation of the the interpulse and
therefore the interpulse is not plotted in Fig. \ref{B1822-09}. There
is drifting detected in the trailing component of the main pulse, but
it is not clear what exact range in the 2DFS shows drifting causing
the large uncertainty on the $P_3$ value. The observation is
consistent with the `$B$'-mode drift found by \cite{fwm81} at 1620 MHz
with a $P_3\simeq11P_0$.  There is also a feature at 0.02 cpp, which
could be related to the $P_3\simeq40P_0$ `$Q$'-mode drift found by
\cite{fwm81}. Contrary to their results, in our observation there is
no evidence that this feature is drifting. This could be because our
observation was much shorter. \\
{\bf B1845$-$01$^\ast$}: The 2DFS of this pulsar (Fig. \ref{B1845-01}) shows a
broad drifting feature confirming the detection of drifting in this
pulsar by \cite{hw87} at 1414 MHz.\\
{\bf B1919+21}: Both components of this pulsar are clearly drifting and
almost all power in the 2DFS is in the drift feature
(Fig. \ref{B1919+21}). The feature of the leading component shows
horizontal structure. The centroid of the whole feature in the leading
component gives $P_2 = -\degrees{13}\pm3$ with the same $P_3$ value. The
reason for this horizontal structure in the drift feature is, like PSR
B0809+74, that there is a subpulse phase step in the drift bands. This
observation confirms the reported phase step by \cite{pw86} seen
at 1420 MHz.
{\bf B1929+10$^\ast$}: The LRFS peaks at low frequencies
(Fig. \ref{B1929+10}), comparable to what was found by \cite{nuwk82}
at 0.43, 1.7 and 2.7 GHz. \citealt{ohs77} suggested, using 430 MHz
data, that this pulsar drifts. The 2DFS of our observation shows two
broad features with opposite drift sense with two different $P_3$
values.  The most clear drift feature is between the dashed lines and
the other feature is directly above this feature up to $1/P_3\simeq
0.3$. Also \cite{bac73} has seen two features in the LRFS of this
pulsar at 606 MHz and the short period feature appeared to have a
negative drift and the long period fluctuations appeared to be
longitude stationary. A negative drift sense is detected for the short
period feature, but the long period feature shows a positive
drift. Both drift features are arising from the leading half of the
pulse profile. An explanation for the observed behavior is that this
pulsar is a drift mode changer showing different $P_3$ values
with opposite drift senses. There is also an indication for a
$P_3\simeq2P_0$ modulation. There is a strong very narrow spike around
$P_3=5P_0$, which could be caused by a few strong pulses. \\
{\bf B1933+16$^\ast$}: This pulsar shows subpulse modulation over the
whole $P_3$ range (Fig. \ref{B1933+16}). It was found by \cite{bac73}
that there is no preferred subpulse drift sense at 430 MHz, however
regular drifting with $P_3\simeq2.2P_0$ has been reported by
\cite{ohs77} at 430 MHz. We can confirm that there is preferred
positive drifting in a broad feature near the $P_3=2$ alias border.\\
{\bf B1944+17$^\ast$}: This pulsar shows a clear broad drift feature in the
2DFS (Fig. \ref{B1944+17}) and the drifting can clearly be seen by eye
in the pulse-stack. The feature is broad because this pulsar shows
drift mode changes (\citealt{dchr86} at both 430 and 1420 MHz). The
$P_3=13P_0$ `$A$'-mode drift and the $P_3=6.4P_0$ `$B$'-mode drift are
visible in the 2DFS at 0.075 and 0.16 cpp respectively. We see also
evidence for a feature in a different alias mode, although much weaker
than the main feature ($P_2 = \degrees{12.4}\pm0.7 = 30$ cpp and
$P_3 = 22.2\pm14 P_0$). It could be that the zero drift `$C$'-mode
(\citealt{dchr86}) is a drift mode for which the drift sense is
changing continuously.\\
{\bf B2016+28$^\ast$}: The 2DFS of the trailing part of the pulse profile
shows a very broad drifting feature (Fig. \ref{B2016+28}), which is
caused by drift mode changes (e.g. \citealt{ohs77a} at both 430 and
1720 MHz). The leading part of the pulse profile shows the same broad
drift feature, but also a much stronger slow drift mode. This slow
drift mode is probably also seen in the trailing part of the pulse
profile, but less pronounced. The drift bands can be seen by eye in
the pulse-stack.\\
{\bf B2020+28$^\ast$}: The LRFS shows a strong even-odd modulation, similar
to the 1.4 GHz observation of \cite{nuwk82}. At 430 MHz \cite{brc75}
found that both components show an even-odd modulation, but no
systematic drift direction was detected in the leading component. In
our observation the 2DFS of both components of this pulsar contains a
broad drifting feature with opposite drift sense close to the alias
border (Fig. \ref{B2020+28}). There is no evidence that the feature
extends over the alias border, although the feature is not
clearly separated  from the alias border.\\
{\bf B2021+51$^\ast$}: This pulsar is clearly drifting
(Fig. \ref{B2021+51}), consistent with e.g. \cite{ohs77} at 1720
MHz. The drifting is detectable over the whole $P_3$ range. The $P_2$
and $P_3$ values that are given in table \ref{Table_section} are for the region
in the 2DFS between the dashed lines. The centroid of the whole 2DFS
gives $P_2=\degrees{47}\pm4$ and $P_3=4.7\pm0.1 P_0$. It is clear
that the drift rate changes by a large factor during the observation,
which was also observed by \cite{ohs77}. It was suggested by
\cite{ohs77} that maybe the apparent drift direction changes
sporadically. In our observation there is no clear evidence that the
alias mode is changing.\\
{\bf B2044+15$^\ast$}: This observation is contaminated by
interference, however the 2DFS of the trailing component of the pulse
profile convincingly shows a broad drifting feature. Only the 2DFS of
the trailing component shows features and is plotted in
Fig. \ref{B2044+15}. Our observation confirms the drifting found by
\cite{bac81} at 430 MHz.\\

\subsubsection{New drifters (classes ``Dif'' and ``Dif$^\ast$'')}

{\bf B0037+56$^\ast$}: The 2DFS of this pulsar\footnotemark[5]
shows a clear drift feature (Fig. \ref{B0037+56}) which appears to
be extended over the alias border. The drift bands are visible
to the eye in the pulse-stack and the apparent change of drift sense
is also visible. There is also a $P_3=2P_0$ modulation present.\\
{\bf B0052+51}: The trailing component of this pulsar has a broad
drift feature in its 2DFS (Fig. \ref{B0052+51}). There is a hint of
drifting with an opposite drift sign in the first component, but this
is not significant. The spectra also show a $P_3\simeq2P_0$
modulation.\\
{\bf B0136+57$^\ast$}: The drift feature is only detected in the leading part
of the pulse profile and it appears that the feature extends to
the horizontal axis (Fig. \ref{B0136+57}).\\
{\bf B0138+59$^\ast$}: The drift feature in the 2DFS is broad and close to
the horizontal axis (Fig. \ref{B0138+59}). The drift feature
is confirmed in a second observation we made.\\
{\bf B0450+55$^\ast$}: Most of the power of the 2DFS is in the drifting
feature (Fig. \ref{B0450+55}) and the drift bands are visible to the
eye in the pulse-stack. The drift feature is extended to both
alias borders. The leading component of this pulsar shows drifting in
the opposite direction. \\
{\bf B0523+11}: This pulsar has a weak drift feature in the 2DFS of
the trailing component (Fig. \ref{B0523+11}). In the 2DFS of the
leading component there is also a feature with the same $P_3$ value,
but in that feature there is no significant offset from the vertical
axis measured. This means significant drifting is detected in the
trailing component, and longitude stationary subpulse modulation with
the same $P_3$ value in the leading component.  No drifting has been
found at 430 MHz by \cite{bac81}, but our observation shows that this
pulsar is a drifter.\\
{\bf B0525+21$^\ast$}:  Subpulse modulation without apparent drift as
well as some correlation between the subpulses of the two components
of the pulse profile has been detected for this pulsar by \cite{bac73}
at 318 MHz and \cite{tmh75} at 400 MHz. We find that the two
components show broad features to which opposite drift senses can be
associated (Fig. \ref{B0525+21}). The features are possibly extended
toward the $P_3=2P_0$ alias border.\\
{\bf B0919+06$^\ast$}: The power in the 2DFS is over the whole $P_3$
range is measured to be significantly offset from the vertical axis
(Fig. \ref{B0919+06}). The power in the 2DFS peaks toward the
horizontal axis and especially this low frequency excess is offset
from the vertical axis. This can clearly be seen in the bottom plot
which shows a ``shoulder'' at the  left side of the peak. No drifting has
been reported for this pulsar by \cite{bac81} at 430 MHz.\\
{\bf B1039$-$19}: Both components of this pulsar show a clear,
broad drift feature in its 2DFS with the same drift sense
(Fig. \ref{B1039-19}).\\
{\bf B1508+55$^\ast$}: The subpulse modulation of this pulsar has been found
to be unorganized and without a preferred drift sense or a particular
$P_3$ value (\citealt{th71} at 147 MHz). In the 2DFS of our
observation there is a broad drift feature present
(Fig. \ref{B1508+55}), which is offset from the vertical axis
over the whole $P_3$ range.\\
{\bf B1604$-$00$^\ast$}: There are very broad features in both parts of the
pulse profile and both components are drifting with the same drift
sense (Fig. \ref{B1604-00}).\\
{\bf B1738$-$08$^\ast$}: The 2DFS of both halves of the pulse profile have a
broad drift feature with the same drift sense
(Fig. \ref{B1738-08}). In the trailing component there is maybe also a
second weak drift feature present with $P_3\simeq2 P_0$ and the same drift
sense. The average $P_3$ values appear to be significantly different
in the two components, which could be because of different drift mode
changes in the two components. The drifting can be seen by eye in
the pulse-stack.\\
{\bf B1753+52}: The trailing part of the pulse profile shows a broad
drift feature in its 2DFS (second 2DFS in Fig. \ref{B1753+52})
and the rest of the pulse profile (first 2DFS) probably has the
same drift sense.\\
{\bf B1819$-$22}: The 2DFS of this pulsar very clearly shows a
drift feature (Fig. \ref{B1819-22}), which is broadened by mode
changes.  A part of the pulse-stack is shown in Fig.
\ref{Classes_fig}.  A full single pulse analysis will follow in a
later paper.\\
{\bf J1830$-$1135$^\ast$}: The 2DFS of this pulsar with a very long pulse
period (6.2 seconds) shows a drift feature at the $P_3=2 P_0$ alias
border at +100 cpp, which is possibly double peaked
(Fig. \ref{J1830-1135}).\\
{\bf B1857$-$26}: The components at both edges of the pulse profile
are drifting with the same drift sense, which can be seen in
Fig. \ref{B1857-26} as an excess of power in the 2DFS at positive
$P_2$ values. The drift bands are sometimes visible to the eye in the
pulse-stack.  The center part of the pulse profile does not show
drifting in its 2DFS and is therefore not plotted. This pulsar is
known to be a nuller (\citealt{rit76,big92a}), but no drifting is
reported in the literature.\\
{\bf B1900+01$^\ast$}: Drifting is clearly seen over the whole $P_3$ range of
the 2DFS and the top part of the 2DFS is double peaked
(Fig. \ref{B1900+01}). The alias mode of this
pulsar probably changes during the observation.\\
{\bf B1911$-$04$^\ast$}: The low frequency modulation, which is generated in
the trailing part of the pulse profile, is double peaked
(Fig. \ref{B1911-04}). This could indicate the presence of a
subpulse phase jump or swing or that the drift sense changes
constantly during the observation. There seems to exist a preferred drift
sense. \\
{\bf B1917+00$^\ast$}: This pulsar shows a broad drifting component in
its 2DFS, which is visible in the bottom plot as an excess of power at
positive $P_2$ (Fig. \ref{B1917+00}). According to \cite{ran86} a much
longer $P_3\simeq50$ value without a measured $P_2$ was reported in a
preprint by L.A. Nowakowski and T.H. Hankins, but to the best of our
knowledge the paper was never published.\\
{\bf B1952+29$^\ast$}: The drifting in both components of this pulsar is
clearly visible to the eye in the pulse-stack and in the 2DFS
(Fig. \ref{B1952+29}). The drift sense is the same for both
components. \\
{\bf B1953+50$^\ast$}: This pulsar shows a very clear broad drifting feature
in its 2DFS (Fig. \ref{B1953+50}) at low frequencies (right peak at
$P_2=70$ cpp).\\
{\bf B2053+36$^\ast$}: This pulsar has a broad drift feature in its
2DFS at low frequencies which is double peaked (Fig. \ref{B2053+36}).
This could indicate that the drift sense is changing constantly during
the observation, which is supported by the fact that the feature is
extended towards zero frequencies. However also a subpulse phase jump
or swing could produce the double peaked feature. Subpulse modulation
without a drift sense has been reported for this pulsar at 430 MHz by
\cite{bac81}.\\
{\bf B2110+27$^\ast$}: This pulsar shows drifting over the whole $P_3$
range in its 2DFS (Fig. \ref{B2110+27}). The lower part of the 2DFS is
clearly double peaked, which could suggest that the alias mode is
constantly changing during the observation. The upper part of the 2DFS
is not convincingly double peaked. In the pulse-stack drifting is
visible to the eye. The drift bands are probably distorted by nulls,
causing the drift feature in the 2DFS to be extended over the whole
$P_3$ range. Short sequences of drift bands with negative drifting and
a $P_3\simeq6P_0$ can be seen in the pulse-stack as well as some single
drift bands with an opposite drift. A few apparent drift reversals are
visible in the sequence, although nulling makes it difficult to
identify them. \\
{\bf B2111+46}: It has been reported that this pulsar shows subpulse
drift with a positive and negative drift sense, but without either
dominating (\citealt{tmh75} at 400 MHz). We also see subpulse
modulation over the whole $P_3$ range without a preferred drift
direction in the middle and trailing components, but there is some
systematic drift in the leading component of this pulsar
(Fig. \ref{B2111+46}).\\
{\bf B2148+63$^\ast$}: The 2DFS of this pulsar shows broad, triple, well
separated features (Fig. \ref{B2148+63}). The values of $P_2$ and
$P_3$ in table \ref{Table_section} are for the feature as a whole. The centroids of
the individual peaks give $P_2 = -\degrees{8.0}\pm0.2$, $P_3 =
2.4\pm0.3 P_0$ and $P_2 = \degrees{8.3}\pm0.3$, $P_3 = 2.2\pm0.2
P_0$. The most likely interpretation of the 2DFS is that the apparent
drift direction is constantly changing via its $P_3=2P_0$ alias border
(see Fig. \ref{FakeFig} for the expected 2DFS in this scenario). All
other interpretations seems unlikely, because the feature is clearly
extended toward the $P_3=2P_0$ alias border, both sides of the feature
are separated from the vertical axis, this separation is the same on
both sides and one side of the feature contains more power. The latter
indicates that negative drifting dominates in this pulsar.\\
{\bf B2154+40$^\ast$}: This pulsar shows a very broad drift feature in the
2DFS of the leading component of the pulse profile, but no
significant drift is detected in the trailing component
(Fig. \ref{B2154+40}). The feature is probably extended toward the
alias border.\\
{\bf B2255+58}: The very clear drift feature in the 2DFS of this
pulsar (Fig. \ref{B2255+58}) shows horizontal structure (like
observed for instance for PSR B0809+74 and PSR B0320+39). From
the modulation phase profile it follows that the drift bands make a
subpulse phase step of about \degrees{140} in the middle of the pulse
profile.  The longitude resolved modulation index shows a minimum at
the position of the subpulse phase step, as is also observed for PSR
B0809+74 and PSR B0320+39.  The phase step in the drift bands can be
seen by eye in the pulse-stack.\\
{\bf B2324+60$^\ast$}: This pulsar shows a broad drift feature in its 2DFS at
the alias border (Fig. \ref{B2324+60}) and some drift bands can be
seen in the pulse-stack. There is also a strong $P_3\simeq200P_0$
feature detected.\\
{\bf J2346$-$0609$^\ast$}: The 2DFS of the trailing component of the pulse
profile has a drift feature close to the alias border
(Fig. \ref{J2346-0609}). The feature is not clear enough to state if
the drift feature crosses the alias border during the
observation. The spectra also show some low frequency modulation,
especially in the leading component.\\
{\bf B2351+61$^\ast$}: The 2DFS of this pulsar is double peaked at low
frequencies (Fig. \ref{B2351+61}), which could indicate that the
drift direction may constantly change during the observation. Also the
presence of a subpulse phase step or swing could produce this
feature. The centroid is significantly offset from zero, so there
exists a preferred drift sense during the observation.\\

\subsection{Longitude stationary drifters (class ``Lon'')}

{\bf B0402+61}: The 2DFS of the trailing component of this pulsar
shows a broad feature without a preferred drift direction
(Fig. \ref{B0402+61}). The
2DFS and LRFS of the leading component is featureless.\\
{\bf B1846$-$06}: The 2DFS of this pulsar shows a broad feature with a
{\bf positive} value for $P_2$ (Fig. \ref{B1846-06}). The same drift sense
seems to be detected at low frequencies in the 2DFS.\\
{\bf B1937$-$26}: There is no significant preferred drift sense detected
in the 2DFS of this pulsar (Fig. \ref{B1937-26}), but there seems
to be a broad double peaked feature at the alias border.\\
{\bf B1946+35}: The LRFS and 2DFS shows a strong low frequency feature in
both components of this pulsar (Fig. \ref{B1946+35}). No significant
offset from the vertical axis has been detected in the 2DFS.\\
{\bf B2011+38}: The broad feature in the 2DFS of this pulsar may have
a preferred negative value for $P_2$ (Fig. \ref{B2011+38}). The 2DFS
and LRFS of this pulsar increases towards low frequencies and peaks at
$P_3=30\pm15P_0$.\\
{\bf B2106+44}: The 2DFS and LRFS of this pulsar peaks towards
low frequencies and seems to have a positive value for $P_2$
(Fig. \ref{B2106+44}).\\

\subsection{Unconfirmed known drifters}

{\bf B0540+23}: Sporadic bursts of both positive and negative
drift have been reported by \cite{ash82} at 400 MHz and by
\cite{now91} at 430 MHz. No preferred drift direction is detected
in the 2DFS of this pulsar (Fig. \ref{B0540+23}), but in the
pulse-stack short drift bands are seen with different drift senses
confirming the previous reported drifting. Because this pulsar does
not show a preferred drift direction in our observation, this pulsar
is not classified as a drifter in our paper.\\
{\bf B0611+22}: Our observation does not show any features in the 2DFS
of this pulsar (Fig. \ref{B0611+22}), something that has also been
reported by \cite{brc75} at 430 MHz. Drifting with
$P_3=50-100P_0$ has been reported by \cite{fb80} at 430 MHz, who
have analyzed successive integrated pulse profiles. It is not clear if
this kind of drifting is related to subpulse drifting, because in
their analysis the subpulses are not directly measured.\\
{\bf B0656+14}: The modulation index of this pulsar shows a sharp peak
at the leading edge of the pulse profile (Fig. \ref{B0656+14}), which
is caused by a very bright subpulse. A full investigation of this
phenomenon will be published in a upcoming paper. A preferred
negative drift sense has been reported by \cite{bac81} at 430 MHz. We
do not see a preferred drift sense, but there is low frequency
modulation.\\
{\bf B0820+02}: Positive drifting has been reported by \cite{bac81} at
430 MHz. Our observation probably lacks the S/N to confirm this
\ref{B0820+02} \\
{\bf B0950+08}: Drifting has been reported for this pulsar
(e.g. \citealt{bac73} and \citealt{wol80}) with a variable
$P_3\simeq6.5$. This drifting is not visible in the 2DFS of our
observation (Fig. \ref{B0950+08}), but subpulse modulation is
seen over the whole $P_3$ range without a preferred drift sense. The
interpulse has no measured modulation and is not plotted. The
observing frequency of \cite{bac73} was 430 MHz, so it could be that
the drifting of this pulsar is only visible at low observing
frequencies. The observing frequency of \cite{wol80} is not mentioned
in their paper.\\
{\bf B1112+50}: This pulsar is known to show nulling and pulse profile
mode switching and in one of these modes drifting subpulses are
reported (e.g. \citealt{wsw86} at 1412 MHz). There is no clear drift
feature detected in the 2DFS of our observation, but subpulse
modulation is seen over the whole $P_3$ range
(Fig. \ref{B1112+50}).\\
{\bf B1612+07}: Negative subpulse drift has been reported by
\cite{bac81} at 430 MHz for this
pulsar\footnotemark[5]. The 2DFS of our observation is
featureless (Fig. \ref{B1612+07}), which could be because a too low
S/N.\\
{\bf B1918+19}: This pulsar is shown to be a drifter with at least
four drift modes at 430 MHz (\citealt{hw87}). There are no features in
the 2DFS of our observation (Fig. \ref{B1918+19}), which could be
because a too low S/N.\\
{\bf B2315+21}: Drifting with a negative drift sense has been reported
for this pulsar at 430 MHz by \cite{bac81}. Our spectra
(Fig. \ref{B2315+21}) do not show any sign of drifting, what could be
because a too low S/N.

\subsection{Pulsars with low frequency modulation}

A modulation index could be derived from the observations of the
following pulsars and their spectra show an excess of power
toward the horizontal axis (i.e. a ``red'' feature). This means that
there is some correlation between successive pulses, but no
quasiperiodicity.
\begin{center}
\begin{tabular}[htb]{|l|l|l|l|}
\hline
B0011+47 & B0355+54 & B0740$-$28 & B0756$-$15\\
B1706$-$16 & B1754$-$24 & B1800$-$21 & B1804$-$08\\
J1808$-$0813 & B1821$-$19 & B1826$-$17 & B1839+56\\
B1905+39 & B1907+10 & B1914+13 & B1924+16\\
B2323+63 & B2327$-$20 & & \\
\hline
\end{tabular}
\end{center}
{\bf B1804$-$08}: The 2DFS of this pulsar possably shows a broad
drift feature which is generated primarily by the trailing component
of the pulse profile (Fig. \ref{B1804-08}), but after randomizing
the order of the pulses this feature turned out to be not
significant.\\
{\bf B1924+16}: The 2DFS of this pulsar shows a hint of a broad
drifting feature (Fig. \ref{B1924+16}), but scrambling the pulse
stack showed that this drifting is not significantly detected.\\

\subsection{Pulsars with a flat spectrum}

A modulation index could be derived from the observations of the
following pulsars and their spectra show no clear features over the
vertical range. This means that the subpulse modulation appears
disordered or that the S/N ratio is too low.
\begin{center}
\begin{tabular}[htb]{|l|l|l|l|}
\hline
B0105+65 & B0144+59 & B0154+61 & B0353+52\\
B0450$-$18 & B0458+46 & B0531+21 & B0559$-$05\\
B0626+24 & B0906$-$17 & J1022+1001 & B1541+09\\
B1600$-$27 & B1649$-$23 & J1713+0747 & B1717$-$16\\
B1730$-$22 & B1732$-$07 & B1736$-$29 & B1737+13\\
B1745$-$12 & B1749$-$28 & B1756$-$22 & B1758$-$29\\
B1805$-$20 & B1811+40 & B1813$-$17 & B1815$-$14\\
B1817$-$13 & B1818$-$04 & B1820$-$11 & B1822$-$14\\
B1829$-$08 & B1830$-$08 & B1831$-$04 & B1831$-$03\\
B1834$-$10 & B1834$-$04 & J1835$-$1106 & J1839$-$0643\\
B1839+09 & B1841$-$05 & B1842$-$04 & B1842+14\\
J1845$-$0743 & B1848+13 & B1849+00 & J1850+0026\\
B1855+02 & B1855+09 & B1859+03 & B1859+07\\
B1900+05 & B1907+00 & B1910+20 & B1911+13\\
B1914+09 & B1920+21 & B1935+25 & B1943$-$29\\
B2000+32 & B2003$-$08 & B2002+31 & B2022+50\\
J2145$-$0750 & B2148+52 & B2217+47 & B2224+65\\
B2306+55 & B2334+61 & &\\
\hline
\end{tabular}
\end{center}
{\bf B0144+59}: 
Two vertical bands are detected in the 2DFS of the trailing component
(best visible in the bottom plot of Fig. \ref{B0144+59}). This bands
may also be present (although weakly) in the middle component.  A
$P_2\simeq\pm\degrees{2.3}$ (160 cpp) subpulse separation can be
associated with this feature, but no particular $P_3$ value. This
indicates that there is a quasiperiodic intensity modulation in the
pulses with a period of about 1.3 ms, but there is no correlation in
the positions of the subpulses from pulse to pulse.  The same features
are seen in another observation we made of this pulsar. We see the
same kind of phenomenon (a bit more clear) for PSR B2217+47. The
leading component does not show any features and is therefore not
plotted. The spectra shown are calculated using transforms of only 32
pulses in order to reduce the resolution. This makes it more easy to
see the features by eye.\\
{\bf B0531+21}: Both the 2DFS of the main and interpulse of the
Crab pulsar does not show any sign of drifting (Fig. \ref{B0531+21}).
A very large modulation index measured is measured ($m=5$), which
is caused by its giant pulses (\citealt{sr68}).\\
{\bf B0626+24}: The 2DFS of this pulsar shows subpulse modulation over
the whole $P_3$ range (Fig. \ref{B0626+24}). Subpulse modulation
without a drift sense has been reported by \cite{bac81} at 430 MHz.\\
{\bf J1022+1001}: This millisecond pulsar is known to show subpulse
modulation (\citealt{es03b}). The power in the 2DFS of the trailing
component peaks toward $P_3=2P_0$ (Fig. \ref{J1022+1001}), consistent
with the analysis of \cite{es03b} at 1380 MHz (we have used the same
data). The 2DFS shown in the figure is that of the trailing
component.\\
{\bf B1541+09}: This pulsar (Fig. \ref{B1541+09}) is observed to have
a low frequency excess and exhibits mode changes and organized, but
short, drifts in both directions (\citealt{now96} at 430 MHz).\\
{\bf J1713+0747}: This pulsar shows subpulse modulation
(\citealt{es03b} at 1190 and 1700 MHz), but the quality of our
observation is too poor to confirm this (Fig. \ref{J1713+0747}).\\
{\bf B1736$-$29}: The interpulse of this pulsars is not plotted, because
no features are detected in its spectra and no modulation index has
been measured (Fig. \ref{B1736-29}).\\
{\bf B1737+13}: This pulsar shows a clear $P_3$=11-14$P_0$ longitude
stationary subpulse modulation, but no drifting has been detected
(\citealt{rws88} at 1412 MHz). No features appear in the spectra of
our observation (Fig. \ref{B1737+13}). No drifting has been detected
for this pulsar by \cite{bac81} at 430 MHz.\\
{\bf B1749$-$28}: This pulsar shows flat featureless spectra
(Fig. \ref{B1749-28}). A flat featureless fluctuation spectrum has
also been observed for this pulsar at a lower observing frequency
 by \citealt{th71}.\\
{\bf B1818$-$04}: The power in the 2DFS is 
possibly double peaked (Fig. \ref{B1818-04}). This could be
because of the presence of a subpulse phase step or swing or because
of drift reversals. It has been reported that the subpulse modulation
is not well organized (\citealt{th71} and \cite{tmh75} both at 400
MHz).\\
{\bf B1839+09}: Subpulse modulation without any drift sense has been
detected by \cite{bac81} at 430 MHz. No features appear in the spectra
of our observation (Fig. \ref{B1839+09}).\\
{\bf B1842$-$04}: When we first analyzed the spectra of this
pulsar, an extremely bright and surprisingly sharp longitude
stationary $P_3=3.00P_0$ subpulse modulation feature appeared. Folding
the data with a three times longer pulse period revealed that the
pulse period of this pulsar, as reported by \cite{clj+92}, is wrong by
a factor three. It turns out that the correct pulse period of this
pulsar has appeared in the literature (\citealt{hlk+04}) without
a comment about this discrepancy. Private communication with G. Hobbs
revealed that this deviation is first discovered, although apparently not
reported in the literature, by the Parkes multibeam survey. Using the
correct pulse period the spectra are featureless
(Fig. \ref{B1842-04}).\\
{\bf B1842+14}: Subpulse modulation without a drift sense has been
detected by \cite{bac81} at 430 MHz. The spectra of our observation is
featureless (Fig. \ref{B1842+14}).\\
{\bf J1850+0026}: The shown 2DFS of this pulsar
(Fig. \ref{J1850+0026}) is of the trailing peak.\\
{\bf B2022+50}: The interpulse of this pulsar (which also does not
show any features in its spectum) is not plotted
(Fig. \ref{B2022+50}).\\
{\bf J2145$-$0750}: Weak quasi-periodicities around $0.22$ and $0.45$
cpp are visible in the LRFS and 2DFS of the leading component of this
millisecond pulsar (Fig. \ref{J2145-0750}), as has been reported by
\cite{es03b} at 860 and 1380 MHz. We have used the same 21 cm data as
has been used by \cite{es03b}.\\
{\bf B2217+47}: There is no preferred drift sense detected in the
feature in the 2DFS of this pulsar (Fig. \ref{B2217+47}), which would
confirm the observation of \citealt{th71} at 147 MHz. However the 2DFS
shows two vertical bands smeared over the whole $P_3$ range.
This modulation is primarily generated in the right part of the pulse
profile. The interpretation is, like for PSR B0144+59, that
there is a quasiperiodic intensity modulation in the pulses with a
period of about 2.5 ms, but there is no correlation in the positions
of the subpulses from pulse to pulse.\\

\begin{figure}[t]
\rotatebox{270}{\resizebox{!}{0.99\hsize}{\includegraphics[angle=0]{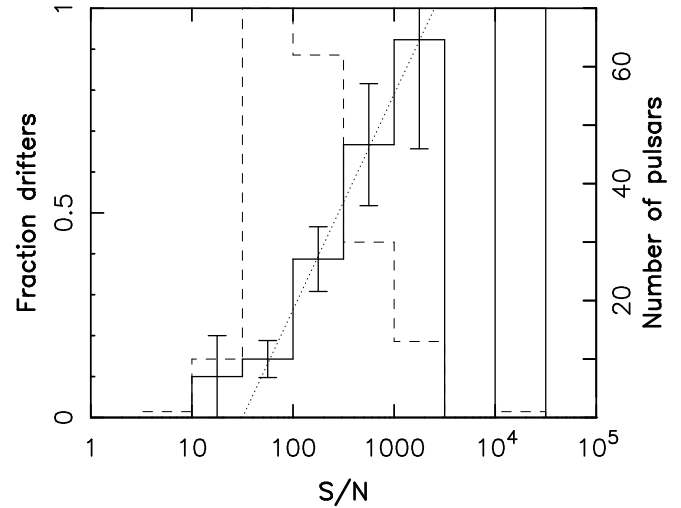}}}
\caption{\label{s2n_hist} The fraction of pulsars we observe to show
the drifting phenomenon (solid line) and the number of pulsars
(dashed line)
versus the measured S/N ratio of the observation. The root-mean-square
(RMS) is calculated as an estimate for the error (if the bin contains
more than one observation). The dotted line is a fit for the S/N
dependence of the chance to detect drifting subpulses.} 
\end{figure}

\begin{figure*}[tb]
\begin{center}
\rotatebox{270}{\resizebox{0.57\hsize}{!}{\includegraphics[angle=0]{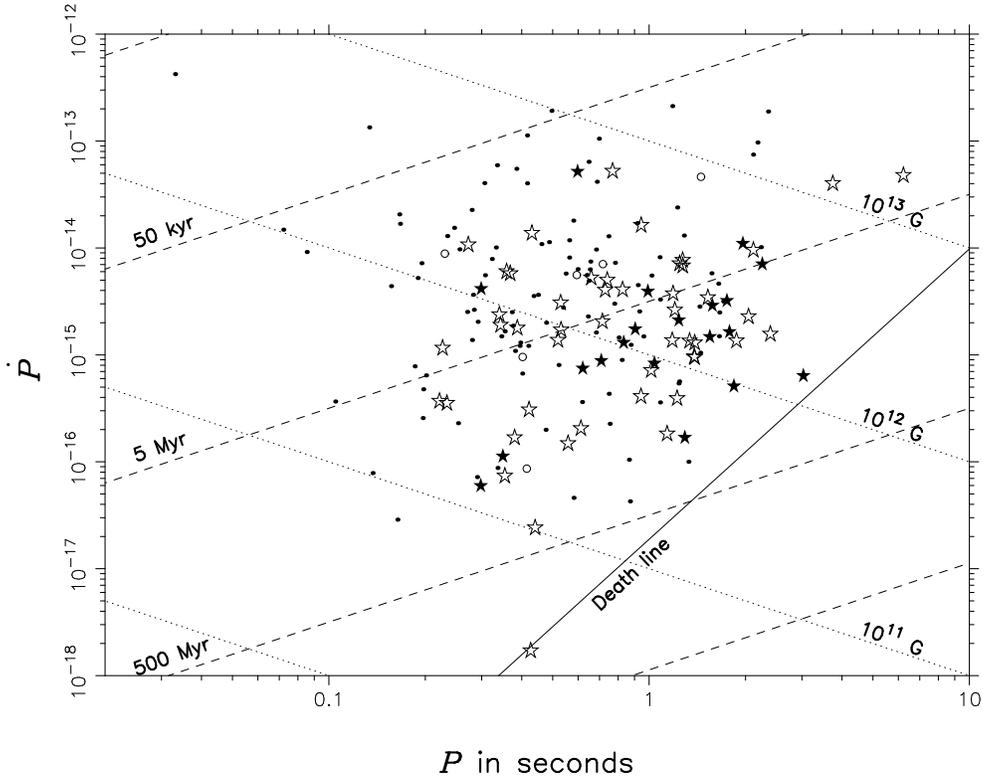}}}
\caption{\label{PPdot}The $P$-$\dot P$ diagram of the analyzed pulsars
(including the low S/N observations), where $P$ is the pulse
period and $\dot P$ its time derivative. The non drifting
pulsars are the dots, the diffuse (Dif and Dif$^\ast$) drifters
are the open stars, the coherent drifters are the filled stars and the
pulsars showing longitude stationary subpulse modulation are the open
circles. Lines of equal surface magnetic field strength and
characteristic ages are plotted, as well as a death line
(\citealt{cr93a}). The millisecond pulsars are not plotted to make the
plot more readable. }
\end{center}
\end{figure*}

\subsection{Pulsars without a measured modulation index}
No modulation index could be measured from our observations of these
pulsars, so no subpulse modulation could be detected.
\begin{center}
\begin{tabular}[htb]{|l|l|l|l|}
\hline
J0134$-$2937 & B1254$-$10 & J1730$-$2304 & B1744$-$24A\\
J1757$-$2223 & B1758$-$23 & J1812$-$2102 & B1821$-$11\\
B1821+05 & J1828$-$1101 & B1832$-$06 & J1852$-$2610\\
J1852+0305 & B1903+07 & B1915+13 & B1916+14\\
B1937+21 & & & \\
\hline
\end{tabular}
\end{center}
{\bf B1821+05}: Subpulse modulation without a drift sense has been
reported by \cite{bac81} at 430 MHz. No features are seen in the 2DFS
(Fig. \ref{B1821+05}) and no modulation index could be measured for
this pulsar. This is probably because of the low S/N of our
observation.\\
{\bf B1915+13}: No features are seen in the spectra of this pulsar by
\cite{brc75} at 430 MHz. In our observation there are also no features
(Fig. \ref{B1915+13}), which could be because of the low S/N of our
observation.\\
{\bf B1937+21}: The spectra of this pulsar are featureless and
there is no modulation index measured. Only the 2DFS of the main pulse
is plotted in Fig. \ref{B1937+21}.\\

\section{Statistics}

\subsection{The numbers}

The selection of our sample of pulsars is based only on the
predicted S/N in a reasonable observing time. While this sample is
luminosity biased, it is not biased on pulsar type or any particular
pulsar characteristics. This allows us, first of all, to address the
very basic question: what fraction of the pulsars show the drifting
phenomenon?

Of the \NrPulsars analyzed pulsars \NrDrifters pulsars show the
drifting phenomenon, indicating that at least one in three pulsars
drift. This is however a lower limit for a number of reasons. First of
all, not all the observations have the expected S/N. This could be
because of interference during the observation, interstellar
scintillation, digitization effects, or because the flux or
pulse width for some pulsars was wrong in the database used. The
latter was confirmed by measuring the pulse with directly from our own
observation. Also there are \NrCandidates pulsars which show
longitude stationary subpulse modulation.
Longitude stationary subpulse modulation could indicate that there is
drifting, but without a preferred drift sense and therefore it could
be related to the same phenomenon.

\begin{figure*}[t]
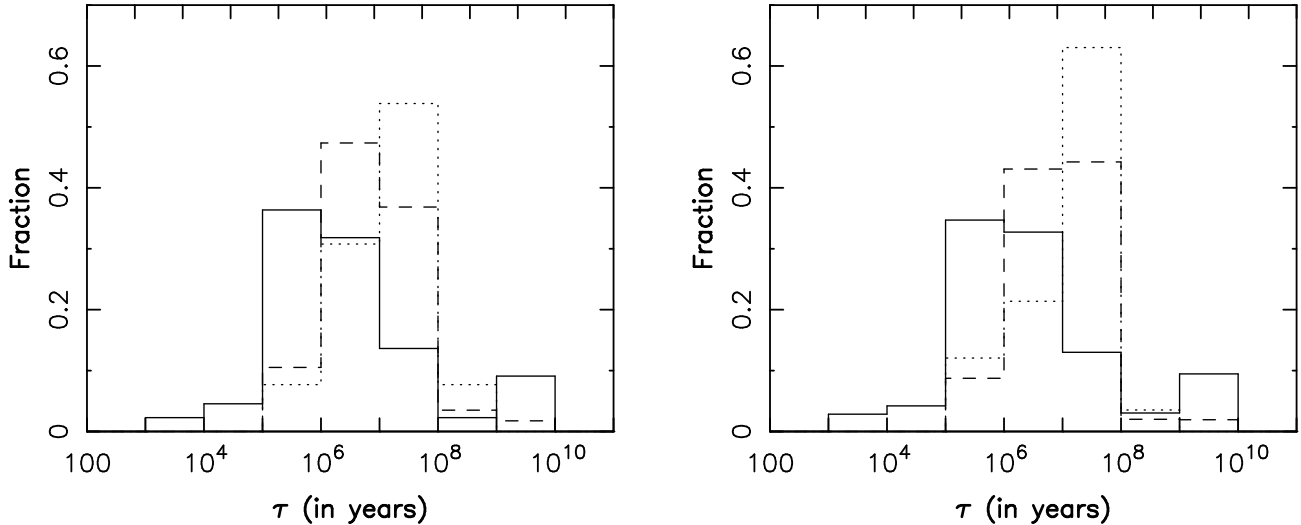

\begin{center}
\rotatebox{270}{\resizebox{!}{0.45\hsize}{\includegraphics[angle=0]{age_hist2.ps}}}\hspace{0.05\hsize}
\rotatebox{270}{\resizebox{!}{0.45\hsize}{\includegraphics[angle=0]{age_hist_snr_corrected.ps}}}
\end{center}
\caption{\label{age_hist}The left panel shows the histogram of
the characteristic ages of the analyzed pulsars with a S/N $\geq100$.
The solid line is the age distribution of the non-drifting pulsars,
the dashed line shows all the drifters and the dotted line shows the
coherent drifters. The right panel shows the ``S/N versus age
bias'' corrected histogram.} 
\end{figure*}

Because many pulsars in our sample were found to be drifting, the
sensitivity of our method to detect drifting could be checked. The S/N
of the observations is determined by comparing the root-mean-square
(RMS) of the off-pulse region of the pulse profile with the power in
the pulse. The width of the pulse was automatically determined by
trying different pulse-widths and maximizing the resulting S/N.  In
Fig. \ref{s2n_hist} the fraction of pulsars that show the drifting
phenomenon is plotted versus the S/N ratio of the observation. One can
see that the probability of detecting drifting is higher for
observations with a higher S/N. The method used is also working for
observations with a low S/N, but not in all cases. This could be
because, in order to detect drifting in observations with a low S/N,
the fraction of pulse energy that is in the drifting subpulses should
be high. Also the fraction of time the pulsar shows drifting subpulses
during the observation should be high and the drifting must be
reasonably coherent.

To make the statistics more independent of the S/N ratio of the
observations, we will do the statistics for the observations with a
S/N $\geq 100$. For \NrPulsarsSNR pulsars in our sample this criterion
is met. By excluding observations with a low S/N a bias toward
long period pulsars (which are observed longer to get enough pulses)
and well studied pulsars (for which long archival data was available)
may be introduced. Therefore all the statistics are checked including
the low S/N observations.

Besides the S/N of the observations, the amount of scatter-broadening
could also influence the probability to detect drifting subpulses. If
the amount of scatter broadening is wider than the subpulse separation
$P_2$, then the sensitivity to detect drifting subpulses will be
severely diminished. There are a number of pulsars in our sample that
seem to show scatter broadening. It is however difficult to
distinguish between a pulse profile that is scatter broadened and a
profile which has an intrinsic exponential tail like shape. Most
pulsars that probably show scatter broadening have a S/N below the
threshold value of 100, so they are therefore excluded from the
statistical analysis (PSRs B1758$-$23, B1817$-$13, B1822$-$14,
J1828$-$1101, B1832$-$06 and B1849+00). There is only one pulsar that
seems to show scatter broadening and has a S/N above the threshold
value (PSR B1815$-$14).
Because the low number of pulsars that are scatter broadened, it seems
very likely that their influence on the statistics can be
neglected. Scatter broadening will be more of an issue for our
subsequent paper, which will focus on a lower frequency study of the
subpulse modulation properties of radio pulsars.

Of the pulsars with high enough S/N observations, \NrDriftersSNR
are detected to be drifters (\DriftPercentageSNR\%) and
\NrCandidatesSNR pulsars show longitude stationary subpulse
modulation (\CandidatePercentageSNR\%). From Fig. \ref{s2n_hist} it
is clear that the real drift percentage could even be higher than
\DriftPercentageSNR\%. This number is consistent with \cite{ash82} and
\cite{bac81}, who found about the same number based on a smaller
sample of pulsars.
There are many reasons why drifting is not expected to be detected for
all pulsars. For instance for some pulsars the line of sight cuts the
magnetic pole centrally and therefore longitude stationary
subpulse modulation is expected. Also, refractive distortion in the
pulsar magnetosphere (e.g. \citealt{pet00,wsv+03,fq04}) or
nulling will disrupt the drift bands, making it difficult or even
impossible to detect drifting. A $P_3=50-100P_0$ has been reported by
\cite{fb80} for PSR B0611+22, indicating that the $P_3$ value for some
pulsars could be very large. In that case longer observations are
needed to detect this drifting and distinguishing it from interstellar
scintillation could become a problem. Some pulsars are known to show
organized drifting subpulses in bursts. In that case some of our
observations could be too short to contain enough drift bands to
detect the drifting.

With a lower limit of one in two it is clear that drifting is at the
very least a common phenomenon for radio pulsars. This implies that
the physical conditions required for the emission mechanism of radio
pulsars to work cannot be very different than the physical conditions
required for the drifting mechanism. Therefore it could well be that
the drifting phenomenon is an intrinsic property of the emission
mechanism, although for some pulsars it is difficult or even
impossible to detect.

\subsection{The drifting phenomenon and the $P$-$\dot P$ diagram}

The unbiased sample of pulsars not only allows us to determine what
fraction of the pulsars show the drifting phenomenon, but also to
correlate the drifting phenomenon with other pulsar parameters. Two
directly measurable and therefore important physical parameters of the
pulsar are the pulse period and its time derivative (spin-down
parameter). From the position of a pulsar in the $P$-$\dot P$ diagram
and assuming magnetic dipole braking, an estimate of the age and the
magnetic field strength can be obtained. Therefore it is useful to try
to correlate the drifting phenomenon with the position of the pulsars
in the $P$-$\dot P$ diagram (Fig. \ref{PPdot}). All the analyzed
pulsars with a measured $\dot P$ are in this diagram\footnote{Except
PSR B1744$-$24A, which has a negative $\dot{P}$.} and the coherent
drifters, diffuse drifters and pulsars showing longitude stationary
subpulse modulation are plotted with different symbols to identify any
trends between position and classification of the pulsars. To make the
plot more readable the millisecond pulsars are not plotted.

The $P$-$\dot P$ diagram reveals that the pulsars that show the
drifting phenomenon are more likely to be found closer to the death
line
and this is more pronounced for the coherent drifters. This suggests
that the population of pulsars that show the drifting phenomenon is on
average older (the pulsar age is defined as $\tau=\frac{1}{2}P/\dot
P$) than the population of pulsars that do not show drifting. This
confirms the result of \cite{ash82}, who also found that drifters are
on average older. Moreover it seems that drifting is more coherent for
older pulsars. This trend is more visible in the pulsar age histograms
(left panel of Fig. \ref{age_hist}), where the nondrifters, the
coherent drifters and all the drifters (both coherent and diffuse) are
plotted separately. This correlation seems to suggest an evolutionary
trend that the subpulse modulation is disordered for the youngest
pulsars and gets more and more organized into drifting subpulses as
the pulsar ages.

The significance of this trend can be determined with the
Kolmogorov-Smirnov test (KS-test), which tells us how likely it is
that two distributions are statistically different. It follows that
the age distribution of the drifters is only \AgeDriftNonDriftperc
likely to be the same as the age distribution of pulsars not showing
the drifting phenomenon. Thus the drifters and nondrifters have
significantly different age distributions.
The KS-test is also used to find out if the coherent drifters have a
separate age distribution. It follows that the coherent drifters are
only \AgeCohNondriftperc and \AgeCohNoCohperc likely to have the same
age distribution as the nondrifting and the diffuse drifting pulsars
respectively.  Therefore the pulsars which drift coherently are likely
to have a separate age distribution.
Although likely, the difference in the age distribution of the
coherent drifters is not detected to be significantly different from
the drifters. Nevertheless it is intriguing to think that drifting
becomes more and more coherent for pulsars with a higher age.  A
larger sample of pulsars is needed to check whether this is
significant. The same trend is found when the low S/N
observations are included in our sample. In that case the drifters and
coherent drifters are respectively \AgeDriftNonDriftpercALL and
\AgeCohNondriftpercALL likely to have the same age distribution as the
nondrifters, confirming the assertion that the drifters and
non-drifters have different age distributions.

The S/N of the observations used in the left panel of
Fig. \ref{age_hist} may be different in each of the age bins, thereby
introducing a ``S/N versus age'' bias. To correct for this effect, for
each distribution and for each age bin the median of the S/N of the
observations was calculated. This median S/N was used to estimate what
the chance was of detecting drifting (the dotted line in
Fig. \ref{s2n_hist}). The distributions of the drifters and coherent
drifters were divided by this chance (because a low S/N implies that
there was only a low chance of detecting drifting) and the
distribution of the non drifters was divided by one minus this chance
(because a low S/N implies a high chance of not detecting
drifting). This gives the ``S/N versus age bias'' corrected age
distributions (right panel of Fig. \ref{age_hist}). As one can see
this correction does not lead to a qualitatively different result.

\begin{figure}[t]
\rotatebox{270}{\resizebox{!}{0.99\hsize}{\includegraphics[angle=0]{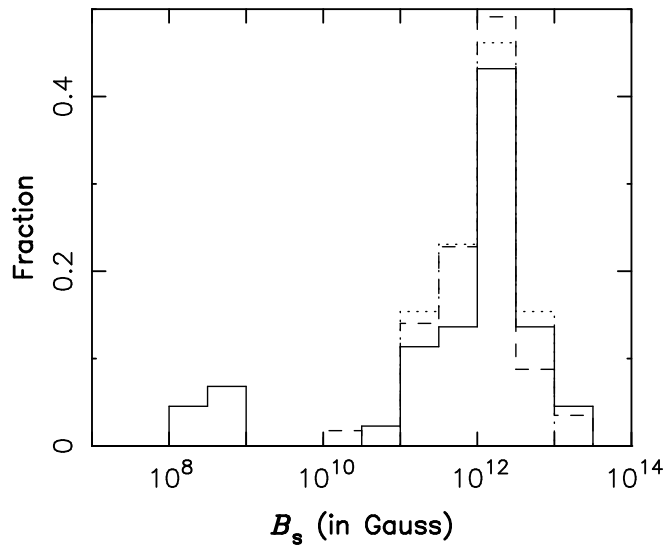}}}
\caption{\label{B_hist} The surface magnetic field strength
histogram of the pulsars which do not show the drifting phenomenon
(solid line), those which do show the drifting phenomenon (dashed
line) and those which drift coherently (dotted line).
Only observations with a S/N $\geq100$ are included. 
} 
\end{figure}

A possible explanation for the age dependence of the drifting
phenomenon is that the drift bands are more distorted for younger
pulsars. One mechanism to distort the drift bands is nulling
(e.g. \citealt{th71,vkr+02,jv04}) and the fraction of time that
pulsars spend in their nulling state (the nulling fraction) is known
to be correlated with the pulsar age. However it has been found by
\cite{rit76} that the nulling fraction is on average higher for older
pulsars, which is confirmed in later studies
(e.g. \citealt{lw95,big92a}). Although the correlation with other
pulsar parameters seems to be stronger, these studies prove that
nulling cannot explain this correlation.

Another possible scenario is that the alignment of the magnetic dipole
axis with the rotation axis has something to do with the observed
trend. Observations seem to show that the angle $\alpha$ between the
magnetic axis and the rotation axis is on average smaller for older
pulsars (e.g. \citealt{tm98}), indicating that the magnetic axis and
the rotation axis becomes more aligned or anti-aligned for older
pulsars. This angle is likely to be an important physical parameter in
the mechanism that drives the drifting phenomenon (for instance the
classical \citealt{rs75} model can only be applied for an
anti-parallel magnetic axis). In this scenario as the pulsar gets
older, the rotation axis and the magnetic axis grows more aligned,
which makes the drifting mechanism more effective or regular.  This
trend is also consistent with the fact that pulsars with a regular
drift pattern tend to have small values for $\alpha$
(\citealt{wri03,ran93b}). However, we have found that the interpulse
pulsar PSR B1702$-$19 is a coherent drifter, suggesting that
coherent drifters can have a large $\alpha$ (an $\alpha$ value of
\degrees{85} and \degrees{90} have been found by \citealt{ran93b} and
\citealt{lm88} respectively). Also the pulse morphology seems to
evolve when the pulsar ages (\citealt{ran83,bgi84}), such that
core single stars are on average younger than pulsars with more
complex profiles. This could make drifting subpulses more likely to
be detected in older pulsars. In the non-radial pulsations model this
trend can also be explained, because the appearance of narrow drifting
subpulses is favored in pulsars with an aligned magnetic axis
(\citealt{cr04}).

An estimate for the component of the surface magnetic field of
pulsars perpendicular to the rotation axis can be directly derived
from the position of the pulsar in the $P$-$\dot P$ diagram
($B_s=10^{12}\sqrt{10^{15}P\dot P}$ Gauss). The histograms of the
magnetic field strengths of the three different groups of pulsars
(Fig. \ref{B_hist}) do not show a clear trend, which is confirmed by
the KS-test.
It follows that the magnetic field strength distribution of the
nondrifters has a chance of $\sim$\BNonCohpers and
$\sim$\BNonDriftperc to be statistically the same as the
distributions of all the drifters (both the coherent and diffuse
drifters) and the coherent drifters respectively. This means that the
magnetic field strength distributions are not significantly
different. If the low S/N pulsars are included, the magnetic field
distributions are more likely to be the same than to be different.

It seems that the drifting phenomenon is only weakly correlated with,
or even independent of magnetic field strength.  This is consistent
with the large fraction of pulsars that are found to show the drifting
phenomenon, because the drifting phenomenon is too common to require
very special physical conditions.

\subsection{The drifting phenomenon and the modulation index}
\label{ModulationSection}

The drifting phenomenon is a form of subpulse modulation, so the
longitude-resolved modulation index $m_i$
(Eq. \ref{modulation_equation}) is an obvious parameter to try to
correlate with the drifting phenomenon. Because the longitude-resolved
modulation index can vary a lot with pulse longitude, as can be seen
in the figures in appendix \ref{Figures_ref}, it is a somewhat
arbitrary what one should call \em the \rm modulation index. The
longitude-resolved modulation index of many pulsars do show a minimum
in the middle of the pulse profile where the total intensity is
relatively high. This means that if the S/N of an observation
increases, the average modulation index will also increase.  This
is because then the modulation index can be measured at pulse
longitudes farther away from the peak intensity of the pulse profile
where the longitude-resolved modulation index tends to be higher. We
have therefore chosen \em the \rm modulation index $m$ to be the
longitude-resolved modulation index $m_i$ at the pulse longitude bin
$i$ where $m_i$ has its minimum value. This definition should make the
modulation index a more S/N independent number than for instance
the average of the longitude-resolved modulation index. The same
definition is used by \cite{jg03} to measure the modulation index.

\begin{figure}[tb]
\begin{center}
\rotatebox{270}{\resizebox{!}{0.94\hsize}{\includegraphics[angle=0]{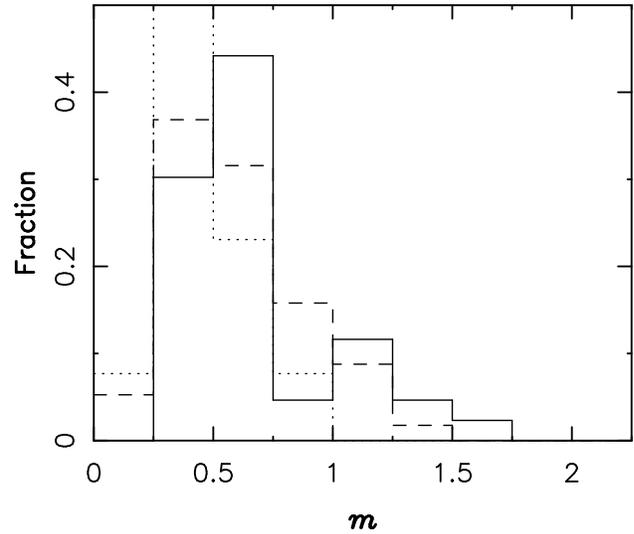}}}
\end{center}
\caption{\label{mod_hist} The modulation index distribution of the
pulsars that do not show the drifting phenomenon (solid line), that do
show the drifting phenomenon (dashed line)
and of the pulsars that drift coherently (dotted line).
Observations with a S/N $<100$ and PSR B0531+21 (with a measured
$m=5$) are not included this plot.} 
\end{figure}

Modulation index histograms are shown in Fig. \ref{mod_hist}. Readily
apparent is the trend that pulsars that show the drifting phenomenon
more coherently have on average a lower modulation index. There seems
to be no significant difference in the modulation index of the pulsars
that do and do not show the drifting phenomenon. The significance of
this trend is also checked with the KS-test and the modulation index
distribution of the drifters is $\sim$\MMinDrNonDr as likely to be the
same as the distribution of the nondrifting pulsars.  The coherent
drifting distribution is only \MMinCohNonDr and \MMinCohNonCoh likely
to be the same as the nondrifters and the diffuse drifting
distributions respectively. These numbers are too high to state that
the modulation index distribution of the coherent drifters are
significantly different.

While the trend seen in Fig. \ref{mod_hist} is not shown to be
statistically significant, it is intriguing and a larger sample of
pulsars is needed to determine its true significance. If the
correlation is proven to be significant it would indicate that pulsars
that show coherently drifting subpulses have on average a lower
modulation index. Although this trend may appear counterintuitive
because drifting subpulses imply subpulse modulation, it can be
explained.

From the summation in Eq. \ref{sigma_equation} (and demonstrated in
Fig. \ref{ScrambleFig}) it follows that the modulation index is
independent of whether the subpulses appear randomly or organized and
from Eq. \ref{modulation_equation} it follows that the modulation
index is even independent of the drift band separation $P_3$ in the
case of a coherent drifter. However, if the number of subpulses per
pulse is large, the subpulses could possibly overlap causing the
intensity to change less from pulse to pulse resulting in a lower
modulation index (\citealt{jg03}). It must be noted that since sparks
cannot physically overlap on the polar cap, this only works if there
is significant broadening in the mapping from polar cap to the
radiation beam pattern. Another parameter influencing the modulation
index is the width of the subpulse intensity distribution. If the
subpulses have a narrow subpulse intensity distribution, e.g. the
subpulses have more equal intensities, the modulation index will also
be lower. A clear example of this effect is the huge measured
modulation index of PSR B0531+21 ($m=5$), caused by its giant pulses
(\citealt{sr68}).

To explain the trend that pulsars that show the drifting phenomenon
coherently have on average a lower modulation index, these pulsars
must either have on average more subpulses per pulse or the subpulse
intensity distribution must be more narrow. In the sparking gap model
it seems reasonable that the subpulses of the coherent drifters have
more equal intensities. Coherent drifting could indicate that the
electrodynamical conditions in the sparking gap are stable, which
could be the reason why the subpulses have on average more equal
intensities.  Also the presence of subpulse phase steps results in a
minimum in the longitude resolved modulation index
(\citealt{esv03,es03c}). This effect can be seen in the longitude
resolved modulation index of PSR B0320+39, PSR B0809+74, PSR B1919+21
and the new drifter PSR B2255+58. Also PSR B0818$-$13 shows a minimum
in its longitude resolved modulation index at the position of its
subpulse phase swing. It is argued by \cite{esv03} that the local
reduction of the modulation index accompanied by a rapid swing in the
modulation phase profile are the result of interference between two
superposed drifting subpulse signals that are out of phase. It is not
unlikely that interference can only occur if the drifting is coherent,
which could explained the trend.

It should also be noted that the modulation index of a purely sinusoidal
subpulse signal results in a modulation index of
$1/\sqrt{2}$. Subpulse patterns with different waveforms or drift band
shapes will generally have larger modulation indices. Many pulsars
have a modulation index which is significantly lower than this
value. This implies that the pulsar emission has both a subpulse
signal and a non-varying component, which could indicate the presence
of superposed out of phase subpulse signals.

Another explanation for this trend would be that refraction is
perhaps more dominant for pulsars that do not show coherently drifting
subpulses. The pulse morphology could well be influenced by refractive
properties of the pulsar magnetosphere
(\citealt{lp98,pet00,wsv+03,fq04}), so it could be that for some
pulsars the organized drifting subpulses are more refractively
distorted than for others. For those pulsars the coherent drifting is
distorted in this scenario, causing the subpulses to appear more
disordered in the pulse window. Moreover it is expected that the
intensities of the individual subpulses varies more because of lensing
(e.g. \citealt{pet00,fq04}) and possible focusing of the radio
emission (\citealt{wsv+03}), causing the modulation index to be higher
in those pulsars.

\begin{figure}[tb]
\rotatebox{270}{\resizebox{!}{0.99\hsize}{\includegraphics[angle=0]{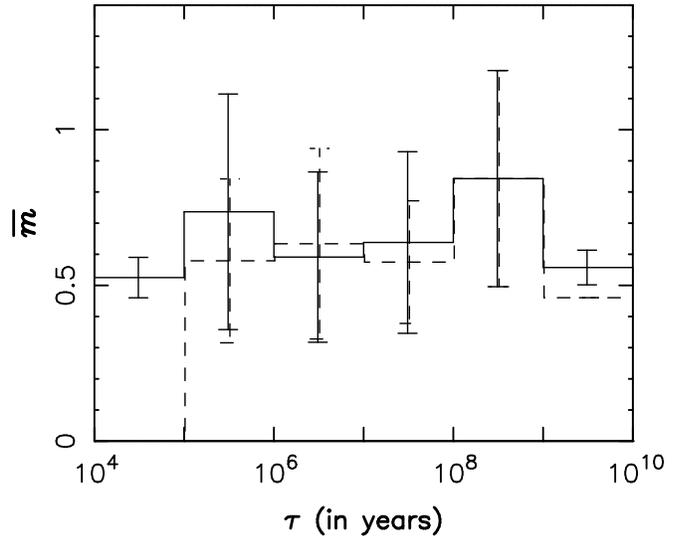}}}
\caption{\label{age_mod_hist}
The average modulation index versus pulsar age
histogram for the pulsars showing the drifting phenomenon (dashed
line) and all the pulsars with a measured modulation index (solid
line). The RMS is calculated as an estimate for the error (if the bin
contains more than one pulsar). Pulsars with a S/N $< 100$ and
PSR B0531+21 ($\tau=1240$, $m=5$) are not included in this plot.
}
\end{figure}

If a correlation between the drifting phenomenon and both the pulsar
age and the modulation index exist, there could also be a correlation
between pulsar age and modulation index. The modulation index versus
pulsar age histogram is plotted in Fig.  \ref{age_mod_hist} and it is
clear that no significant correlation is found, indicating that the
modulation index is the same for pulsars with different ages. This
seems to suggest that a high pulsar age and a low modulation index are
two independent factors affecting the likelihood that a pulsar will
exhibit coherently drifting subpulses.

\begin{figure*}[tb]
\begin{center}
\rotatebox{270}{\resizebox{!}{0.7\hsize}{\includegraphics[angle=0]{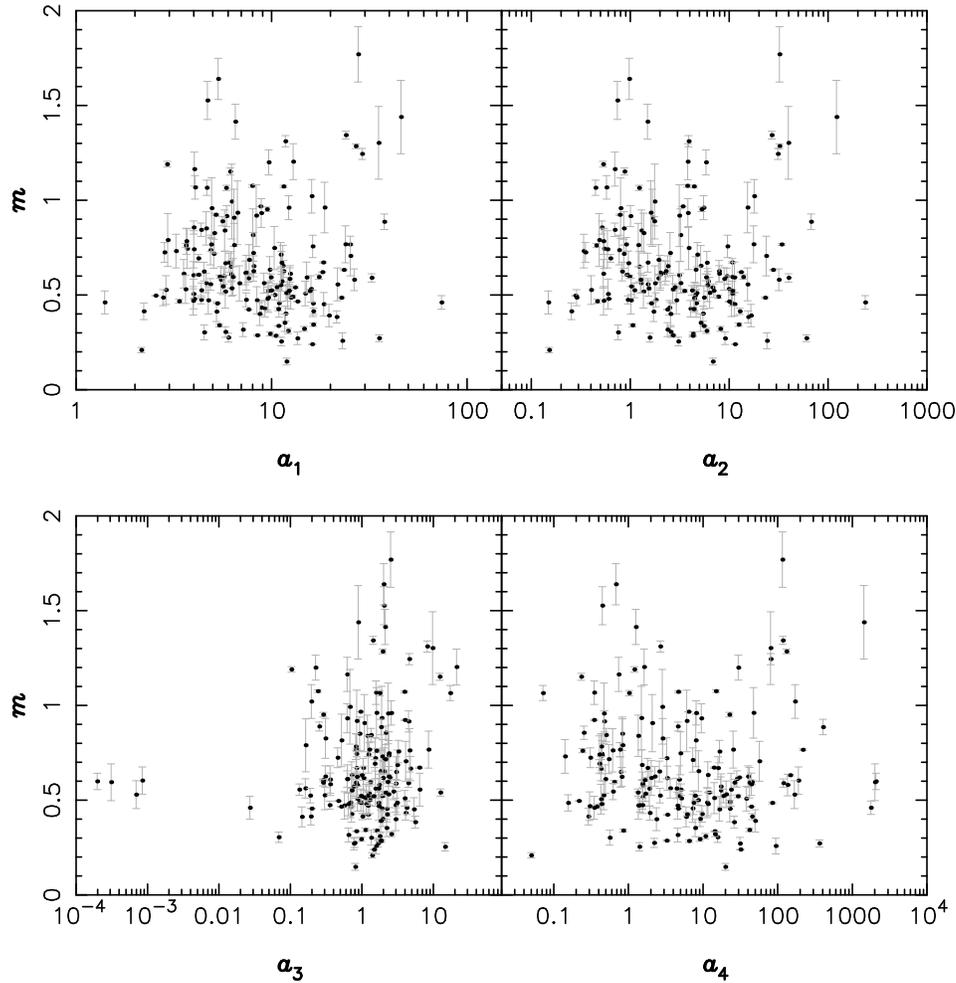}}}
\end{center}
\caption{\label{Complexity}The modulation index for all analyzed
pulsars which have a measured modulation index (except PSR B0531+21)
versus the four complexity parameters as described in the text. 
} 
\end{figure*}

The modulation index of core type emission is observed by
\citealt{wab+86} to be in general lower than that of conal type of
emission. This is also a consequence of the \citealt{gs00} model. In
the sparking gap model, the drifting phenomenon is associated with
conal emission and therefore expected to be seen in pulsars with an on
average higher modulation index. The modulation index distributions of
the drifters and nondrifters are likely to be the same, so drifting
phenomenon appears not to be an exclusively conal phenomenon (as
suggested by \citealt{ran86}). If well organized coherent drifting is
an exclusively conal phenomenon, it is expected that coherent drifters
have an on average a higher modulation index, exactly opposite to the
observed trend.

Also, if drifting is an exclusively conal phenomenon, no drifting is
expected for pulsars classified as ``core single stars''. Although
this may be true for many cases there are some exceptions. The
interpulse of PSR B1702$-$19 is classified as a core single star
(\citealt{ran90} and references therein) and shows a clear and narrow
$P_3$ feature. The diffuse drifter B2255+58 is another good example of
a drifter that is classified as a core single star. Most of the core
single stars that show drifting are diffuse Dif$^\ast$ drifters (PSR
B0136+57, B0823+26, B1642$-$03, B1900+01, B1911$-$04, B1953+50 and
B2053+36). The coherent drifter PSR B1844$-$04 has been classified as
a core single or a triple profile. This means that it is questionable
if the lack of of ordered subpulse modulation is a useful criterion to
identify core emission (as suggested by \citealt{ran86}). It must be
noted that the classification can be frequency dependent, so core
single pulsars at low frequencies could show conal emission at higher
frequencies. The many core single stars that appear to be drifting
stresses the importance of being unbiased on pulsar type when studying
the drifting phenomenon.

\subsection{Complexity parameter}

In the framework of the sparking gap model (e.g.
\citealt{rs75,gs00,gmg03}) the subpulses are generated (indirectly) by
discharges in the polar gap (i.e. sparks). Each individual spark
should emit nearly steady, unmodulated radiation, so the modulation of
the pulsar emission is due to the changing positions of the subpulses
in the pulse window and the number of visible sparks in different
pulses. The more sparks there are visible in the pulse window, the
less the intensity will change from pulse to pulse because the
subpulses could overlap. The number of sparks that fits on the polar
cap is quantified by the complexity parameter (\citealt{gs00}) and
therefore one expects an anti-correlation between the modulation index
$m$ (which is a measure for how much the intensity varies from pulse
to pulse) and this complexity parameter (\citealt{jg03}). As
noted in the previous subsection, this only works if there is
significant broadening in the mapping from polar cap to the radiation
beam pattern.

The complexity parameter is a function of the pulse period and its
derivative and its precise form depends on the model one assumes for
the pulsar emission. By correlating the modulation index of a sample
of pulsars with various complexity parameters as predicted by
different emission models one could try to distinguish which model
best fits the data (\citealt{jg03}). We have correlated the
modulation indices in our sample of pulsars with the complexity
parameter of four different emission models as derived by \cite{jg03}
and \cite{gs00}:
\begin{equation}
\begin{array}{lllll}
a_1&=&5(\dot P/10^{-15})^{2/7}(P/1\mathrm{s})^{-9/14}, &a_2=&\sqrt{\dot PP^{-3}}\\
a_3&=&\sqrt{P\dot P}, &a_4=&\sqrt{\dot P/P^{-5}}
\end{array}
\end{equation}
These are the complexity parameters for the sparking gap model
(\citealt{gs00}), continuous current outflow instabilities
(\citealt{as79}; \citealt{ha01a}), surface magnetohydrodynamic wave
instabilities (\citealt{lou01e}) and outer magnetospheric
instabilities (\citealt{jg03}) respectively. 

\begin{figure}[tb]
\begin{center}
\rotatebox{270}{\resizebox{!}{0.99\hsize}{\includegraphics[angle=0]{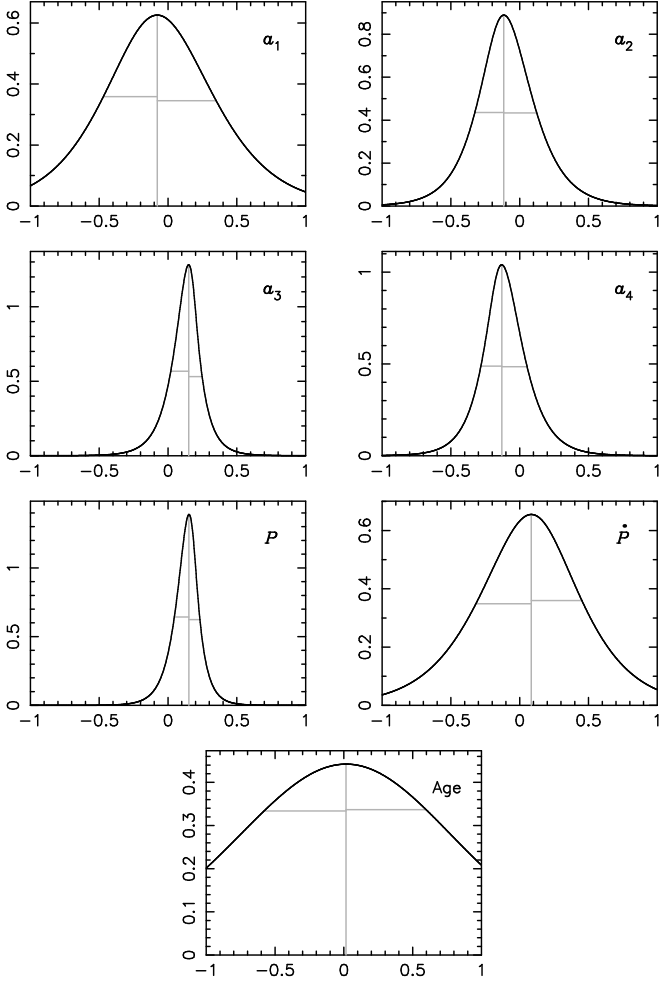}}}
\end{center}
\caption{\label{Probability}The probability functions $P(\rho)$
of the correlation coefficient $\rho$ for the four complexity
parameters, the pulse period, the spin-down parameter and the age of
the pulsar. The position of the maximum as well as the 1-$\sigma$
widths are indicated as well.} 
\end{figure}

Physically, $a_1$ is proportional to the polar cap radius $r_p$
divided by the gap height $h$ as predicted in the sparking gap model
and is therefore a measure for the number of sparks across the polar
cap. The parameter $a_2$ is proportional to the acceleration
parameter, which is the surface magnetic field strength $B_\mathrm{s}$
divided by $P^2$. This acceleration parameter is proportional to total
current outflow from the polar cap, and roughly to the circulation
time of the sparks expressed in pulse periods. The acceleration
parameter is also proportional to the square root of the spin down
energy loss rate. Finally $a_3$ and $a_4$ are proportional to
respectively the magnetic field strength at the surface and at the
light cylinder.

According to \cite{jg03} the anti-correlation between the modulation
index and the complexity parameter will be masked by viewing angle
issues in conal emission, so one might get better results by including
only pulsars that are known to emit core emission.  Because, as
discussed in Sect.  \ref{ModulationSection}, it is not clear how
drifting subpulses relate to the morphological classification of
the pulsar, all pulsars with a measured modulation index are included
in our sample. The modulation index is chosen to be the minimum in the
longitude-resolved modulation index (like in \citealt{jg03}), which
should give the best estimate for the modulation index of the core
emission if present in the pulse profile.

The modulation index versus the four complexity parameters plots
are shown in Fig. \ref{Complexity}. To find out if there exist an
(anti-)correlation without fitting a specific function to the data a
rank-order correlation is used. This means that the rank of the values
among all the other values is used rather than the values itself. This
implies that the correlation coefficient is identical for the set of
points ($x_i$, $y_i$) and ($F(x_i)$, $G(y_i)$), as long as the
functions $F$ and $G$ are monotonic functions. This means that for
instance, because $a_2$ is proportional to the square root of the spin
down energy loss rate, the correlation coefficient of $a_2$ and the
modulation index will be same as the correlation coefficient of the
the spin down energy loss rate and the modulation index.

Following \cite{jg03} we have used the Spearman rank-ordered
correlation coefficient $\rho$ and its significance parameter $\Delta$
(\citealt{ptv+92}). A problem arises when one wants to include the
uncertainties of the data points, because then the rank of the values
is not uniquely defined anymore. As one can see in
Fig. \ref{Complexity}, the errorbars are overlapping each other, so
they should be included in the analysis. The significance parameter
$\Delta$ does not include the uncertainties on the data points and is
therefore not directly usable to estimate the significance of $\rho$.

\begin{table}[!tb]
\begin{center}
\begin{tabular}{|l|r@{$\;$}l|}
\hline
Parameter & correlation&coefficient ($\rho$)\\
\hline
$a_1$, $r_\mathrm{p}/h$ & $-0.07\!$&$^{+0.4}_{-0.4}$\\[3pt]
$a_2$, $B_\mathrm{s}/P^2$  & $-0.11\!$&$^{+0.3}_{-0.3}$\\[3pt]
$a_3$, $B_\mathrm{s}$ & $0.14\!$&$^{+0.12}_{-0.16}$\\[3pt]
$a_4$, $B_\mathrm{lc}$ & $-0.12\!$&$^{+0.23}_{-0.20}$\\[3pt]
$P$  & $0.14\!$&$^{+0.12}_{-0.15}$\\[3pt]
$\dot P$ & $0.1\!$&$^{+0.4}_{-0.4}$\\[3pt]
Age & $0.0\!$&$^{+0.6}_{-0.6}$\\[3pt]
\hline
\end{tabular}
\end{center}
\caption{\label{ProbabilityTable}The correlation coefficients and
their significance as derived from Fig. \ref{Probability}.}
\end{table}

To include the uncertainties in the analysis we have used a Monte
Carlo approach. The data points are replaced with Gaussian
distributions with a width corresponding to the 1-$\sigma$
uncertainties of the measurements. In each integration step a point is
randomly picked from these distributions. Instead of calculating
$\rho$ and $\Delta$ directly from the data points, we calculate them
for the randomly choosen points. So for each integration step we
randomly select a set of points for which we get a $\rho$ and
$\Delta$. The probability distribution $P(\rho)$ is calculated by
averaging the Gaussian distributions centered around the calculated
values of $\rho$ with a 1-$\sigma$ width $\Delta$. The calculated
probablility distributions are plotted in Fig. \ref{Probability}. The
position of the peak of the probability distribution corresponds to
the most likely value of the correlation coefficient and the
1-$\sigma$ width of the peak is a measure for the significance of the
correlation coefficient.

The results of this analysis are tabulated in table
\ref{ProbabilityTable}. Based on a sample of 12 pulsars, \cite{jg03}
concluded that the sparking gap model ($a_1$) showed the highest
anti-correlation and that the surface magnetohydrodynamic wave
instabilities ($a_3$) is unlikely. Also in this enlarged sample, $a_3$
shows the least evidence for an anti-correlation (it is even more
likely that the modulation index is positively correlated with
$a_3$). The strongest anti-correlations are found for $a_2$ and $a_4$,
which corresponds respectively to continuous current outflow
instabilities and outer magnetospheric instabilities. However, none of
the correlations are significantly inconsistent with an
anti-correlation, and therefore none of the models can be ruled out
based on these observations.

One can also see that the modulation index is uncorrelated with the
age of the pulsar, which is consistent with
Fig. \ref{age_mod_hist}. There is a hint that the modulation index is
weakly correlated with the pulse period and the surface magnetic field
strength $B_\mathrm{s}$.

\subsection{Properties of drift behavior}

The value of $P_3$ is observed to be independent of the observing
frequency (\citealt{ikl+93}), but the value of $P_2$ could vary a
little (e.g. \citealt{es03b}). Moreover observations show that
measuring a value for $P_2$ can be far from trivial
(e.g. \citealt{es03c}) and it is only a meaningful parameter if the
drift bands are linear. This means that correlating $P_3$ with other
pulsar parameters is the most direct way to find out if the drift rate
depends on any physical parameters of the pulsar. The strongest
correlation is expected to be found when $P_2$ is constant for
different pulsars. Such a correlation would be a very important
observational restriction on pulsar emission models.

\begin{figure}[tb]
\rotatebox{270}{\resizebox{!}{0.99\hsize}{\includegraphics[angle=0]{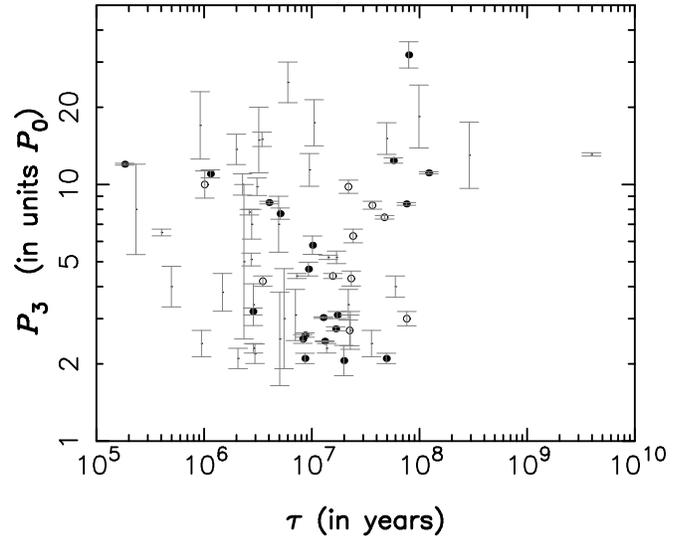}}}
\caption{\label{p3age} The measured value of the vertical drift
band separation $P_3$ versus the pulsar age
of all the pulsars showing the drifting phenomenon. The coherent
drifters are the filled circles, the Dif drifters (with drift feature
clearly separated from the alias borders) are the open circles and the
Dif$^\ast$ drifters are the small dots.
} 
\end{figure}

A significant correlation between $P_3$ and the pulsar age has been
reported in the past (\citealt{wol80,ash82,ran86}). As one can
see in Fig.  \ref{p3age} there is no clear correlation present in our
data, which is confirmed by $\chi^2$-fitting. The figure looks
qualitatively the same as Fig. 4 of \cite{ran86}, although she only
plots ``conal $P_3$ values''. As discussed in Sect.
\ref{ModulationSection}, it is not clear if one should make this
distinction.

There is no correlation found between $P_3$ and the magnetic field
strength, contrary to what was reported by \cite{wol80} and
\cite{ash82}. Also there seems to be no correlation between $P_3$ and
the pulse period (consistent with \citealt{wol80}, contrary to the
tendency reported by \citealt{bac73}). The evidence for a pulsar
subpopulation located close to the $P_3=2P_0$ Nyquist limit
(\citealt{wri03,ran86}) also seems to be weak.

In a sparking gap model one would expect that the spark-associated
plasma columns drift because of an $\mathbf{E}\times \mathbf{B}$
drift, which depends on both the pulse period and its derivative
(e.g. \citealt{rs75,gs00}). The absence of any correlation between
$P_3$ and a physical pulsar parameter is difficult to explain in this
model, unless many pulsars in our sample are aliased. If a pulsar is
aliased a higher $\mathbf{E}\times \mathbf{B}$ drift can result in a
lower $P_3$ value and visa versa, making $P_3$ not a direct measure of
the $\mathbf{E}\times \mathbf{B}$ drift. In the plot a distinction is
therefore made between the coherent drifters, the diffuse Dif
drifters and the Dif$^\ast$ drifters (the latter are probably more
likely to be aliased), but this separation does not reveal a
significant correlation. Also if $P_2$ is highly variable from pulsars
to pulsar, any correlation with $P_3$ is expected to be weaker.

There are a number of pulsars found that show evidence for drift
reversals. This kind of behavior is known for PSR B2303+30 which shows
drift reversals around $P_3=2P_0$ (\citealt{rwr05}). This is confirmed
in the 2DFS of our observation and two other other pulsars that very
clearly show the same kind of behavior in its 2DFS are found: PSR
B2148+63 and PSR B2310+42. Another pulsar that is known to show drift
reversals is PSR B0826$-$34 (\citealt{ggk+04,elg+05}). This pulsar
continuously changes the apparent drift direction via longitude
stationary subpulse modulation. There are a few more pulsars found
which possibly show the same kind of subpulse behavior: PSRs B0037+56,
B1642$-$03, B1944+17, B2110+27, B2351+61. The evidence for drift
reversals is less clear than for the afore mentioned $P_3\simeq2P_0$
pulsars. If pulsars are proven to show drift reversals via longitude
stationary subpulse modulation and one believes that the physical
drift direction of the emission entities cannot change drift
direction, it would imply that the the drifting in both drift
directions is aliased.

A correlation between the drift direction and the pulsar spin-down was
reported by \cite{rl75}, such that a high spin-down is correlated with
positive drifting. The same trend was also found by \cite{ash82} and
\cite{bac81}, although its significance was less. In our sample there
is no significant correlation between the drift direction and the
pulsar spin-down. Also there is no significant difference between the
number of positive and negative drifters.

\label{SubpulsePhase}
The value of $P_3$ is a much better defined parameter than $P_2$
if the drift bands are non linear. Although this makes it difficult to
find any correlations with $P_2$, the fact that the drift bands can be
non linear is very interesting by itself. The drift bands of for
instance PSR B0809+74 (\citealt{es03c,pw86,wbs81}) and PSR B0320+39
(\citealt{esv03,es03c}) show subpulse phase steps and we find that the
new drifter PSR B2255+58 also shows a phase step. Although the
carousel model predicts curved drift bands under certain geometries,
it cannot explain sharp discontinuities of this type. A way out could
be that the observed discontinuities are caused by local
irregularities of the magnetic field (\citealt{wbs81}).

It is argued by \cite{esv03} that the local reduction of the
modulation index accompanied by a rapid swing in the modulation phase
profile are the result of interference between two superposed drifting
subpulse signals that are out of phase. In the non-radial pulsations
model subpulse phase steps could be explained (\citealt{cr04}), but it
has trouble explaining how the modulation phase profile can be anything
but piecewise linear. Visually curved drift bands are expected by
applying subpulse amplitude windowing (pulse longitude dependent
subpulse intensities), as pointed out by \citealt{cr04}. However it is
observed that the phase profile of, for example, PSR B0818$-$13 makes
a swing in the middle of the pulse profile. The phase profile is not
sensitive for subpulse amplitude windowing and therefore the subpulse
phase swing cannot be explained by the non-radial pulsations model.

\section{Summary and conclusions}

Subpulse modulation is shown to be very common for radio pulsars. Of
the \NrPulsars analyzed pulsars {\NrPulsarsModulation} are shown to
exhibit subpulse modulation.
The measured upper limits on the modulation index of many of the
pulsars that do not have a measured modulation index were high,
indicating that pulsars without any subpulse modulation are probably
rare. The number of pulsars that are known to show the drifting
phenomenon is significantly expanded by \NrNewDrifters.
Our sample of pulsars is not biased on pulsar type or any particular
pulsar characteristics, which allows us to do meaningful statistics on
the drifting phenomenon.

As the drifting phenomenon is thought to be exclusively a conal
phenomenon, the modulation index of the drifters is expected to be on
average higher than the modulation index of the nondrifters.
The absence of such a correlation (and possible opposite trend for the
coherent drifters) seems to suggest that drifting is not exclusively
related to conal emission.  Furthermore a number of pulsars
classified as core single stars are found to show drifting,
which stresses the importance to be unbiased on pulsar type when
studying the drifting phenomenon.

Of the \NrPulsars analyzed pulsars \NrDrifters are shown to exhibit
the drifting phenomenon (of which \NrConvinedPulsars drift coherently
or have drift features clearly separated from the alias
borders), which means that at least one in three pulsars show
drifting.
If the observations that had little chance of detecting the drifting
phenomenon because of an insufficient S/N are ignored, it is shown
that at least some 55\% of the pulsars drift.
This implies that the physical conditions required for the drifting
mechanism to work cannot be very different than the required physical
conditions for the emission mechanism of radio pulsars, which is
consistent with the absence of a strong correlation between the
drifting phenomenon and the magnetic field strength. It could well be
that the drifting phenomenon is an intrinsic property of the emission
mechanism, although drifting could in some cases be very difficult or
even impossible to detect.

The set of modulation indices of our sample of pulsars is not
shown to be inconsistent with four complexity parameters as derived
for different emission models. Therefore none of the models can be
ruled out based on the present observations. Other correlations are
found which should be explained by emission models.
The population of pulsars that show the drifting phenomenon are on
average older than the population of pulsars that do not show drifting
and it seems that drifting is more coherent for older pulsars.

Although significant correlations between $P_3$ and the pulsar age,
the magnetic field strength and the pulse period have been reported
previously, we find no such correlations in our enlarged sample. In a
sparking gap model one would expect that the subpulses drift because
of an $\mathbf{E}\times \mathbf{B}$ drift, which depends on both the
pulse period and its derivative.
The absence of a correlation between $P_3$ and any physical pulsar
parameter is difficult to explain in such a model, unless many pulsars
in our sample are aliased or if $P_2$ is highly variable from
pulsar to pulsar.

No significant correlation is found between the modulation index and
the pulsar age. This seems to suggest that a high pulsar age and a low
modulation index are two independent factors for pulsars that affect
the likelihood of them exhibiting coherently drifting subpulses. The
evolutionary trend found seems to suggest that the mechanism that
generates the drifting subpulses gets more and more stable as the
pulsar ages, which could be because the magnetic axis and the rotation
axis becomes more aligned for older pulsars.

The presence of subpulse phase steps results in a minimum in the
longitude resolved modulation index. If subpulse phase steps are
exclusively (or at least more likely) to occur in pulsars with
coherently drifting subpulses, the modulation index of coherent
drifters is expected to be on average lower. This is indeed the trend
the we observe. It is argued by \cite{esv03} that the local reduction
of the modulation index accompanied by a rapid swing in the phase
angle are the result of interference between two superposed drifting
subpulse signals that are out of phase. It is not unlikely that
interference can only occur if the drifting is coherent.
Many pulsars are shown to have a modulation index which is
significantly lower than what is expected for a purely sinusoidal
subpulse signal. This implies the presence of a non-varying component
in the pulsar signal, which could be caused by superposed out of phase
subpulse signals.

Another possible scenario to explain the trend is that coherent
drifting indicates that the electrodynamical conditions in the
sparking gap are stable, which could cause the subpulses to have more
equal intensities. Another explanation for this trend would be that
refraction in the magnetosphere is stronger for pulsars that do not
show the drifting phenomenon coherently. In that scenario the
organized drifting subpulses are refracted in the magnetosphere,
causing the coherent drifting to be distorted. Furthermore it is
expected that refraction would cause the subpulses to appear more
disordered in the pulse window and that the intensity distribution of
the subpulses becomes broadened because of lensing and possible
focusing of the radio emission.

\acknowledgements The authors are grateful for the comments of the
referee of this manuscript, which have led to many improvements in
this paper. We are also thankful for the valuable comments and
suggestions of Geoff Wright and Joanna Rankin and we would like to
thank the staff of the WSRT for their support with scheduling and
assisting with the observations. The Westerbork Synthesis Radio
Telescope is operated by the ASTRON (Netherlands Foundation for
Research in Astronomy) with support from the Netherlands Foundation
for Scientific Research NWO.

\bibliographystyle{aa}
%\bibliography{21cm}

\clearpage
\onecolumn
\begin{longtable}{l|l||r|c||r|r||r|r|r@{$\;$}l|r@{}l||l}
Pulsar & Class  & $P_0$ (s) & $\dot P$ & Pulses & S/N\hspace{1.0mm} &  $m$\hspace{5mm} & $m_\mathrm{thresh}$ & $P_2$ & (deg)& $P_3$ & $\;(P_0)$&  Figure\\
\hline
\hline
\endfirsthead
B0011+47 &    & 1.2407 & $ 5.6\cdot10^{-16}$ & 1564 & 105  & 0.7$\pm$0.1 & 0.31 & & & & &  \ref{B0011+47}\\
B0031$-$07 & Dif  & 0.9430 & $ 4.1\cdot10^{-16}$ & 1204 & 54  & 1.2$\pm$0.1 & 0.51 & $ -40\!$&$^{+2}_{-50}$ & $ 8.3$&$\pm0.3$ &  \ref{B0031-07}\\
B0037+56 & Dif$^\ast$   & 1.1182 & $ 2.9\cdot10^{-15}$ & 2049 & 131  & 0.80$\pm$0.06 & 0.22 & $ 7\!$&$^{+10}_{-2}$ & $22$&$\pm3$ &  \ref{B0037+56}\\
B0052+51 & Dif  & 2.1152 & $ 9.5\cdot10^{-15}$ & 1112 & 112  & 0.92$\pm$0.06 & 0.18 & $ -75\!$&$^{+50}_{-50}$ & $4$&$\pm2$ &  \ref{B0052+51}\\
 &  &    & &  & & &  & $ 30\!$&$^{+70}_{-7}$ & $5$&$\pm1$ &  \\
B0105+65 &    & 1.2837 & $ 1.3\cdot10^{-14}$ & 1346 & 71  & 0.43$\pm$0.04 & 0.33 & & & & &  \ref{B0105+65}\\
J0134$-$2937 &    & 0.1370 & $ 7.8\cdot10^{-17}$ & 4096 & 35  &  & 0.67 & & & & &  \ref{J0134-2937}\\
B0136+57 & Dif$^\ast$  & 0.2725 & $ 1.1\cdot10^{-14}$ & 3432 & 450  & 0.49$\pm$0.01 & 0.09 & $ -70\!$&$^{+15}_{-80}$ & $ 6.5$&$\pm0.2$ &  \ref{B0136+57}\\
B0138+59 & Dif$^\ast$  & 1.2229 & $ 3.9\cdot10^{-16}$ & 2560 & 381  & 0.47$\pm$0.01 & 0.09 & $ 300\!$&$^{+140}_{-110}$ & $15$&$\pm4$ &  \ref{B0138+59}\\
B0144+59 &   & 0.1963 & $ 2.6\cdot10^{-16}$ & 13534 & 151  & 1.20$\pm$0.07 & 0.36 &  & & & &  \ref{B0144+59}\\
B0148$-$06 & Coh   & 1.4647 & $ 4.4\cdot10^{-16}$ & 1400 & 135  & 0.59$\pm$0.04 & 0.22 & $ -12.5\!$&$^{+0.4}_{-1.9}$ & $ 14.2$&$\pm0.2$ &  \ref{B0148-06}\\
 &  &    & &  & & &  & $ -45\!$&$^{+7}_{-20}$ & $ 14.7$&$\pm0.3$ &  \\
B0149$-$16 & Coh   & 0.8327 & $ 1.3\cdot10^{-15}$ & 1024 & 62  & 0.67$\pm$0.05 & 0.42 & $ -9\!$&$^{+12}_{-1}$ &$ 5.8$&$\pm0.5$  &  \ref{B0149-16}\\
B0154+61 &    & 2.3517 & $ 1.9\cdot10^{-13}$ & 769 & 138  & 1.2$\pm$0.1 & 0.15 & & & & &  \ref{B0154+61}\\
B0301+19 & Dif$^\ast$  & 1.3876 & $ 1.3\cdot10^{-15}$ & 1263 & 129  & 0.84$\pm$0.04 & 0.23 & $ -35\!$&$^{+7}_{-55}$ & $ 5.2$&$\pm0.3$ &  \ref{B0301+19}\\
B0320+39 & Coh  & 3.0321 & $ 6.4\cdot10^{-16}$ & 7169 & 443  & 0.21$\pm$0.01 & 0.14 & $ 18\!$&$^{+5}_{-3}$ & $ 8.4$&$\pm0.1$ &  \ref{B0320+39}\\
B0329+54 & Dif$^\ast$  & 0.7145 & $ 2.0\cdot10^{-15}$ & 19969 & 19064  & 0.423$\pm$0.001 & 0.01 & $ -200\!$&$^{+30}_{-150}$ & $3$&$\pm2$ &  \ref{B0329+54}\\
 &  &    & &  & & &  & $ 70\!$&$^{+13}_{-6}$ & $5$&$\pm1$ &  \\
B0353+52 &    & 0.1970 & $ 4.8\cdot10^{-16}$ & 8866 & 117  & 0.63$\pm$0.06 & 0.28 & & & & &  \ref{B0353+52}\\
B0355+54 &    & 0.1564 & $ 4.4\cdot10^{-15}$ & 11264 & 1133  & 0.77$\pm$0.01 & 0.07 & & & & &  \ref{B0355+54}\\
B0402+61 & Lon   & 0.5946 & $ 5.6\cdot10^{-15}$ & 12150 & 156  & 0.7$\pm$0.1 & 0.37 & & & & &  \ref{B0402+61}\\
 &  &    & &  & & &  & ?& & $23$&$\pm4$ &  \\
B0450+55 & Dif$^\ast$  & 0.3407 & $ 2.4\cdot10^{-15}$ & 2664 & 1095  & 0.49$\pm$0.01 & 0.06 & $ 200\!$&$^{+70}_{-120}$ & $ 20$&$\pm10$ &  \ref{B0450+55}\\
 &  &    & &  & & &  & $ -70\!$&$^{+8}_{-25}$ & $10$&$\pm1$ &  \\
B0450$-$18 &    & 0.5489 & $ 5.8\cdot10^{-15}$ & 1537 & 362  & 0.31$\pm$0.01 & 0.13 & & & & &  \ref{B0450-18}\\
B0458+46 &    & 0.6386 & $ 5.6\cdot10^{-15}$ & 1440 & 92  & 0.46$\pm$0.07 & 0.43 & & & & &  \ref{B0458+46}\\
B0523+11 & Dif  & 0.3544 & $ 7.4\cdot10^{-17}$ & 14849 & 206  & 0.56$\pm$0.05 & 0.34 & $ 30\!$&$^{+10}_{-11}$ & $ 3.0$&$\pm0.2$ &  \ref{B0523+11}\\
B0525+21 & Dif$^\ast$  & 3.7455 & $ 4.0\cdot10^{-14}$ & 1029 & 429  & 1.15$\pm$0.02 & 0.05 & $ -20\!$&$^{+2}_{-9}$ & $ 3.8$&$\pm0.7$ &  \ref{B0525+21}\\
 &  &    & &  & & &  & $ 50\!$&$^{+55}_{-10}$ & $ 3.7$&$\pm0.4$ &  \\
B0531+21 &    & 0.0331 & $ 4.2\cdot10^{-13}$ & 178951 & 294  & 5.31$\pm$0.04 & 0.28 & & & & &  \ref{B0531+21}\\
B0540+23 &    & 0.2460 & $ 1.5\cdot10^{-14}$ & 3073 & 773  & 1.29$\pm$0.01 & 0.04 & & & & &  \ref{B0540+23}\\
B0559$-$05 &    & 0.3960 & $ 1.3\cdot10^{-15}$ & 2049 & 73  & 0.63$\pm$0.07 & 0.46 & & & & &  \ref{B0559-05}\\
B0609+37 & Coh  & 0.2980 & $ 5.9\cdot10^{-17}$ & 2561 & 139  & 0.56$\pm$0.03 & 0.22 & $ -20\!$&$^{+4}_{-18}$ & $ 32$&$\pm4$ &  \ref{B0609+37}\\
B0611+22 &    & 0.3350 & $ 5.9\cdot10^{-14}$ & 2560 & 139  & 0.59$\pm$0.02 & 0.26 & & & & &  \ref{B0611+22}\\
B0621$-$04 & Coh  & 1.0391 & $ 8.3\cdot10^{-16}$ & 1536 & 38  & 0.85$\pm$0.09 & 0.47 & $ 25\!$&$^{+14}_{-16}$ & $ 2.055$&$\pm0.001$ &  \ref{B0621-04}\\
B0626+24 &    & 0.4766 & $ 2.0\cdot10^{-15}$ & 1025 & 248  & 0.29$\pm$0.01 & 0.14 & & & & &  \ref{B0626+24}\\
B0628$-$28 & Dif$^\ast$  & 1.2444 & $ 7.1\cdot10^{-15}$ & 4267 & 1064  & 0.59$\pm$0.02 & 0.07 & $ 30\!$&$^{+80}_{-6}$ & $ 7$&$\pm1$ &  \ref{B0628-28}\\
B0656+14 &    & 0.3849 & $ 5.5\cdot10^{-14}$ & 2065 & 292  & 1.24$\pm$0.03 & 0.10 & & & & &  \ref{B0656+14}\\
%B0727$-$18 & Dif$^\ast$   & 0.5102 & $ 1.9\cdot10^{-14}$ & 3632 & 97  & 0.89$\pm$0.04 & 0.34 & $ -160\!$&$^{+60}_{-110}$ & $ 22$&$\pm5$ &  \ref{B0727-18}\\
B0740$-$28 &    & 0.1668 & $ 1.7\cdot10^{-14}$ & 3585 & 380  & 0.27$\pm$0.02 & 0.12 & & & & &  \ref{B0740-28}\\
B0751+32 &  Dif$^\ast$  & 1.4423 & $ 1.1\cdot10^{-15}$ & 774 & 60  & 0.89$\pm$0.08 & 0.31 & $ -30\!$&$^{+15}_{-300}$ & $ 5$&$\pm5$ &  \ref{B0751+32}\\
B0756$-$15 &    & 0.6823 & $ 1.6\cdot10^{-15}$ & 1224 & 61  & 0.62$\pm$0.07 & 0.42 & & & & &  \ref{B0756-15}\\
B0809+74 & Coh  & 1.2922 & $ 1.7\cdot10^{-16}$ & 13092 & 2635  & 0.496$\pm$0.001 & 0.04 & $ -16\!$&$^{+1}_{-16}$ & $ 11.1$&$\pm0.1$ &  \ref{B0809+74}\\
B0818$-$13 & Coh  & 1.2381 & $ 2.1\cdot10^{-15}$ & 2251 & 345  & 0.34$\pm$0.01 & 0.12 & $ -6.5\!$&$^{+0.2}_{-0.7}$ & $ 4.7$&$\pm0.2$ &  \ref{B0818-13}\\
B0820+02 &    & 0.8649 & $ 1.0\cdot10^{-16}$ & 971 & 42  & 0.53$\pm$0.08 & 0.41 & & & & &  \ref{B0820+02}\\
B0823+26 & Dif$^\ast$  & 0.5307 & $ 1.7\cdot10^{-15}$ & 1596 & 2635  & 0.967$\pm$0.001 & 0.01 & $ 55\!$&$^{+40}_{-7}$ & $7$&$\pm2$ &  \ref{B0823+26}\\
B0834+06 & Dif$^\ast$  & 1.2738 & $ 6.8\cdot10^{-15}$ & 1032 & 199  & 0.47$\pm$0.05 & 0.14 & $ 20\!$&$^{+55}_{-9}$ & $ 2.2$&$\pm0.2$ &  \ref{B0834+06}\\
 &  &    & &  & & &  & $ 40\!$&$^{+140}_{-4}$ & $ 2.1$&$\pm0.2$ &  \\
B0906$-$17 &    & 0.4016 & $ 6.7\cdot10^{-16}$ & 2115 & 46  & 0.8$\pm$0.2 & 0.62 & & & & &  \ref{B0906-17}\\
B0919+06 & Dif$^\ast$  & 0.4306 & $ 1.4\cdot10^{-14}$ & 8255 & 823  & 0.620$\pm$0.002 & 0.04 & $ -150\!$&$^{+13}_{-75}$ & $ 4.0$&$\pm0.8$ &  \ref{B0919+06}\\
B0950+08 &    & 0.2531 & $ 2.3\cdot10^{-16}$ & 1311 & 422  & 1.08$\pm$0.02 & 0.27 & & & & &  \ref{B0950+08}\\
J1022+1001 &    & 0.0165 & $ 4.3\cdot10^{-20}$ & 211891 & 251  & 0.60$\pm$0.07 & 0.50 & & & & &  \ref{J1022+1001}\\
B1039$-$19 & Dif  & 1.3864 & $ 9.4\cdot10^{-16}$ & 929 & 160  & 0.51$\pm$0.08 & 0.18 & $ 25\!$&$^{+9}_{-4}$ & $ 4.3$&$\pm0.3$ &  \ref{B1039-19}\\
 &  &    & &  & & &  & $ 9.0\!$&$^{+3.5}_{-0.1}$ & $ 4.3$&$\pm0.1$ &  \\
B1112+50 &    & 1.6564 & $ 2.5\cdot10^{-15}$ & 1599 & 147  & 1.5$\pm$0.2 & 0.27 & & & & &  \ref{B1112+50}\\
B1133+16 & Dif$^\ast$  & 1.1879 & $ 3.7\cdot10^{-15}$ & 1514 & 1156  & 1.4$\pm$0.1 & 0.04 & $ 200\!$&$^{+55}_{-90}$ & $3$&$\pm2$ &  \ref{B1133+16}\\
 &  &    & &  & & &  & $ 130\!$&$^{+120}_{-15}$ & $3$&$\pm1$ &  \\
\caption{\label{Table_section}The details of all the analysed pulsars. The classification of the pulsar in the second column, where ``Coh'' is a coherent drifter, ``Dif'' and ``Dif$^\ast$'' are diffuse drifters with or without drift features which are clearly separated from the alias borders and ``Lon'' are pulsars showing longitude stationary subpulse modulation. The next columns are the pulse period, its dimensionless time derivative, the number of pulses in the observation, the signal to noise ratio, the minimum in the longitude resolved modulation index, the minimum detectable modulation index, the horizontal and vertical driftband separation and the figure number.}\\
\clearpage
Pulsar & Class  & $P_0$ (s) & $\dot P$ & Pulses & S/N\hspace{1.0mm} &  $m$\hspace{5mm} & $m_\mathrm{thresh}$ & $P_2$ & (deg)& $P_3$ & $\;(P_0)$&  Figure\\
\hline
\hline
B1237+25 & Dif$^\ast$  & 1.3824 & $ 9.6\cdot10^{-16}$ & 1265 & 2121  & 0.480$\pm$0.002 & 0.04 & $ -20\!$&$^{+2}_{-3}$ & $ 2.7$&$\pm0.1$ &  \ref{B1237+25}\\
 &  &    & &  & & &  & $ 16\!$&$^{+1}_{-1}$ & $ 2.7$&$\pm0.1$ &  \\
B1254$-$10 &    & 0.6173 & $ 3.6\cdot10^{-16}$ & 1389 & 14  &  & 1.58 & & & & &  \ref{B1254-10}\\
B1508+55 & Dif$^\ast$  & 0.7397 & $ 5.0\cdot10^{-15}$ & 4808 & 568  & 0.52$\pm$0.01 & 0.09 & $ -160\!$&$^{+40}_{-80}$ & $ 5$&$\pm5$ &  \ref{B1508+55}\\
J1518+4904 & Dif   & 0.0409 & $ 2.7\cdot10^{-20}$ & 31723 & 287  & 0.36$\pm$0.01 & 0.13 & $ 55\!$&$^{+25}_{-4}$ & $ 2.6$&$\pm0.1$ &  \ref{J1518+4904}\\
B1540$-$06 & Coh  & 0.7091 & $ 8.8\cdot10^{-16}$ & 5121 & 293  & 0.27$\pm$0.03 & 0.14 & $ 24\!$&$^{+9}_{-4}$ & $ 3.03$&$\pm0.02$ &  \ref{B1540-06}\\
 &  &    & &  & & &  & $ -16\!$&$^{+12}_{-45}$ & $ 3.02$&$\pm0.02$ &  \\
B1541+09 &    & 0.7484 & $ 4.3\cdot10^{-16}$ & 1126 & 93  & 0.47$\pm$0.07 & 0.41 & & & & &  \ref{B1541+09}\\
B1600$-$27 &    & 0.7783 & $ 3.0\cdot10^{-15}$ & 1077 & 34  & 0.7$\pm$0.1 & 0.71 & & & & &  \ref{B1600-27}\\
B1604$-$00 & Dif$^\ast$  & 0.4218 & $ 3.1\cdot10^{-16}$ & 2012 & 1958  & 0.61$\pm$0.07 & 0.02 & $ 55\!$&$^{+100}_{-3}$ & $ 3.4$&$\pm0.5$ &  \ref{B1604-00}\\
 &  &    & &  & & &  & $ 70\!$&$^{+35}_{-5}$ & $ 3.1$&$\pm0.5$ &  \\
B1612+07 &    & 1.2068 & $ 2.4\cdot10^{-15}$ & 1926 & 56  & 0.68$\pm$0.05 & 0.45 & & & & &  \ref{B1612+07}\\
B1642$-$03 & Dif$^\ast$  & 0.3877 & $ 1.8\cdot10^{-15}$ & 3118 & 656  & 0.337$\pm$0.002 & 0.06 & $ 60\!$&$^{+9}_{-4}$ & $ 15$&$\pm1$ &  \ref{B1642-03}\\
B1649$-$23 &    & 1.7037 & $ 3.2\cdot10^{-15}$ & 1025 & 38  & 1.0$\pm$0.2 & 0.61 & & & & &  \ref{B1649-23}\\
J1650$-$1654 & Coh  & 1.7496 & $ 3.2\cdot10^{-15}$ & 1002 & 55  & 0.7$\pm$0.1 & 0.49 & $ 9\!$&$^{+7}_{-1}$ & $ 2.59$&$\pm0.05$ &  \ref{J1650-1654}\\
B1702$-$19 & Coh  & 0.2990 & $ 4.1\cdot10^{-15}$ & 11071 & 591  & 0.34$\pm$0.01 & 0.13 & $ -80\!$&$^{+6}_{-70}$ & $ 11.0$&$\pm0.4$ &  \ref{B1702-19}\\
 &  &    & &  & & &  & ?& & $ 11.0$&$\pm0.4$ &  \\
B1706$-$16 &    & 0.6531 & $ 6.3\cdot10^{-15}$ & 1309 & 377  & 0.71$\pm$0.06 & 0.12 & & & & &  \ref{B1706-16}\\
J1713+0747 &    & 0.0046 & $ 8.5\cdot10^{-21}$ & 791257 & 181  & 0.60$\pm$0.04 & 0.40 & & & & &  \ref{J1713+0747}\\
B1717$-$16 &    & 1.5656 & $ 5.8\cdot10^{-15}$ & 1120 & 104  & 0.65$\pm$0.08 & 0.22 & & & & &  \ref{B1717-16}\\
B1717$-$29 & Coh  & 0.6204 & $ 7.5\cdot10^{-16}$ & 1794 & 17  & 1.0$\pm$0.2 & 1.24 & $ -9.6\!$&$^{+3}_{-0.6}$ & $ 2.45$&$\pm0.02$ &  \ref{B1717-29}\\
B1730$-$22 &    & 0.8717 & $ 4.3\cdot10^{-17}$ & 880 & 76  & 0.41$\pm$0.04 & 0.32 & & & & &  \ref{B1730-22}\\
J1730$-$2304 &    & 0.0081 & $ 2.0\cdot10^{-20}$ & 103617 & 76  &  & 1.52 & & & & &  \ref{J1730-2304}\\
B1732$-$07 &    & 0.4193 & $ 1.2\cdot10^{-15}$ & 1826 & 79  & 0.43$\pm$0.04 & 0.33 & & & & &  \ref{B1732-07}\\
B1736$-$29 &    & 0.3229 & $ 7.9\cdot10^{-15}$ & 2606 & 26  & 1.0$\pm$0.2 & 0.89 & & & & &  \ref{B1736-29}\\
B1737+13 &    & 0.8031 & $ 1.5\cdot10^{-15}$ & 1037 & 30  & 0.9$\pm$0.2 & 0.73 & & & & &  \ref{B1737+13}\\
B1738$-$08 & Dif$^\ast$  & 2.0431 & $ 2.3\cdot10^{-15}$ & 859 & 153  & 0.86$\pm$0.04 & 0.27 & $ 60\!$&$^{+20}_{-7}$ & $ 5.2$&$\pm0.1$ &  \ref{B1738-08}\\
 &  &    & &  & & &  & $ 9.5\!$&$^{+4}_{-0.5}$ & $  4.4$&$\pm0.3$ &  \\
B1744$-$24A &    & 0.0116 & $ -3.4\cdot10^{-20}$ & 309244 & 34  &  & 2.21 & & & & &  \ref{B1744-24A}\\
B1745$-$12 &    & 0.3941 & $ 1.2\cdot10^{-15}$ & 2054 & 75  & 0.49$\pm$0.08 & 0.49 & & & & &  \ref{B1745-12}\\
B1749$-$28 &   & 0.5626 & $ 8.1\cdot10^{-15}$ & 1285 & 898  & 0.540$\pm$0.004 & 0.04 & & & &  &  \ref{B1749-28}\\
B1753+52 & Dif  & 2.3914 & $ 1.6\cdot10^{-15}$ & 784 & 60  & 0.73$\pm$0.09 & 0.50 & $ 100\!$&$^{+30}_{-60}$ & $ 11$&$\pm5$ &  \ref{B1753+52}\\
 &  &    & &  & & &  & $ 9.7\!$&$^{+3}_{-0.7}$ & $ 6.3$&$\pm0.4$ &  \\
B1754$-$24 &    & 0.2341 & $ 1.3\cdot10^{-14}$ & 3589 & 113  & 0.58$\pm$0.06 & 0.27 & & & & &  \ref{B1754-24}\\
B1756$-$22 &    & 0.4610 & $ 1.1\cdot10^{-14}$ & 1561 & 103  & 0.46$\pm$0.07 & 0.19 & & & & &  \ref{B1756-22}\\
J1757$-$2223 &    & 0.1853 & $ 7.8\cdot10^{-16}$ & 4629 & 9  &  & 2.04 & & & & &  \ref{J1757-2223}\\
B1758$-$23 &    & 0.4158 & $ 1.1\cdot10^{-13}$ & 2014 & 87  &  & 0.48 & & & & &  \ref{B1758-23}\\
B1758$-$29 &    & 1.0819 & $ 3.3\cdot10^{-15}$ & 1067 & 19  & 0.9$\pm$0.2 & 1.10 & & & & &  \ref{B1758-29}\\
B1800$-$21 &    & 0.1336 & $ 1.3\cdot10^{-13}$ & 6210 & 134  & 0.46$\pm$0.04 & 0.27 & & & & &  \ref{B1800-21}\\
B1804$-$08 &    & 0.1637 & $ 2.9\cdot10^{-17}$ & 5241 & 414  & 0.30$\pm$0.03 & 0.12 & & & & &  \ref{B1804-08}\\
B1805$-$20 &    & 0.9184 & $ 1.7\cdot10^{-14}$ & 1056 & 102  & 0.51$\pm$0.04 & 0.23 & & & & &  \ref{B1805-20}\\
J1808$-$0813 &    & 0.8760 & $ 1.2\cdot10^{-15}$ & 974 & 73  & 0.52$\pm$0.09 & 0.41 & & & & &  \ref{J1808-0813}\\
B1811+40 &    & 0.9311 & $ 2.5\cdot10^{-15}$ & 7200 & 109  & 0.56$\pm$0.1 & 0.53 & & & & &  \ref{B1811+40}\\
J1812$-$2102 &    & 1.2234 & $ 2.4\cdot10^{-14}$ & 1435 & 63  &  & 0.65 & & & & &  \ref{J1812-2102}\\
B1813$-$17 &    & 0.7823 & $ 7.3\cdot10^{-15}$ & 1096 & 50  & 0.7$\pm$0.1 & 0.58 & & & & &  \ref{B1813-17}\\
B1815$-$14 &    & 0.2915 & $ 2.0\cdot10^{-15}$ & 2579 & 175  & 0.27$\pm$0.03 & 0.18 & & & & &  \ref{B1815-14}\\
B1817$-$13 &    & 0.9215 & $ 4.5\cdot10^{-15}$ & 1031 & 46  & 0.65$\pm$0.06 & 0.41 & & & & &  \ref{B1817-13}\\
B1818$-$04 &   & 0.5981 & $ 6.3\cdot10^{-15}$ & 7681 & 353  & 0.40$\pm$0.04 & 0.16 & & & & &  \ref{B1818-04}\\
B1819$-$22 & Dif  & 1.8743 & $ 1.4\cdot10^{-15}$ & 1096 & 258  & 0.76$\pm$0.01 & 0.10 & $ -9.1\!$&$^{+0.1}_{-1.0}$ & $ 9.8$&$\pm0.6$ &  \ref{B1819-22}\\
B1820$-$11 &    & 0.2798 & $ 1.4\cdot10^{-15}$ & 3021 & 97  & 0.61$\pm$0.04 & 0.35 & & & & &  \ref{B1820-11}\\
B1821+05 &    & 0.7529 & $ 2.3\cdot10^{-16}$ & 1080 & 24  &  & 0.99 & & & & &  \ref{B1821+05}\\
B1821$-$11 &    & 0.4358 & $ 3.6\cdot10^{-15}$ & 1568 & 59  &  & 0.37 & & & & &  \ref{B1821-11}\\
B1821$-$19 &    & 0.1893 & $ 5.2\cdot10^{-15}$ & 4105 & 222  & 0.632$\pm$0.004 & 0.08 & & & & &  \ref{B1821-19}\\
B1822$-$09 & Dif$^\ast$  & 0.7690 & $ 5.2\cdot10^{-14}$ & 1180 & 409  & 0.672$\pm$0.005 & 0.24 & $ 50\!$&$^{+110}_{-15}$ & $ 8$&$\pm4$ &  \ref{B1822-09}\\
B1822$-$14 &    & 0.2792 & $ 2.3\cdot10^{-14}$ & 3061 & 79  & 1.8$\pm$0.2 & 0.48 & & & & &  \ref{B1822-14}\\
B1826$-$17 &    & 0.3071 & $ 5.6\cdot10^{-15}$ & 2764 & 361  & 0.59$\pm$0.01 & 0.11 & & & & &  \ref{B1826-17}\\
J1828$-$1101 &    & 0.0721 & $ 1.5\cdot10^{-14}$ & 24167 & 64  &  & 0.98 & & & & &  \ref{J1828-1101}\\
B1829$-$08 &    & 0.6473 & $ 6.4\cdot10^{-14}$ & 1311 & 132  & 0.56$\pm$0.09 & 0.22 & & & & &  \ref{B1829-08}\\
B1830$-$08 &    & 0.0853 & $ 9.2\cdot10^{-15}$ & 9412 & 124  & 1.4$\pm$0.2 & 0.22 & & & & &  \ref{B1830-08}\\
J1830$-$1135 & Dif$^\ast$  & 6.2216 & $ 4.8\cdot10^{-14}$ & 1004 & 140  & 1.1$\pm$0.1 & 0.21 & $ 25\!$&$^{+25}_{-7}$ & $ 2.1$&$\pm0.2$ &  \ref{J1830-1135}\\
\caption{continued.}\\
\clearpage
Pulsar & Class  & $P_0$ (s) & $\dot P$ & Pulses & S/N\hspace{1.0mm} &  $m$\hspace{5mm} & $m_\mathrm{thresh}$ & $P_2$ & (deg)& $P_3$ & $\;(P_0)$&  Figure\\
\hline
\hline
B1831$-$03 &    & 0.6867 & $ 4.2\cdot10^{-14}$ & 1236 & 88  & 0.45$\pm$0.06 & 0.26 & & & & &  \ref{B1831-03}\\
B1831$-$04 &    & 0.2901 & $ 7.2\cdot10^{-17}$ & 11255 & 114  & 0.41$\pm$0.05 & 0.36 & & & & &  \ref{B1831-04}\\
B1832$-$06 &    & 0.3058 & $ 4.0\cdot10^{-14}$ & 4557 & 33  &  & 0.98 & & & & &  \ref{B1832-06}\\
B1834$-$04 &    & 0.3542 & $ 1.7\cdot10^{-15}$ & 4114 & 73  & 0.57$\pm$0.05 & 0.41 & & & & &  \ref{B1834-04}\\
B1834$-$10 &    & 0.5627 & $ 1.2\cdot10^{-14}$ & 1286 & 151  & 0.32$\pm$0.01 & 0.12 & & & & &  \ref{B1834-10}\\
J1835$-$1106 &    & 0.1659 & $ 2.1\cdot10^{-14}$ & 5125 & 108  & 0.89$\pm$0.04 & 0.42 & & & & &  \ref{J1835-1106}\\
B1839+09 &    & 0.3813 & $ 1.1\cdot10^{-15}$ & 2056 & 79  & 0.48$\pm$0.06 & 0.41 & & & & &  \ref{B1839+09}\\
B1839+56 &    & 1.6529 & $ 1.5\cdot10^{-15}$ & 1031 & 53  & 1.07$\pm$0.06 & 0.44 & & & & &  \ref{B1839+56}\\
B1839$-$04 & Coh  & 1.8399 & $ 5.1\cdot10^{-16}$ & 1025 & 179  & 0.49$\pm$0.04 & 0.15 & $ -35\!$&$^{+5}_{-15}$ & $ 12.4$&$\pm0.3$ &  \ref{B1839-04}\\
 &  &    & &  & & &  & $ 120\!$&$^{+20}_{-20}$ & $ 12.4$&$\pm0.3$ &  \\
J1839$-$0643 &    & 0.4495 & $ 3.6\cdot10^{-15}$ & 3590 & 50  & 0.51$\pm$0.08 & 0.47 & & & & &  \ref{J1839-0643}\\
B1841$-$04 & Coh  & 0.9910 & $ 3.9\cdot10^{-15}$ & 1551 & 69  & 0.62$\pm$0.09 & 0.44 & $ -13\!$&$^{+11}_{-5}$ & $ 8.5$&$\pm0.1$ &  \ref{B1841-04}\\
B1841$-$05 &    & 0.2557 & $ 9.7\cdot10^{-15}$ & 3077 & 115  & 0.26$\pm$0.04 & 0.26 & & & & &  \ref{B1841-05}\\
B1842+14 &    & 0.3755 & $ 1.9\cdot10^{-15}$ & 3073 & 77  & 0.67$\pm$0.06 & 0.46 & & & & &  \ref{B1842+14}\\
B1842$-$04 &    & 0.4868 & $ 1.1\cdot10^{-14}$ & 1556 & 68  & 0.53$\pm$0.07 & 0.45 & & & & &  \ref{B1842-04}\\
B1844$-$04 & Coh  & 0.5978 & $ 5.2\cdot10^{-14}$ & 1240 & 115  & 0.38$\pm$0.03 & 0.26 & $ 80\!$&$^{+70}_{-45}$ & $12$&$\pm1$ &  \ref{B1844-04}\\
B1845$-$01 & Dif$^\ast$  & 0.6594 & $ 5.3\cdot10^{-15}$ & 1031 & 213  & 0.28$\pm$0.01 & 0.11 & $ 90\!$&$^{+170}_{-35}$ & $14$&$\pm2$ &  \ref{B1845-01}\\
J1845$-$0743 &    & 0.1047 & $ 3.7\cdot10^{-16}$ & 8193 & 115  & 1.02$\pm$0.09 & 0.26 & & & & &  \ref{J1845-0743}\\
B1846$-$06 &  Lon  & 1.4513 & $ 4.6\cdot10^{-14}$ & 1208 & 128  & 1.31$\pm$0.03 & 0.24 & ? & & $3$&$\pm1$ &  \ref{B1846-06}\\
B1848+13 &    & 0.3456 & $ 1.5\cdot10^{-15}$ & 1544 & 42  & 0.55$\pm$0.09 & 0.55 & & & & &  \ref{B1848+13}\\
B1849+00 &    & 2.1802 & $ 9.7\cdot10^{-14}$ & 1030 & 94  & 0.25$\pm$0.02 & 0.17 & & & & &  \ref{B1849+00}\\
J1850+0026 &    & 1.0818 & $ 3.6\cdot10^{-16}$ & 1029 & 41  & 0.61$\pm$0.08 & 0.38 & & & & &  \ref{J1850+0026}\\
J1852+0305 &    & 1.3261 & $ 1.0\cdot10^{-16}$ & 1323 & 10  &  & 2.21 & & & & &  \ref{J1852+0305}\\
J1852$-$2610 &    & 0.3363 & $ 8.8\cdot10^{-17}$ & 2526 & 42  &  & 0.50 & & & & &  \ref{J1852-2610}\\
B1855+02 &    & 0.4158 & $ 4.0\cdot10^{-14}$ & 2051 & 77  & 0.7$\pm$0.1 & 0.49 & & & & &  \ref{B1855+02}\\
B1855+09 &    & 0.0054 & $ 1.8\cdot10^{-20}$ & 158193 & 175  & 0.6$\pm$0.1 & 0.46 & & & & &  \ref{B1855+09}\\
B1857$-$26 & Dif  & 0.6122 & $ 2.0\cdot10^{-16}$ & 1339 & 360  & 0.47$\pm$0.02 & 0.20 & $ 80\!$&$^{+80}_{-9}$ & $ 7.5$&$\pm0.2$ &  \ref{B1857-26}\\
 &  &    & &  & & &  & $ 180\!$&$^{+20}_{-45}$ & $ 7.3$&$\pm0.2$ &  \\
B1859+03 &   & 0.6555 & $ 7.5\cdot10^{-15}$ & 1552 & 182  & 0.35$\pm$0.05 & 0.12 & & & & &  \ref{B1859+03}\\
B1859+07 &    & 0.6440 & $ 2.3\cdot10^{-15}$ & 1030 & 46  & 0.47$\pm$0.08 & 0.52 & & & & &  \ref{B1859+07}\\
B1900+01 & Dif$^\ast$  & 0.7293 & $ 4.0\cdot10^{-15}$ & 1171 & 327  & 0.56$\pm$0.05 & 0.09 & $ 30\!$&$^{+30}_{-3}$ & $ 3.4$&$\pm0.7$ &  \ref{B1900+01}\\
B1900+05 &    & 0.7466 & $ 1.3\cdot10^{-14}$ & 1025 & 54  & 0.49$\pm$0.06 & 0.40 & & & & &  \ref{B1900+05}\\
J1901$-$0906 & Coh  & 1.7819 & $ 1.6\cdot10^{-15}$ & 785 & 147  & 0.47$\pm$0.08 & 0.13 & $ -11\!$&$^{+1}_{-8}$ & $ 6.9$&$\pm0.3$ &  \ref{J1901-0906}\\
 &  &    & &  & & &  & $ -5.6\!$&$^{+0.1}_{-0.7}$ & $ 3.1$&$\pm0.1$ &  \\
B1903+07 &    & 0.6480 & $ 4.9\cdot10^{-15}$ & 2246 & 32  &  & 0.72 & & & & &  \ref{B1903+07}\\
B1905+39 &    & 1.2358 & $ 5.4\cdot10^{-16}$ & 1032 & 50  & 0.78$\pm$0.07 & 0.57 & & & & &  \ref{B1905+39}\\
B1907+00 &    & 1.0169 & $ 5.5\cdot10^{-15}$ & 1030 & 40  & 0.63$\pm$0.06 & 0.44 & & & & &  \ref{B1907+00}\\
B1907+10 &    & 0.2836 & $ 2.6\cdot10^{-15}$ & 3023 & 95  & 0.51$\pm$0.07 & 0.44 & & & & &  \ref{B1907+10}\\
B1910+20 &    & 2.2330 & $ 1.0\cdot10^{-14}$ & 1025 & 43  & 0.67$\pm$0.07 & 0.39 & & & & &  \ref{B1910+20}\\
B1911+13 &    & 0.5215 & $ 8.0\cdot10^{-16}$ & 2055 & 56  & 0.32$\pm$0.04 & 0.26 & & & & &  \ref{B1911+13}\\
B1911$-$04 & Dif$^\ast$  & 0.8259 & $ 4.1\cdot10^{-15}$ & 2082 & 616  & 0.287$\pm$0.004 & 0.08 & $ 70\!$&$^{+40}_{-30}$ & $15$&$\pm5$ &  \ref{B1911-04}\\
B1914+09 &    & 0.2703 & $ 2.5\cdot10^{-15}$ & 6145 & 97  & 0.59$\pm$0.08 & 0.52 & & & & &  \ref{B1914+09}\\
B1914+13 &    & 0.2818 & $ 3.6\cdot10^{-15}$ & 4096 & 185  & 0.41$\pm$0.02 & 0.19 & & & & &  \ref{B1914+13}\\
B1915+13 &    & 0.1946 & $ 7.2\cdot10^{-15}$ & 1541 & 41  &  & 0.46 & & & & &  \ref{B1915+13}\\
B1916+14 &    & 1.1810 & $ 2.1\cdot10^{-13}$ & 1226 & 23  &  & 0.79 & & & & &  \ref{B1916+14}\\
B1917+00 & Dif$^\ast$  & 1.2723 & $ 7.7\cdot10^{-15}$ & 4106 & 135  & 0.69$\pm$0.08 & 0.27 & $ 70\!$&$^{+7}_{-25}$ & $ 7.8$&$\pm0.2$ &  \ref{B1917+00}\\
B1918+19 &    & 0.8210 & $ 9.0\cdot10^{-16}$ & 1550 & 60  & 0.58$\pm$0.07 & 0.48 & & & & &  \ref{B1918+19}\\
B1919+21 & Dif  & 1.3373 & $ 1.3\cdot10^{-15}$ & 1033 & 207  & 0.30$\pm$0.04 & 0.18 & $ -3.4\!$&$^{+0.3}_{-0.2}$ & $ 4.4$&$\pm0.1$ &  \ref{B1919+21}\\
 &  &    & &  & & &  & $-11\!$&$^{+1}_{-1}$ & $ 4.4$&$\pm0.1$ &  \\
B1920+21 &    & 1.0779 & $ 8.2\cdot10^{-15}$ & 1025 & 74  & 0.40$\pm$0.06 & 0.30 & & & & &  \ref{B1920+21}\\
B1924+16 &   & 0.5798 & $ 1.8\cdot10^{-14}$ & 2514 & 87  & 0.76$\pm$0.06 & 0.44 & &  & &  &  \ref{B1924+16}\\
B1929+10 & Dif$^\ast$  & 0.2265 & $ 1.2\cdot10^{-15}$ & 1324 & 681  & 0.466$\pm$0.002 & 0.04 & $ 90\!$&$^{+140}_{-8}$ & $ 9.8$&$\pm0.8$ &  \ref{B1929+10}\\
 &  &    & &  & & &  & $ -160\!$&$^{+10}_{-100}$ & $ 4.4$&$\pm0.1$ &  \\
B1933+16 & Dif$^\ast$  & 0.3587 & $ 6.0\cdot10^{-15}$ & 2613 & 2081  & 0.240$\pm$0.001 & 0.02 & $ 300\!$&$^{+130}_{-50}$ & $ 2.4$&$\pm0.3$ &  \ref{B1933+16}\\
B1935+25 &    & 0.2010 & $ 6.4\cdot10^{-16}$ & 7173 & 66  & 0.58$\pm$0.06 & 0.47 & & & & &  \ref{B1935+25}\\
B1937+21 &    & 0.0016 & $ 1.1\cdot10^{-19}$ & 600160 & 339  &  & 0.19 & & & & &  \ref{B1937+21}\\
B1937$-$26 &  Lon  & 0.4029 & $ 9.6\cdot10^{-16}$ & 2129 & 91  & 0.93$\pm$0.08 & 0.31 & ?& & $ 2.5$&$\pm0.5$ &  \ref{B1937-26}\\
B1943$-$29 &    & 0.9594 & $ 1.5\cdot10^{-15}$ & 1319 & 43  & 0.8$\pm$0.1 & 0.56 & & & & &  \ref{B1943-29}\\
B1944+17 & Dif$^\ast$  & 0.4406 & $ 2.4\cdot10^{-17}$ & 2056 & 278  & 1.19$\pm$0.02 & 0.17 & $ -11.7\!$&$^{+0.3}_{-0.3}$ & $ 13$&$\pm5$ &  \ref{B1944+17}\\
B1946+35 &  Lon  & 0.7173 & $ 7.1\cdot10^{-15}$ & 1161 & 176  & 0.43$\pm$0.01 & 0.12 & ?& & $33$&$\pm2$ &  \ref{B1946+35}\\
\caption{continued.}\\
\clearpage
Pulsar & Class  & $P_0$ (s) & $\dot P$ & Pulses & S/N\hspace{1.0mm} &  $m$\hspace{5mm} & $m_\mathrm{thresh}$ & $P_2$ & (deg)& $P_3$ & $\;(P_0)$&  Figure\\
\hline
\hline
B1952+29 & Dif$^\ast$  & 0.4267 & $ 1.7\cdot10^{-18}$ & 1536 & 312  & 0.46$\pm$0.06 & 0.07 & $ -190\!$&$^{+15}_{-170}$ & $ 13.1$&$\pm0.2$ &  \ref{B1952+29}\\
 &  &    & &  & & &  & $ -40\!$&$^{+6}_{-30}$ & $ 12.5$&$\pm0.4$ &  \\
B1953+50 & Dif$^\ast$  & 0.5189 & $ 1.4\cdot10^{-15}$ & 1635 & 464  & 0.9$\pm$0.2 & 0.08 & $ 5.3\!$&$^{+16}_{-0.2}$ & $ 25$&$\pm7$ &  \ref{B1953+50}\\
B2000+32 &    & 0.6968 & $ 1.1\cdot10^{-13}$ & 2060 & 90  & 0.8$\pm$0.1 & 0.41 & & & & &  \ref{B2000+32}\\
B2000+40 & Coh  & 0.9051 & $ 1.7\cdot10^{-15}$ & 768 & 109  & 0.53$\pm$0.04 & 0.28 & $ -7.3\!$&$^{+0.8}_{-0.6}$ & $ 2.5$&$\pm0.1$ &  \ref{B2000+40}\\
 &  &    & &  & & &  & $ -17\!$&$^{+3}_{-8}$ & $ 2.5$&$\pm0.1$ &  \\
B2002+31 &    & 2.1113 & $ 7.5\cdot10^{-14}$ & 1115 & 214  & 0.54$\pm$0.02 & 0.12 & & & & &  \ref{B2002+31}\\
B2003$-$08 &    & 0.5809 & $ 4.6\cdot10^{-17}$ & 1285 & 37  & 0.8$\pm$0.2 & 0.74 & & & & &  \ref{B2003-08}\\
B2011+38 &  Lon  & 0.2302 & $ 8.9\cdot10^{-15}$ & 3722 & 271  & 1.34$\pm$0.02 & 0.21 & ?& & $ 30$&$\pm15$ &  \ref{B2011+38}\\
B2016+28 & Dif$^\ast$  & 0.5580 & $ 1.5\cdot10^{-16}$ & 25899 & 511  & 0.61$\pm$0.01 & 0.15 & $ -70\!$&$^{+15}_{-25}$ & $ 20$&$\pm2$ &  \ref{B2016+28}\\
 &  &    & &  & & &  & $ -12\!$&$^{+1}_{-8}$ & $ 4.0$&$\pm0.4$ &  \\
B2020+28 & Dif$^\ast$  & 0.3434 & $ 1.9\cdot10^{-15}$ & 3759 & 1438  & 0.15$\pm$0.02 & 0.04 & $ -55\!$&$^{+5}_{-15}$ & $ 2.5$&$\pm0.2$ &  \ref{B2020+28}\\
 &  &    & &  & & &  & $ 25\!$&$^{+15}_{-2}$ & $ 2.3$&$\pm0.1 $ &  \\
B2021+51 & Dif$^\ast$  & 0.5292 & $ 3.1\cdot10^{-15}$ & 20326 & 1180  & 0.50$\pm$0.03 & 0.08 & $ 10\!$&$^{+15}_{-1}$ & $ 5.1$&$\pm0.3$ &  \ref{B2021+51}\\
B2022+50 &    & 0.3726 & $ 2.5\cdot10^{-15}$ & 2278 & 96  & 0.52$\pm$0.04 & 0.31 & & & & &  \ref{B2022+50}\\
B2043$-$04 & Coh  & 1.5469 & $ 1.5\cdot10^{-15}$ & 1025 & 122  & 0.69$\pm$0.02 & 0.23 & $ 4.5\!$&$^{+5}_{-0.3}$ & $ 2.74$&$\pm0.05$ &  \ref{B2043-04}\\
B2044+15 & Dif$^\ast$  & 1.1383 & $ 1.8\cdot10^{-16}$ & 768 & 51  & 0.72$\pm$0.06 & 0.40 & $ -7.1\!$&$^{+0.5}_{-1.4}$ & $18$&$\pm6$ &  \ref{B2044+15}\\
B2045$-$16 & Coh  & 1.9616 & $ 1.1\cdot10^{-14}$ & 384 & 44  & 0.8$\pm$0.1 & 0.55 & $ 17\!$&$^{+18}_{-2}$ & $ 3.2$&$\pm0.1$ &  \ref{B2045-16}\\
B2053+36 & Dif$^\ast$  & 0.2215 & $ 3.7\cdot10^{-16}$ & 3594 & 173  & 0.59$\pm$0.02 & 0.22 & $ 100\!$&$^{+340}_{-75}$ & $11$&$\pm2$ &  \ref{B2053+36}\\
B2106+44 &  Lon  & 0.4149 & $ 8.6\cdot10^{-17}$ & 2067 & 292  & 0.52$\pm$0.04 & 0.19 & ?& & $ 19$&$\pm7$ &  \ref{B2106+44}\\
B2110+27 & Dif$^\ast$  & 1.2029 & $ 2.6\cdot10^{-15}$ & 1032 & 370  & 1.06$\pm$0.02 & 0.09 & $ 140\!$&$^{+18}_{-15}$ & $ 4.4$&$\pm0.1$ &  \ref{B2110+27}\\
B2111+46 & Dif  & 1.0147 & $ 7.1\cdot10^{-16}$ & 3083 & 258  & 0.62$\pm$0.05 & 0.17 & $ -180\!$&$^{+20}_{-50}$ & $ 2.7$&$\pm0.4$ &  \ref{B2111+46}\\
J2145$-$0750 &    & 0.0161 & $ 3.0\cdot10^{-20}$ & 224536 & 190  & 0.53$\pm$0.07 & 0.16 & & & & &  \ref{J2145-0750}\\
B2148+52 &    & 0.3322 & $ 1.0\cdot10^{-14}$ & 2055 & 89  & 0.39$\pm$0.06 & 0.34 & & & & &  \ref{B2148+52}\\
B2148+63 & Dif$^\ast$  & 0.3801 & $ 1.7\cdot10^{-16}$ & 2256 & 143  & 0.89$\pm$0.02 & 0.26 & $ -35\!$&$^{+7}_{-15}$ & $ 2.4$&$\pm0.3$ &  \ref{B2148+63}\\
B2154+40 & Dif$^\ast$  & 1.5253 & $ 3.4\cdot10^{-15}$ & 1148 & 475  & 0.60$\pm$0.01 & 0.11 & $ 110\!$&$^{+80}_{-10}$ & $ 3.1$&$\pm0.8$ &  \ref{B2154+40}\\
B2217+47 &    & 0.5385 & $ 2.8\cdot10^{-15}$ & 1592 & 191  & 0.52$\pm$0.03 & 0.18 & & & & &  \ref{B2217+47}\\
B2224+65 &    & 0.6825 & $ 9.7\cdot10^{-15}$ & 2548 & 56  & 0.96$\pm$0.07 & 0.42 & & & & &  \ref{B2224+65}\\
B2255+58 & Dif  & 0.3682 & $ 5.8\cdot10^{-15}$ & 2305 & 506  & 0.52$\pm$0.01 & 0.12 & $ 11\!$&$^{+11}_{-1}$ & $10$&$\pm1$ &  \ref{B2255+58}\\
B2303+30 & Coh  & 1.5759 & $ 2.9\cdot10^{-15}$ & 1109 & 96  & 0.72$\pm$0.08 & 0.26 & $ 15.0\!$&$^{+3}_{-0.3}$ & $ 2.1$&$\pm0.1$ &  \ref{B2303+30}\\
B2306+55 &    & 0.4751 & $ 2.0\cdot10^{-16}$ & 3596 & 45  & 0.8$\pm$0.1 & 0.70 & & & & &  \ref{B2306+55}\\
B2310+42 & Coh  & 0.3494 & $ 1.1\cdot10^{-16}$ & 5006 & 2073  & 0.46$\pm$0.01 & 0.11 & $ 60\!$&$^{+20}_{-10}$ & $ 2.1$&$\pm0.1$ &  \ref{B2310+42}\\
 &  &    & &  & & &  & $ 6.0\!$&$^{+0.3}_{-0.3}$ & $ 2.1$&$\pm0.1$ &  \\
B2315+21 &    & 1.4447 & $ 1.0\cdot10^{-15}$ & 1024 & 44  & 0.7$\pm$0.1 & 0.58 & & & & &  \ref{B2315+21}\\
B2319+60 & Coh  & 2.2565 & $ 7.0\cdot10^{-15}$ & 1041 & 1652  & 0.923$\pm$0.004 & 0.03 & $ 70\!$&$^{+60}_{-10}$ & $ 7.7$&$\pm0.4$ &  \ref{B2319+60}\\
 &  &    & &  & & &  & $ 12\!$&$^{+4}_{-1}$ & $ 7.7$&$\pm0.1$ &  \\
B2323+63 &    & 1.4363 & $ 2.8\cdot10^{-15}$ & 1036 & 29  & 1.6$\pm$0.1 & 0.67 & & & & &  \ref{B2323+63}\\
B2324+60 & Dif$^\ast$  & 0.2337 & $ 3.5\cdot10^{-16}$ & 3633 & 289  & 0.95$\pm$0.01 & 0.16 & $ 160\!$&$^{+40}_{-30}$ & $17$&$\pm4$ &  \ref{B2324+60}\\
B2327$-$20 &    & 1.6436 & $ 4.6\cdot10^{-15}$ & 1024 & 125  & 0.55$\pm$0.04 & 0.15 & & & & &  \ref{B2327-20}\\
B2334+61 &    & 0.4953 & $ 1.9\cdot10^{-13}$ & 3085 & 19  & 1.3$\pm$0.2 & 1.07 & & & & &  \ref{B2334+61}\\
J2346$-$0609 & Dif$^\ast$  & 1.1815 & $ 1.4\cdot10^{-15}$ & 770 & 79  & 0.77$\pm$0.04 & 0.22 & $ -90\!$&$^{+25}_{-120}$ & $ 2.3$&$\pm0.1$ &  \ref{J2346-0609}\\
B2351+61 & Dif$^\ast$  & 0.9448 & $ 1.6\cdot10^{-14}$ & 7655 & 661  & 1.07$\pm$0.01 & 0.07 & $ 60\!$&$^{+25}_{-8}$ & $ 17$&$\pm6$ &  \ref{B2351+61}\\
\caption{continued.}\\
\end{longtable}

\clearpage
\twocolumn

\clearpage
\appendix \onecolumn {\Large Astro-ph version is missing 187 figures
due to file size restrictions. Please download appendices from {\tt
http://www.science.uva.nl/$\sim$wltvrede/21cm.pdf}. }
\section{Figures}
\label{Figures_ref}
\begin{figure}[hb!] \caption{
\label{B0011+47}
\label{B0031-07}
\label{B0037+56}
\label{B0105+65}
\label{J0134-2937}
\label{B0136+57}
\label{B0138+59}
\label{B0149-16}
\label{B0154+61}
\label{B0320+39}
}
\caption{
\label{B0353+52}
\label{B0355+54}
\label{B0450-18}
\label{B0458+46}
\label{B0540+23}
\label{B0559-05}
\label{B0609+37}
\label{B0611+22}
\label{B0621-04}
\label{B0626+24}
}
\caption{
\label{B0628-28}
\label{B0656+14}
\label{B0740-28}
\label{B0756-15}
\label{B0809+74}
\label{B0818-13}
\label{B0820+02}
\label{B0823+26}
\label{B0906-17}
\label{B0919+06}
}
\caption{
\label{B0950+08}
\label{J1022+1001}
\label{B1254-10}
\label{B1508+55}
\label{J1518+4904}
\label{B1541+09}
\label{B1600-27}
\label{B1612+07}
\label{B1642-03}
\label{B1649-23}
}
\caption{
\label{J1650-1654}
\label{B1706-16}
\label{J1713+0747}
\label{B1717-29}
\label{B1717-16}
\label{J1730-2304}
\label{B1730-22}
\label{B1732-07}
\label{B1736-29}
\label{B1737+13}
}
\caption{
\label{B1744-24A}
\label{B1745-12}
\label{B1749-28}
\label{B1754-24}
\label{B1756-22}
\label{J1757-2223}
\label{B1758-29}
\label{B1758-23}
\label{B1800-21}
\label{B1804-08}
}
\caption{
\label{B1805-20}
\label{J1808-0813}
\label{B1811+40}
\label{J1812-2102}
\label{B1813-17}
\label{B1815-14}
\label{B1817-13}
\label{B1818-04}
\label{B1819-22}
\label{B1820-11}
}
\caption{
\label{B1821-19}
\label{B1821-11}
\label{B1821+05}
\label{B1822-14}
\label{B1826-17}
\label{J1828-1101}
\label{B1829-08}
\label{J1830-1135}
\label{B1830-08}
\label{B1831-04}
}
\caption{
\label{B1831-03}
\label{B1832-06}
\label{B1834-10}
\label{B1834-04}
\label{J1835-1106}
\label{J1839-0643}
\label{B1839+09}
\label{B1839+56}
\label{B1841-05}
\label{B1841-04}
}
\caption{
\label{B1842-04}
\label{B1842+14}
\label{B1844-04}
\label{J1845-0743}
\label{B1845-01}
\label{B1846-06}
\label{B1848+13}
\label{B1849+00}
\label{J1850+0026}
\label{J1852-2610}
}
\caption{
\label{J1852+0305}
\label{B1855+02}
\label{B1855+09}
\label{B1859+03}
\label{B1859+07}
\label{B1900+01}
\label{B1900+05}
\label{B1903+07}
\label{B1905+39}
\label{B1907+00}
}
\caption{
\label{B1907+10}
\label{B1910+20}
\label{B1911-04}
\label{B1911+13}
\label{B1914+09}
\label{B1914+13}
\label{B1915+13}
\label{B1916+14}
\label{B1917+00}
\label{B1918+19}
}
\caption{
\label{B1920+21}
\label{B1924+16}
\label{B1929+10}
\label{B1933+16}
\label{B1935+25}
\label{B1937-26}
\label{B1937+21}
\label{B1943-29}
\label{B1944+17}
\label{B1953+50}
}
\caption{
\label{B2000+32}
\label{B2002+31}
\label{B2003-08}
\label{B2011+38}
\label{B2021+51}
\label{B2022+50}
\label{B2043-04}
\label{B2044+15}
\label{B2045-16}
\label{B2053+36}
}
\caption{
\label{B2106+44}
\label{B2110+27}
\label{J2145-0750}
\label{B2148+52}
\label{B2148+63}
\label{B2217+47}
\label{B2224+65}
\label{B2255+58}
\label{B2303+30}
\label{B2306+55}
}
\caption{
\label{B2315+21}
\label{B2323+63}
\label{B2324+60}
\label{B2327-20}
\label{B2334+61}
\label{B2351+61}
}
\caption{
\label{B0052+51}
\label{B0144+59}
\label{B0148-06}
\label{B0301+19}
\label{B0329+54}
}
\caption{
\label{B0402+61}
\label{B0450+55}
\label{B0523+11}
\label{B0525+21}
\label{B0531+21}
}
\caption{
\label{B0751+32}
\label{B0834+06}
\label{B1039-19}
\label{B1112+50}
\label{B1133+16}
}
\caption{
\label{B1237+25}
\label{B1540-06}
\label{B1604-00}
\label{B1702-19}
\label{B1738-08}
}
\caption{
\label{B1753+52}
\label{B1822-09}
\label{B1839-04}
\label{B1857-26}
\label{J1901-0906}
}
\caption{
\label{B1919+21}
\label{B1946+35}
\label{B1952+29}
\label{B2000+40}
\label{B2016+28}
}
\caption{
\label{B2020+28}
\label{B2111+46}
\label{B2154+40}
\label{B2310+42}
\label{B2319+60}
}
\caption{
\label{J2346-0609}
}
\end{figure}

%\clearpage
\section{Figures ordered by appearance in text}
\label{Figures_ref2}

\end{document}